\documentclass[iop,revtex4]{emulateapj}
\usepackage{natbib}
\usepackage{morefloats}
\usepackage{epsfig}
\usepackage{epsf} 
\usepackage{rotating}

\slugcomment{Submitted to The Astrophysical Journal Supplement Series}

\shorttitle{A NIR spectroscopic survey of star-forming galaxies at $z\sim1.6$}
\shortauthors{Silverman et al.}

\begin{document}


\title{The FMOS-COSMOS survey of star-forming galaxies at $z\sim 1.6$ III. Survey design, performance, and sample characteristics} 



\author{J.~D.~Silverman\altaffilmark{1}, D.~Kashino\altaffilmark{2}, D.~Sanders\altaffilmark{3}, J.~S.~Kartaltepe\altaffilmark{4,5}, N.~Arimoto\altaffilmark{6}, A.~Renzini\altaffilmark{7}, G.~Rodighiero\altaffilmark{8}, E.~Daddi\altaffilmark{9}, J. Zahid\altaffilmark{3}, T.~Nagao\altaffilmark{10}, L.~J.~Kewley\altaffilmark{3,11}, S.~J.~Lilly\altaffilmark{12}, N.~Sugiyama\altaffilmark{1,2}, I. Baronchelli\altaffilmark{8}, P.~Capak\altaffilmark{13}, C.~M.~Carollo\altaffilmark{12}, J.~Chu\altaffilmark{3}, G.~Hasinger\altaffilmark{3}, O.~Ilbert\altaffilmark{14}, S. Juneau\altaffilmark{9}, M.~Kajisawa\altaffilmark{10}, A.~M.~Koekemoer\altaffilmark{15}, K.~Kovac\altaffilmark{12}, O.~ Le F\`{e}vre\altaffilmark{14}, D.~Masters\altaffilmark{16}, H.~J.~McCracken\altaffilmark{17}, M.~Onodera\altaffilmark{12}, A. Schulze\altaffilmark{1}, N.~Scoville\altaffilmark{18}, V.~Strazzullo\altaffilmark{7}, Y.~Taniguchi\altaffilmark{10}}

\email{john.silverman@ipmu.jp}



\altaffiltext{1}{Kavli Institute for the Physics and Mathematics of the Universe (WPI), The University of Tokyo Institutes for Advanced Study, The University of Tokyo, Kashiwa, Japan 277-8583}
\altaffiltext{2}{Division of Particle and Astrophysical Science, Graduate School of Science, Nagoya University, Nagoya, 464-8602, Japan}
\altaffiltext{3}{Institute for Astronomy, University of Hawaii, 2680 Woodlawn Drive, Honolulu, HI, 96822}
\altaffiltext{4}{National Optical Astronomy Observatory, 950 N. Cherry Ave., Tucson, AZ, 85719}
\altaffiltext{5}{School of Physics and Astronomy, Rochester Institute of Technology, 84 Lomb Memorial Drive, Rochester, NY 14623, USA}
\altaffiltext{6}{Subaru Telescope, 650 North A'ohoku Place, Hilo, Hawaii, 96720, USA}
\altaffiltext{7}{Instituto Nazionale de Astrofisica, Osservatorio Astronomico di Padova, vicolo dell'Osservatorio 5, I-35122, Padova, Italy, EU}
\altaffiltext{8}{Dipartimento di Fisica e Astronomia, Universita di Padova, vicolo Osservatorio, 3, 35122, Padova, Italy}
\altaffiltext{9}{Laboratoire AIM, CEA/DSM-CNRS-Universite Paris Diderot, Irfu/Service d'Astrophysique, CEA Saclay}
\altaffiltext{10}{Graduate School of Science and Engineering, Ehime University, 2-5 Bunkyo-cho, Matsuyama 790-8577, Japan}
\altaffiltext{11}{Research School of Astronomy and Astrophysics, The Australian National University, Cotter Road, Weston Creek, ACT 2611}
\altaffiltext{12}{Institute of Astronomy, ETH Z\"urich, CH-8093, Z\"urich, Switzerland.}
\altaffiltext{13}{Spitzer Science Center, California Institute of Technology, Pasadena, CA 91125, USA}
\altaffiltext{14}{Aix Marseille Universit\'e, CNRS, LAM (Laboratoire d'Astrophysique de Marseille) UMR 7326, 13388, Marseille, France}
\altaffiltext{15}{Space Telescope Science Institute, 3700 San Martin Drive, Baltimore MD 21218, U.S.A.}
\altaffiltext{16}{The Observatories of the Carnegie Institution for Science, 813 Santa Barbara Street, Pasadena, CA 91101, USA}
\altaffiltext{17}{Institut d'Astrophysique de Paris, Universit\'e Pierre et Marie Curie - Paris 6, 98 bis Boulevard Arago, F-75014 Paris, France}
\altaffiltext{18}{California Institute of Technology, MC 105-24, 1200 East California Boulevard, Pasadena, CA 91125}


\begin{abstract}

We present a spectroscopic survey of galaxies in the COSMOS field using the Fiber Multi-Object Spectrograph (FMOS), a near-infrared instrument on the Subaru Telescope.  Our survey is specifically designed to detect the H$\alpha$ emission line that falls within the $H$-band (1.6-1.8 $\mu$m) spectroscopic window from star-forming galaxies with $1.4 < z < 1.7$ and $M_{\rm stellar}\gtrsim10^{10}$ M$_{\odot}$.  With the high multiplex capability of FMOS, it is now feasible to construct samples of over one thousand galaxies having spectroscopic redshifts at epochs that were previously challenging.  The high-resolution mode ($R\sim2600$) effectively separates H$\alpha$ and [NII]$\lambda6585$ thus enabling studies of the gas-phase metallicity and photoionization state of the interstellar medium.  The primary aim of our program is to establish how star formation depends on stellar mass and environment, both recognized as drivers of galaxy evolution at lower redshifts.  In addition to the main galaxy sample, our target selection places priority on those detected in the far-infrared by $Herschel$/PACS to assess the level of obscured star formation and investigate, in detail, outliers from the star formation rate - stellar mass relation.  Galaxies with H$\alpha$ detections are followed up with FMOS observations at shorter wavelengths using the J-long (1.11-1.35 $\mu$m) grating to detect H$\beta$ and [OIII]$\lambda$5008 that provides an assessment of extinction required to measure star formation rates not hampered by dust, and an indication of embedded Active Galactic Nuclei.  With 460 redshifts measured from 1153 spectra, we assess the performance of the instrument with respect to achieving our goals, discuss inherent biases in the sample, and detail the emission-line properties.  Our higher-level data products, including catalogs and spectra, are available to the community.
\end{abstract}





\section{Introduction}

Over the last few years, a clearer picture has started to emerge on how galaxies are built up with cosmic time. These developments are primarily afforded by the remarkable wealth of multi-wavelength imaging that telescopes have provided, from both space and the ground, together with optical spectroscopic surveys (e.g., SDSS, VVDS, zCOSMOS, DEEP2) that have amassed samples ranging from several thousands to over a million galaxies. This continuously expanding database provides a map of the galaxy distribution up to $z\sim2$ (and beyond), and enables a measure of their intrinsic properties such as their stellar mass, ongoing star formation rate (SFR) and information on their stellar populations such as ages and the chemical enrichment of the interstellar medium.  As a consequence, it has been recognized that at least to $z\sim 3$ most galaxies are either star-forming or had their star formation quenched while subsequently evolving in a passive mode.  The fraction of the latter rapidly drops beyond $z\sim1 $ although most galaxies at high masses (M$_{\rm stellar}\gtrsim10^{11}$ M$_{\odot}$) are quenched at least up to $z\sim2$ \citep[e.g.,][]{Williams2009, Cassata2013}.  On the other hand, most star-forming galaxies cluster around a relatively tight SFR-stellar mass relation, that came to be called the {\it main sequence} (MS) of star-forming galaxies \citep[e.g.,][]{Noeske2007,Elbaz2007,Daddi2007,Wuyts2011,Speagle2014}, whereas relatively few outliers, towards high specific SFR from the main sequence  ($\sim 2\%$ by number), appear to be experiencing a starburst episode \citep{Rodighiero2011,Sargent2012}, possibly triggered by a merging event \citep[e.g.,][]{Kartaltepe2012}.  For both the star-forming and quenched populations, the mass functions have been measured with fair accuracy all the way to such high redshifts \citep[e.g.,][]{Bundy2006,Fontana2009,Ilbert2010,Ilbert2013,Grazian2015} and the fraction of quenched galaxies has been mapped as a function of stellar mass and, up to $z\sim 1$, also of local environment \citep{Peng2010,Peng2012,Scoville2013}.

At $z>1$, our understanding of the influence of the local environment on star-forming galaxies, based on spectroscopic surveys, is mainly limited to small samples of groups or massive clusters.  However, studies based on large spectroscopic efforts have begun to construct significant samples of galaxy over-densities at these higher redshift \citep[][]{Gerke2012,Diener2013} using surveys such as zCOSMOS and DEEP2.  Although, there has been a wide gap at $1.4 \le z \le 1.8$, namely the {\it redshift desert} as it came to be called, which historically proved to be the most difficult redshift regime for optical spectroscopic surveys to penetrate \citep[see Figure 13 of][]{Lefevre2015}. This is primarily the result  of strong spectral features (e.g., the 4000 ${\rm \AA}$ break and emission lines such as H$\alpha$, [OIII]$\lambda$5008 and [OII]$\lambda3727$) having moved to the near-IR, hence to measure redshifts optical spectroscopy had to rely only on weak absorption lines detected in rest-frame ultraviolet \citep[e.g.,][]{Lilly2007} or carry out deep and observationally expensive campaigns \citep{Cimatti2008,Lefevre2015}.  Until recently, this wide critical ``slice" of our Universe,  precisely at the peak cosmic epoch in galaxy growth and AGN activity, has not yet been adequately mapped by spectroscopic  surveys.  
  
\begin{figure}
\hspace{0.5cm}
\includegraphics[width=7.5cm,angle=-90]{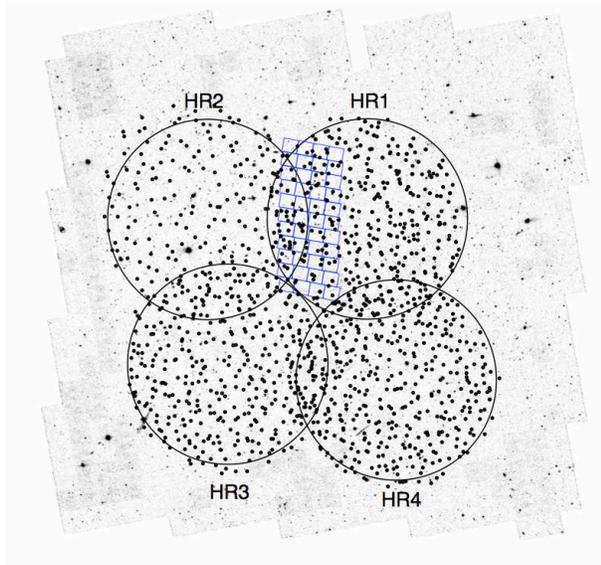}
\caption{Location of the four FMOS footprints overlaid on the HST/ACS mosaic of the COSMOS field.  The area with WFC3 NIR imaging from CANDELS is shown in blue.  Galaxies observed with FMOS using the H-long grating are indicated by small black dots.}
\label{layout}
\end{figure}

Near-infrared instruments on large-aperture telescopes with high multiplex capabilities are providing the means for surveys to fill such a gap in redshift space.  The Fiber Multi-Object Spectrograph \citep[FMOS;][]{Kimura2010} at the prime focus of the Subaru Telescope is designed specifically for this purpose.  With 400 fibers spread over a field-of-view of $30\arcmin$ in diameter, it is now feasible to carry out a redshift survey of a thousand galaxies, practically unfeasible with previous near-infrared instruments.  There also exists a complementary capability with HST/WFC3 in grism mode that benefits from avoiding atmospheric contamination; although, its field-of-view is substantially smaller than that of FMOS thus preventing surveys \citep[e.g.,][]{Wuyts2013,Price2014} from covering a large enough volume thus limiting the dynamic range in environment.  The CANDELS field within COSMOS is an example of this since the area coverage is smaller than a single FMOS pointing (Figure~\ref{layout}) thus not capable of including rare objects (i.e., massive galaxies and AGN).  Equally important, the high spectral resolution of FMOS is suitable to separate H$\alpha$ and [NII]$\lambda6550, 6585$ (not possible with the low resolution grism spectra from HST/WFC3), hence measuring gas phase metallicity and providing an indication of AGN activity when coupled with spectral coverage of H$\beta$ and [OIII]$\lambda$5008.  Other multi-object spectroscopic facilities such as KMOS on ESO's Very Large Telescope \citep{Sharples2013}, MOSFIRE \citep{McLean2012} on Keck, and MOIRCS \citep{Onodera2012} on Subaru nicely complement the wide-field capabilities of FMOS due to their high sensitivity, thus reaching the fainter galaxy population, although over a substantially smaller area.  

In this paper, we present the design and observational results of a large near-infrared spectroscopic survey targeting star-forming galaxies at $1.4 \lesssim z \lesssim 1.7$ in the COSMOS field using Subaru/FMOS.  This is a joint effort based on a ``Subaru Intensive Program'' (PI J. Silverman) approved by NAOJ and a program awarded through the University of Hawaii, Institute for Astronomy (PI D. Sanders).  The early H-band observations presented here include seven nights awarded by NOAJ to carry out a pilot study and a subsequent Intensive Program.  The observations acquired through UH have focused on the J-band followup of the positive H$\alpha$ detections from the H-band observations to provide the full suite of key emission lines of galaxies \citep{Kartaltepe2015}. 

Our broad aim is to advance our understanding of galaxy evolution at this most active epoch by obtaining accurate measurements of SFR, extinction, metallicity, AGN identification and providing a characterization of the environment that only spectroscopic redshifts can accomplish.  Our program makes use of the high-resolution mode to cleanly separate H$\alpha$ and [NII] as detected in the $H$-band.  Follow-up observations are also acquired for a subset of targets to detect H$\beta$ and [OIII]$\lambda$5008 in the $J$-band, thus providing an extinction correction based on the (1) H$\alpha$ to H$\beta$ ratio (i.e., Balmer decrement), (2) an indication of the photoionization state of the ISM, and (3) an improved measurement of metallicity.  While the first results are public \citep{Kashino2013,Zahid2014b,Rodighiero2014}, herein we fully detail the survey design including the observing strategy and target selection.  We explicitly demonstrate inherent biases in the sample such as that induced by a limit on the expected H$\alpha$ flux that was imposed a priori to increase the likelihood of an emission-line detection.  With 1153 FMOS spectra in hand, we present the effectiveness of our target selection as a function of key parameters and provide a catalog of emission-line properties including spectroscopic redshifts with a detailed assessment of their accuracy.  With respect to the latter point, the zCOSMOS deep sample \citep{Lilly2007} allows us to perform an independent check on the FMOS redshifts.  The paper is organized with the following sections: survey design (Section 3), target selection (Sections 4 and 5), induced selection biases (Section 6), observations and data reduction (Section 7), emission-line fitting (Section 8), determination of spectroscopic redshifts (Section 9), description of our catalog (Section 10), and emission-line properties (Section 11).  Finally, we conclude with our plans to acquire a final sample of over 1000 galaxies in COSMOS with $1.4 \lesssim z \lesssim 1.7$ having spectroscopic redshifts (based on FMOS high-resolution spectra) thus enabling us to significantly improve upon the density field and our investigation of galaxy properties as a function of environment.  We also highlight that our earlier and complementary effort to build a large sample of galaxies with IR spectra using FMOS in low-resolution mode in the COSMOS field is well underway \citep{Matsuoka2013,Kartaltepe2015} with the full details of the survey to be presented in a forthcoming paper (Kartaltepe et al. in preparation).  Throughout this work, we assume $H_0=70 $ km s$^{-1}$ Mpc$^{-1}$, $\Omega_{\Lambda}=0.75$, $\Omega_{\rm{M}}=0.25$, AB magnitudes and a Salpeter initial mass function.

\section{Near-infrared spectroscopy with Subaru/FMOS in high-resolution mode}

The Fiber Multi-Object Spectrograph (FMOS; \citealt{Kimura2010}) on the Subaru Telescope is paving the way for surveys of high-redshift galaxies \citep{Yabe2012,Yabe2014,Roseboom2012,Okada2015,Tonegawa2015} and AGNs \citep{Nobuta2012,Matsuoka2013,Brightman2013}.  Such progress is afforded by the ability to simultaneously acquire near-infrared spectra for about 200 galaxies (over a circular region of 30$\arcmin$ in diameter) when implementing the cross-beam switching (CBS) mode, a technique that dithers targets between fiber pairs effectively measuring the sky background spatially close to individual objects and through the same fibers as the science targets.  Each fiber is 1.2$\arcsec$ in diameter ideal for high-redshift targets observed under typical seeing conditions ($\sim0.7\arcsec$) at the summit of Mauna Kea.  An OH-airglow suppression filter \citep{Maihara1994,Iwamuro2001} masks out strong atmospheric emission lines that usually plague the J-band and H-band at the expense of losing $\sim25\%$  of spectral coverage.  This fraction of lost pixels will come into the discussion later to explain the number of non-detections with respect to the H$\alpha$ emission line.

While some of the early studies mentioned above were carried out in a low spectral resolution mode, FMOS has been operational in a high-resolution mode since the spring of 2012 (S12A).  There is an overall improvement in system throughput by a factor of $\sim$3 due to the removal of the VPH grating required for low-resolution observations.  The spectral resolution in high-resolution mode is $\lambda  / \Delta\lambda\approx2600$\footnote{We find that the spectral resolution $\lambda/ \Delta\lambda$ achieved with the H-long grating is closer to 3000 based on the Th-Ar calibration images as opposed to 2600 as reported in \citet{Kimura2010}.} in the H-band that cleanly separates H$\alpha$ and [NII]$\lambda6585$ and has a velocity resolution $FWHM\sim115$ km s$^{-1}$.  Four high-resolution gratings are designed to provide contiguous spectral coverage over the full J (J-short: 0.92-1.12 $\mu$m, J-long: 1.11-1.35 $\mu$m) and H (H-short: 1.40-1.60 $\mu$m, H-long: 1.60-1.80 $\mu$m) bands.  We refer the reader to \citet{Kimura2010} for further details of the instrument and its performance.

\begin{deluxetable}{lll}
\tabletypesize{\scriptsize}
\tablecaption{Location of the FMOS footprints \label{fmos_pointings}}
\tablehead{\colhead{Name}&\colhead{RA}&\colhead{Dec}}
\startdata
HR1&09:59:56.0&+02:22:14\\
HR2&10:01:35.0&+02:24:53\\
HR3&10:01:19.7&+02:00:30\\
HR4&09:59:38.7&+01:58:08
\enddata
\end{deluxetable}

\section{Survey strategy}

The FMOS-COSMOS survey is an effort to acquire a large number of near-infrared spectra of star-forming galaxies with Subaru/FMOS that will yield a final sample of at least $\sim$1000 galaxies, for the high-resolution component, having spectroscopic redshifts between $1.4\lesssim z\lesssim1.7$.  As a result of the performance of FMOS, our survey is essentially an emission-line survey;  as will be demonstrated, the detection of the continuum is out of reach for the majority of the galaxies at these redshifts in a reasonable amount of exposure time.  

We designate the central square degree\footnote{A second intensive program (PI Silverman) is now underway to cover the outer area of the COSMOS field that should be completed in early 2016.} of the COSMOS field as our spectroscopic survey area to benefit from the rich multi-wavelength efforts.  In particular, the zCOSMOS Deep program \citep{Lilly2007} has acquired $\sim$500 quality spectroscopic redshifts within our redshift range of interest.  With a field-of-view of 0.19 square degrees, four FMOS footprints cover the central area (Figure~\ref{layout} and Table~\ref{fmos_pointings}).  Each footprint is observed multiple times to increase the spatial sampling necessary to measure galaxy clustering with projected pair separations reaching down to $\sim$200 kpc (Kashino et al. in preparation).  Galaxies slightly outside the FMOS field-of-view, shown in Figure~\ref{fmos_pointings}, are targeted as a result of the fibers having a considerable angular patrol area.      

\subsection{Initial detection of H$\alpha$}

The H$\alpha$ emission line is the key spectral feature for redshift determination and an indicator of star formation \citep{Kennicutt1998}.  With the H-long grating, we observe a wavelength range of 1.6 to 1.8 $\mu$m, capable of detecting H$\alpha$ for galaxies with $1.43 < z < 1.74$.  This redshift range is chosen specifically to take advantage of the fact that the H-long grating has the highest throughput among the FMOS gratings \citep[see Figure 19 of][]{Kimura2010}.  The detection of [NII]$\lambda$6585 emission lends further confidence in the redshift measurement and is used to estimate the gas-phase metallicity.  The width of the line profile yields insight into the kinematics of the line-emitting gas as a result of our velocity resolution.  For example, velocity-broadened lines may further indicate the presence of galactic outflows or underlying AGN as identified by components having velocity widths substantially higher than expected for star-forming galaxies.  

Our program is designed to complete at least 12 individual FMOS pointings using the H-long grating that covers each footprint with three or more passes.  We observe each FMOS configuration for the maximal amount of time possible in one night with on-source exposure times ranging from 3 to 5 hours.  Broadly speaking, the observational campaign is tuned to identify H$\alpha$ emission in about 1000 galaxies from a total of 2400 spectra based on 200 science targets per FMOS configuration and a success rate (44\%) as assessed in Section~\ref{sec:success}.

\subsection{Spectral coverage of H$\beta$ and [OIII]$\lambda$5008}

A second layer of our survey is to re-observe galaxies (with positive H$\alpha$ detections) using the J-long grating, that covers a wavelength range of 1.11-1.35 $\mu$m, to detect the H$\beta$ and [OIII]$\lambda5008$ emission lines.  To alleviate the substantial demand on Subaru for this purpose, we are collaborating with the Institute for Astronomy at the University of Hawaii Manoa (PI Dave Sanders) to supplement our observations acquired through the intensive program.  Exposure times are comparable to the H-long observations.  The detection of additional emission lines provides a consistency check on the accuracy of the spectroscopic redshifts.  Equally important, the detection of both [OIII]$\lambda$5008 and H$\beta$ are vital ingredients for determining the oxygen enrichment of the interstellar medium \citep{Zahid2014a,Zahid2014b} and identifying galaxies with line ratios indicative of AGN photoionization \citep{Kewley2013a,Kewley2013b,Kartaltepe2015}.  

Specific to these J-long observations, the remaining free fibers are dedicated to a low-redshift ($0.7 \lesssim z \lesssim 1.0$) extension of this study.  Galaxies are targeted with the aim of identifying H$\alpha$+[NII] in emission for (1) those primarily having IR detections with $Spitzer$ (24 or 70 $\mu$m) and $Herschel$ (100 or 160 $\mu$m).  The remaining fibers are placed on X-ray selected AGNs in a manner similar to those described in Section~\ref{agn_select} with the only difference being the lower redshift range.  The full details of this low-redshift sample will be presented in a subsequent paper (Kartaltepe et al. in preparation).        

As demonstrated in subsequent sections, the detection of H$\beta$ is challenging, especially given the sensitivity of FMOS, thus we rely on stacking spectra for many scientific investigations.  \citet{Kashino2013} demonstrate the effectiveness of this approach while binning in stellar mass, star formation rate and color excess to determine whether the average extinction levels are in agreement with estimates from broad-band photometry including the far-infrared emission observed by $Herschel$ \citep{Rodighiero2014}.  Even so, we do have a sizable sample (112) of galaxies with all four lines detected (see Section~\ref{sec:otherlines} for details).
\\
\section{Primary target selection: star-forming galaxies at $z\sim1.6$}

The COSMOS photometric catalogs \citep{Capak2007,McCracken2010,McCracken2012,Ilbert2013} and those including the higher-level derived properties of individual galaxies are used for target selection.  For each galaxy, the broad-band photometry is fit using LePhare \citep{Arnouts2011} with population synthesis models \citep[BC03; see][for further details]{Ilbert2009,Ilbert2013} to derive accurate photometric redshifts that are crucial to establish their intrinsic properties such as stellar mass, SFR and color excess.  Our scheme for target selection of the main sample (i.e., massive, star-forming galaxies) is devised to achieve our scientific goals while maximizing the efficiency with respect to detecting H$\alpha$ in the FMOS IR spectroscopic window.  We list the specific selection limits applied to the parent photometric catalogs (further elaborated below) including refinements that have been implemented for the intensive program.  To reiterate, we use stellar masses and SFRs based on a Salpeter initial mass function.

\begin{itemize}

\item $K_s<23$ (AB mag), a limit set by the depth and completeness of CFHT WIRCam observations \citep{McCracken2010}.

\item $1. 43<z_{\rm phot}< 1.74$; range for which H$\alpha$ falls within the FMOS H-long spectral window. 

\item $M_{\rm stellar}\gtrsim10^{10}$ M$_{\odot}$

\item Observed colors ($B-z$ and $z-K$) satisfy the criteria of being  a star-forming (s)BzK \citep{Daddi2004}.

\item Expected H$\alpha$ flux above a given threshold of detectability (as detailed below)

\end{itemize}        

\begin{figure}
\epsscale{1.2}
\plotone{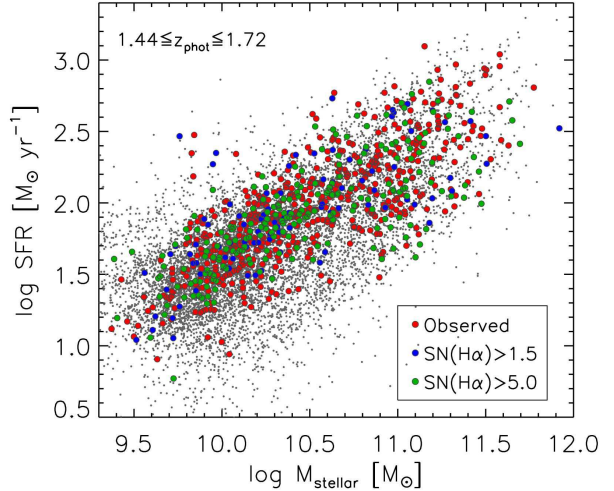}
\caption{SFR as a function of stellar mass for the targeted galaxy sample ($1.44\leq z_{\rm phot} \le 1.72$ and $K_s<23.5$).  All SFRs, shown here, are those returned from a SED fitting routine.  While a stellar mass limit is primarily imposed for the majority of the sample (M$_{\rm stellar}>10^{9.8}$ M$_{\odot}$), we have observed with FMOS some lower mass galaxies as fillers.  Galaxies in the parent catalog are shown as small circles while those observed with FMOS are marked in color indicative of whether a spectroscopic redshift was attained (green: H$\alpha$ S/N $>$ 5, blue: 1.5 $<$ H$\alpha$ S/N $<$ 5, red: H$\alpha$ S/N $<$ 1.5)}
\label{sed-sample}
\end{figure}

In Figure~\ref{sed-sample}, we show the distribution of stellar masses and SFRs for parent galaxy population and those that have been included in the FMOS observing program.  The above requirements on selection are set to ensure that the observed sample sufficiently spans the distribution of the main sequence of star-forming galaxies.  Even so, there are inherent selection biases that are subtle and non-negligible that we illustrate in Section~\ref{text:biases}.  Given our selection limit of $K=23$ and scaling from completeness estimates shown in Figure 2 of \citet{Ilbert2013}, we reach a completeness of 60\% (93\%) at $9.8< log~M_{\rm stellar} < 10.0$ (10.2-10.4).  It is also clear from Figure~\ref{sed-sample} that we have observed some galaxies below the mass limit given above and briefly describe these additional targets in Section~\ref{lowmass}.  On another point worth mentioning, we do not avoid targeting galaxies having successful or unsuccessful attempts to measure a redshift from other spectroscopic efforts such as the zCOSMOS deep program \citep{Lilly2007} in the optical.  Such additional selection criteria would only further complicate the selection function that may lead to inherent biases that are challenging to assess and properly remove.  Our comparisons, as presented below, between the FMOS and zCOSMOS results show a complementarity and further enable us to investigate the reasons for the redshift failures of both surveys.

\subsection{Predicting the H$\alpha$ emission}

We further select galaxies having a prediction of their H$\alpha$ emission-line flux (based on SFRs and dust corrections as derived from the broad-band photometry) above a specific threshold to maximize the success rate of line detection.  Using the FMOS exposure time calculator for an initial estimate, a threshold of $4\times10^{-17}$ erg s$^{-1}$ cm$^{-2}$ is set to detect a line (integrated across a profile of FWHM=300 km s$^{-1}$) with a significance greater than 3$\sigma$.  This emission-line limit is equivalent to a SFR of $\sim10$ M$_{\odot}$ yr$^{-1}$ for galaxies at $z\sim1.6$ \citep{Kashino2013}.  While a consideration of the spatial extent of the targets and the size of the fiber aperture is practical, we neglected to do so for the initial observations given various uncertainties at the time concerning the performance of the instrument.  

For the pilot study, SFRs are assessed using the rest-frame UV luminosity as determined from the $B_J$-band as done in \citet{Daddi2004}.  We restricted the targets to the `very good' sBzK where $\delta log[SFR(UV)]<0.13$ dex as done in \citet{Rodighiero2011} to ensure that our predicted H$\alpha$ fluxes were reliable for the initial performance verification of FMOS in high resolution mode with our sample.  The uncertainty on SFR(UV) includes a contribution from the error on the $B_J$ magnitude and the $B_J-z$ color used to correct for extinction.  This `very good' BzK sample represents $\sim$60\% of the full sBzK population where the missed objects are faint in the $B$-band and have large errors on their color thus likely are the most heavily extinct galaxies.

The impact of dust on the prediction of the emission-line flux is assessed from the color excess E$_{\rm star}(B-V)$ as derived from the rest-frame $B_J-z$ color \citep{Daddi2004} for each galaxy as mentioned above.  In addition, a multiplicative factor $f$ is applied to compensate for the differential extinction between stellar and line-emitting regions  $E_{\rm neb}(B-V)=E_{\rm star}(B-V)/f$.  Initially, we used the canonical value of $f=0.44$, appropriate for local star-forming galaxies \citep{Calzetti2000}.    

\begin{figure}
\epsscale{2.5}
\plottwo{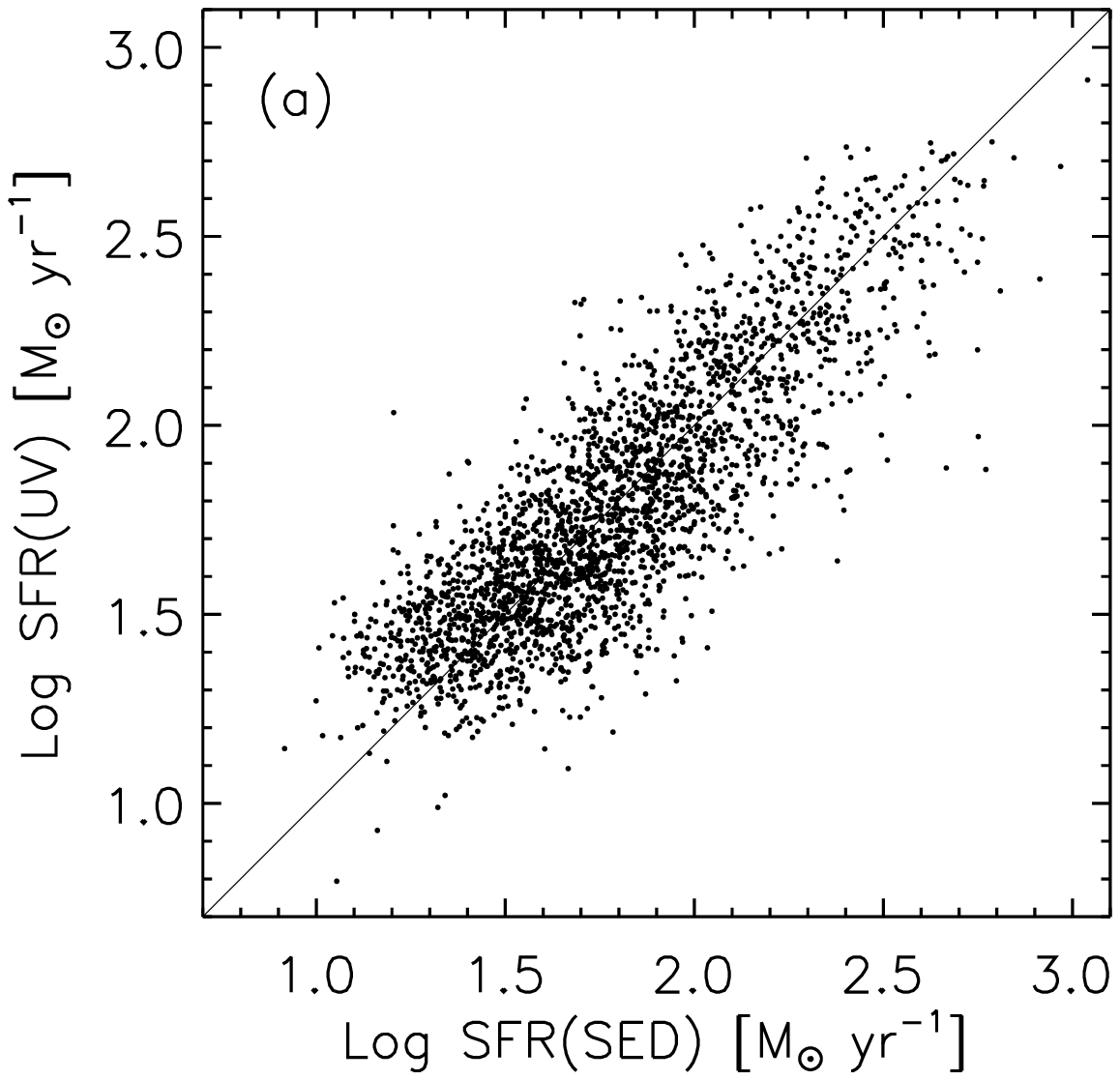}{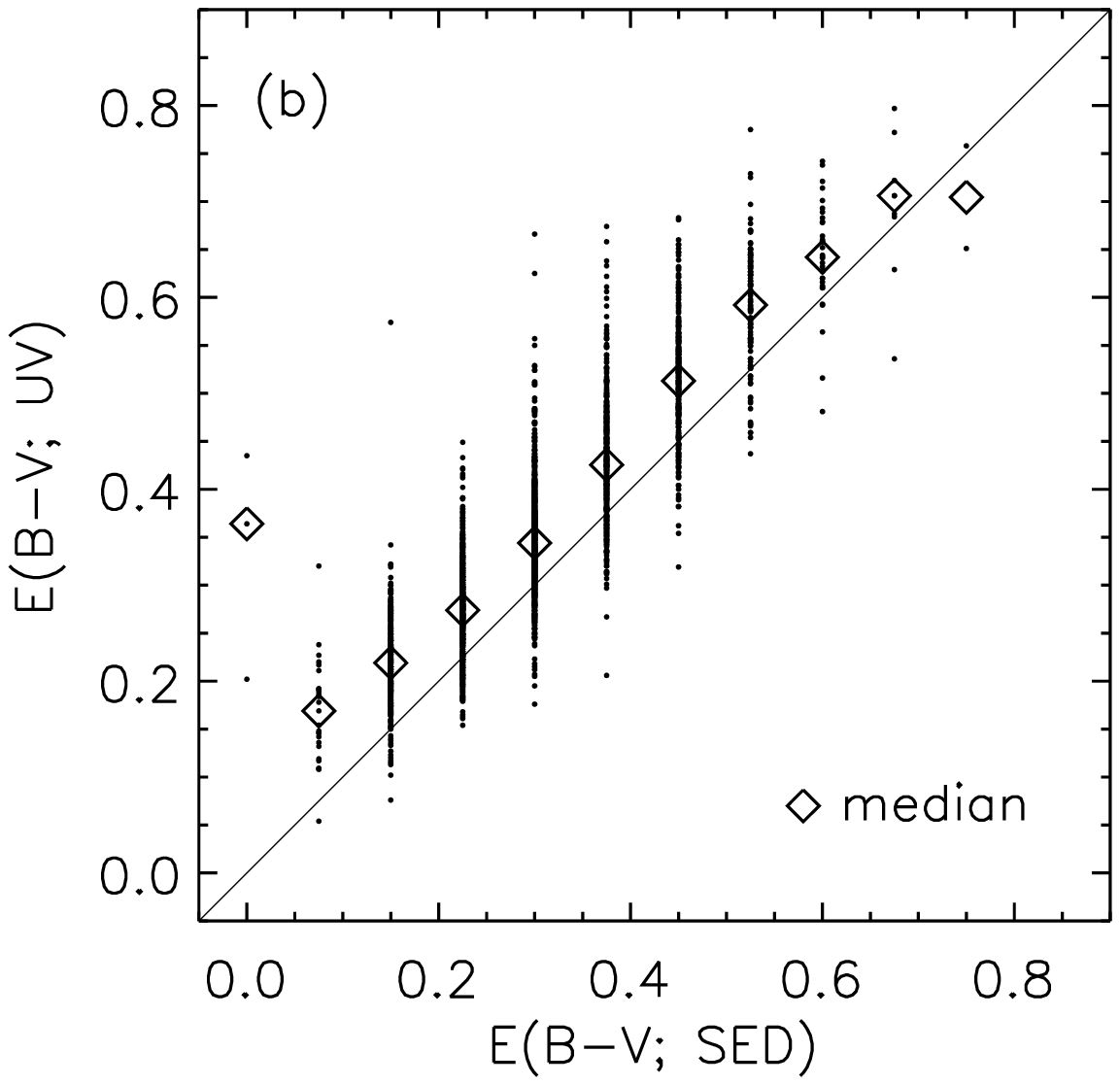}
\caption{$(a)$ Comparison of UV- and SED-based SFRs for the `very good' sBzK sample. $(b)$ Color excess as derived from $B_J-z$ (UV) color and broad-band SED fitting.}
\label{sfr_ext_comparison}
\end{figure}

\subsection{Refining the intensive program}

The pilot program allowed us to test the performance of the instrument and optimize our target selection for the subsequent intensive program.  Based on the quality of the resulting spectra and the availability of updated photometric galaxy catalogs, we implemented a number of changes to the selection of galaxy targets.  The most significant of these are (1) the implementation of a fainter limiting $K-band$ magnitude (23 to 23.5), (2) a higher cut on the predicted H$\alpha$ flux, and (3) an optimisation of the conversion factor of stellar to nebular extinction.  The first two modifications are coupled as a result of the deeper UltraVISTA observations \citep{McCracken2012} that generated a larger sample of galaxies from which we could select targets.  The predicted emission-line flux was raised from $4\times10^{-17}$ erg s$^{-1}$ cm$^{-2}$ to $6.8\times10^{-17}$ erg s$^{-1}$ cm$^{-2}$ {\it with a higher priority of assigning fibers to those with $f_{H\alpha}^{\rm pred}>1\times10^{-16}$ erg s$^{-1}$ cm$^{-2}$}.  The fainter K-band limit of 23.5 results in an increase in the completeness in terms of stellar mass: 80\% (99\%) at $9.8< log~M_{\rm stellar} < 10.0$ (10.2-10.4).  The higher emission-line limit increases the likelihood that we have the sensitivity to detect the light actually coming down the fiber since the majority of our galaxies extend beyond the fiber aperture.  This is effectively an aperture correction that had not been applied to the pilot observations.  As a final consideration, the conversion of stellar to nebular extinction changed based on our early analysis that pointed to a higher $f$-factor than previously implemented (see section~\ref{sec:results}).  For the intensive program, we use $f=0.66$ \citep{Kashino2013} as opposed to 0.44.

We made a number of additional changes to the selection.  With respect to predicting the level of H$\alpha$ emission, the intrinsic SFR for the intensive program is derived from a fit to the broad-band SED in contrast to using solely the $B-band$ luminosity (corrected for extinction).  This is based on the assumption of a constant star-formation history that has been shown to return a SFR similar to that derived from the rest-frame UV band \citep[e.g.,][]{Rodighiero2014}.  As shown in the top panel of Figure~\ref{sfr_ext_comparison}, there is a direct one-to-one relation between the SED- and the UV-based SFRs with scatter at the level of $\sigma=0.20$.  The color excess $E_{ star}(B-V)$ is also derived from the full SED fitting routine as opposed to being based on a single color $B-z$.  There is some concern that the $E_{\rm star}$(B-V) output from the SED fitting may not effectively discriminate between dust extinction and stellar age.  In the lower panel of the same figure, there is a clear correspondence in the the level of extinction assessed from both methods.  However, the stellar extinction based on a single color is slightly higher by about $\sim0.05$ than that returned from the global SED fit, although not severe enough to affect our predictions of the H$\alpha$ flux and impact our redshift success rate.  At the expense of slightly lowering the success rate, we perform no additional screening of the sample, such as considering errors on SFR, to refrain from further removing heavily obscured galaxies.

\begin{figure}
\epsscale{1.1}
\plotone{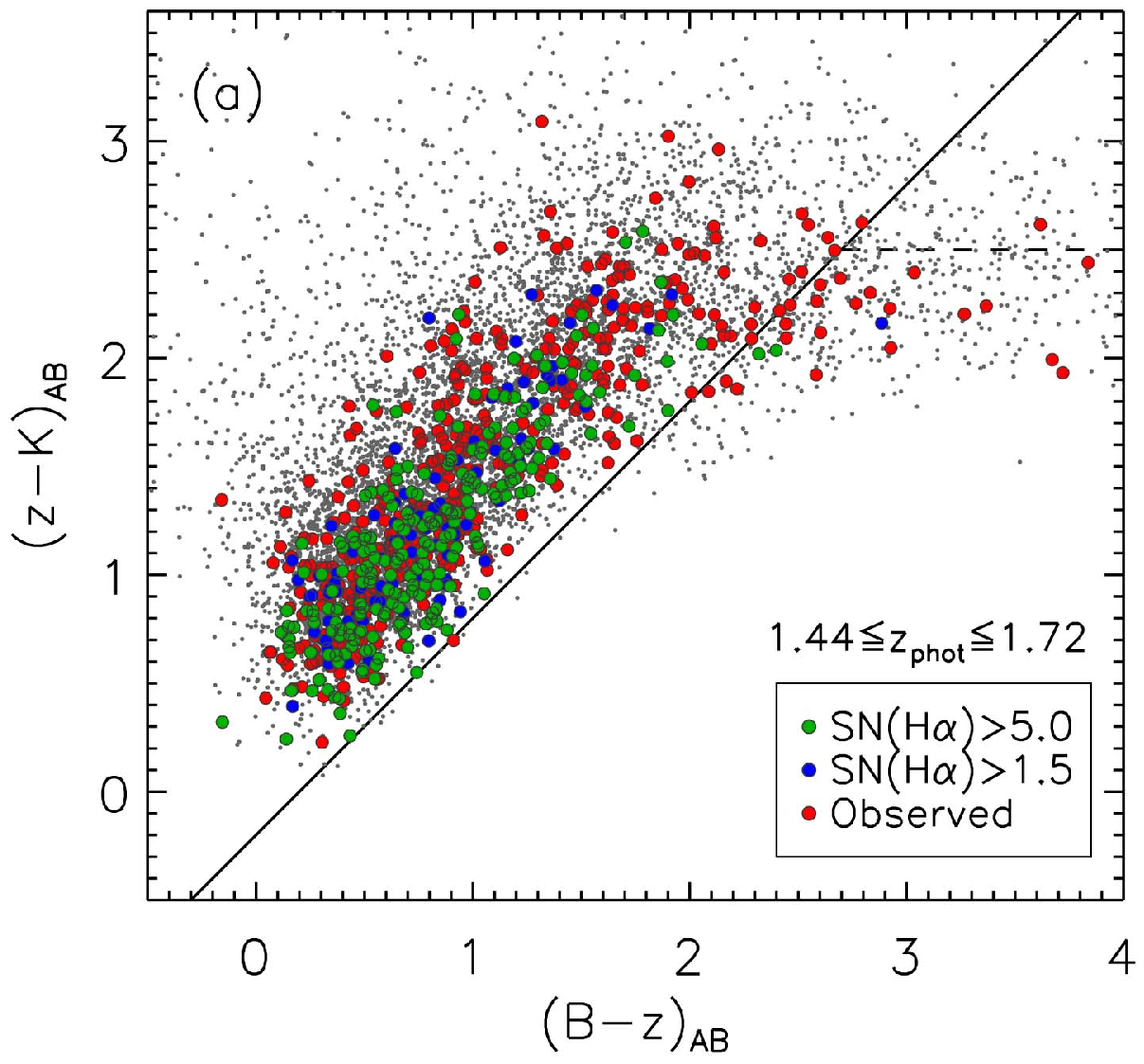}
\plotone{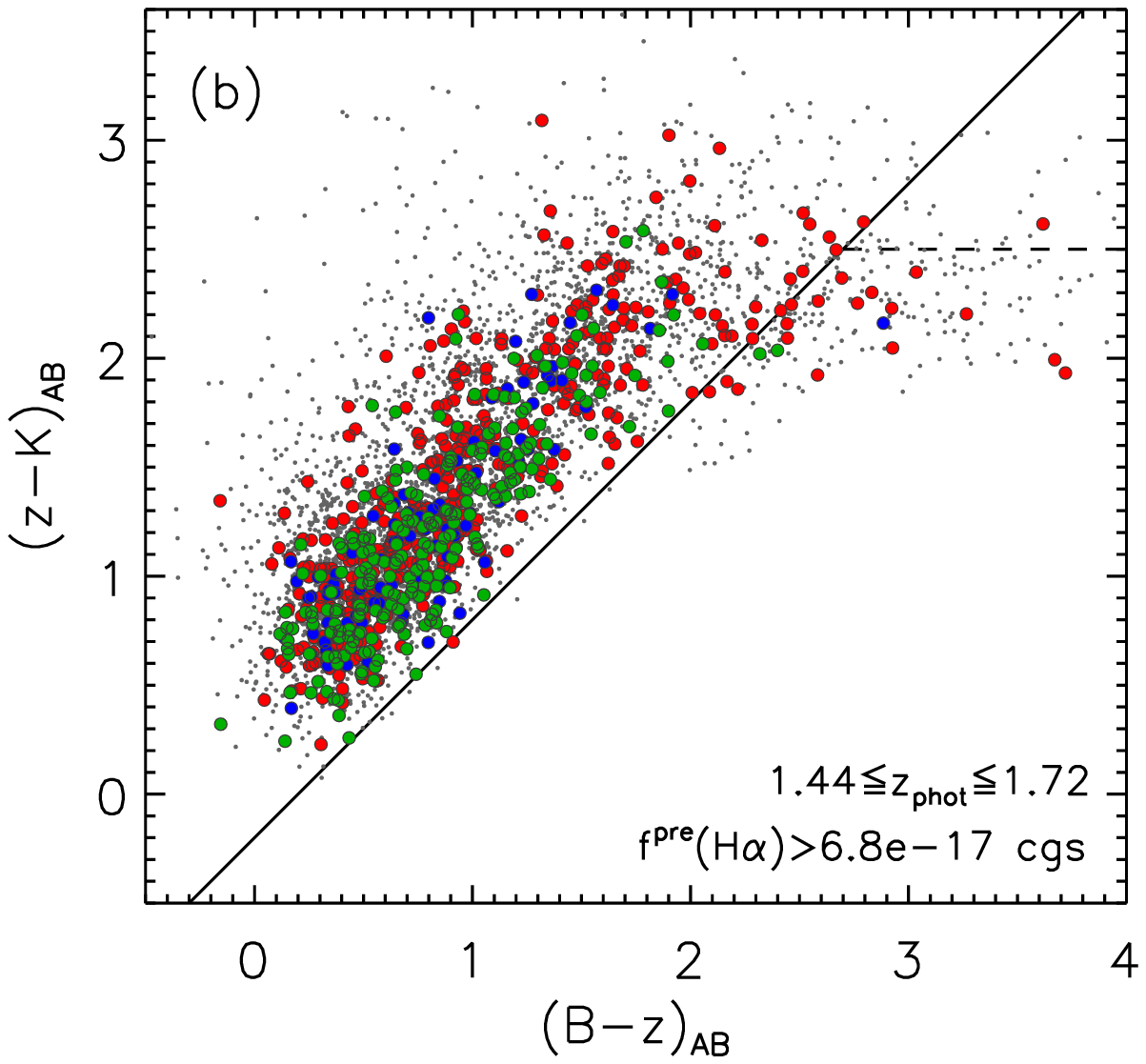}
\caption{SED-selected sample with $1.44\leq z_{\rm phot} \le 1.72$, M$_{\rm stellar}>10^{9.8}$ M$_{\odot}$, and $K_s<23.5$: Galaxies in the parent catalog are shown as small grey dots while those observed with FMOS are marked in color indicative of whether a spectroscopic redshift was attained (same labels as in Figure~\ref{sed-sample}).  The top panel presents the full sample while the bottom panel shows only those galaxies with predicted H$\alpha$ fluxes above our imposed higher limit of $6.8\times10^{-17}$ erg s$^{-1}$ cm$^{-2}$ thus targeted with FMOS.} 
\label{fig:bzk}
\end{figure}

We demonstrate that the SED-selected and sBzK samples are nearly equivalent in their characteristics.  For instance, an overwhelming majority of the SED-selected galaxies (within our redshift and stellar mass range of interest) satisfy the sBzK color selection (as a star-forming galaxy) as shown in Figure~\ref{fig:bzk}.  While there is a population of galaxies with $(B-z)\gtrsim2.5$ that falls out of the sBzK color space and below the region where the passive BzK galaxies lie ($z-K>2.5$), it has not yet been demonstrated that these galaxies have accurate photometric redshifts that would place them within an acceptable redshift range for which the BzK selection method is applicable.  For most of them, we have no firm confirmation of their redshifts through IR or optical spectroscopy.  Finally, the impact of imposing a limit of the expected H$\alpha$ emission-line flux does not obviously remove galaxies of a particular location in the $(B-z)$-$(z-K)$ plane (Fig.~\ref{fig:bzk}$b$).  

\section{Additional targets}

\subsection{FIR sources from the $Herschel$ PACS Evolutionary Probe (PEP) survey}

\begin{figure}
\epsscale{1.2}
\plotone{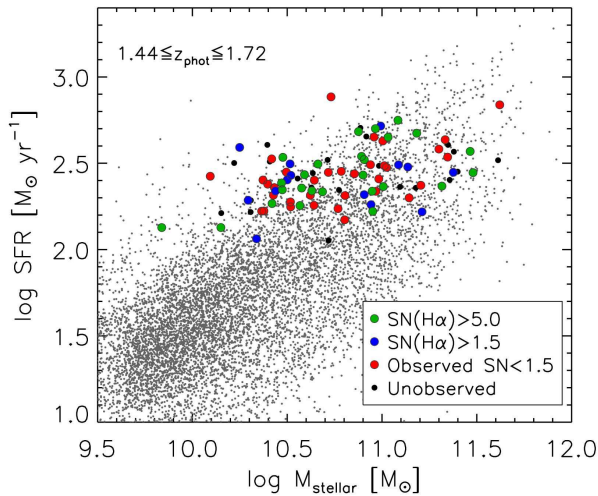}
\caption{$Herschel$/PACS targets shown as a function of SFR and stellar mass.  SFRs for PACS sources are based on 160$\mu$m fluxes.  Galaxies observed with FMOS are marked in color indicative of whether a spectroscopic redshift was attained (same labels as in Figure~\ref{sed-sample}).  $Herschel$/PACS sources that not observed are marked as black circles.  For comparison, we include the larger star-forming galaxy sample (small grey dots) with SFRs derived from SED fitting as described in the text.}
\label{herschel}
\end{figure}

The COSMOS field has been covered with $Herschel$-PACS imaging \citep{Lutz2011} at 100 $\mu$m (8 mJy) and 160 $\mu$m (17 mJy) with 5$\sigma$ detection limits as given in parenthesis.  These flux limits correspond to a SFR of $\sim100$ M$_{\odot}$ yr$^{-1}$ at $z=1.6$.  As further detailed in \citet{Rodighiero2011}, the PACS sources are matched with the IRAC-selected catalog of \citet{Ilbert2009} thus providing broad-band photometry from the UV to IR, accurate photometric redshifts, and stellar masses derived from SED fits.  SFRs are estimated as the sum of the UV and IR luminosity with the latter based on the integral of the best-fit SED in the 8-1000 $\mu$m rest-frame spectral window \citep[see][ for further details]{Rodighiero2010}.  In Figure~\ref{herschel}, we show the SFRs and stellar masses of the $Herschel$-PACS sources between $1.44<z_{\rm phot}<1.72$ and indicate those observed with FMOS.  It is clear that the PACS sources are in the minority, as compared to the full star-forming `main-sequence' population, thus require prioritization with respect to fiber assignment to target these interesting objects with sufficient statistics, especially those elevated well above ($>4\times$) the star-forming main sequence.  The FMOS spectra can aid in the evaluation as to whether AGN contribute to the high levels of FIR emission either through the detection of broadened emission lines or emission line ratios indicative of photoionization from an AGN.

\subsection{X-ray-selected Active Galactic Nuclei (AGN)}
\label{agn_select}

Near-infrared spectroscopic surveys of distant galaxies, hosting rapidly accreting supermassive black holes, are lacking in statistical power.  To alleviate this deficiency, we are using the high multiplex capabilities of FMOS to target the rarer galaxy population with AGN.  Therefore, we observe optical/near-infrared counterparts to X-ray sources from the $Chandra$ coverage of the central square degree of COSMOS \citep{Elvis2009,Puccetti2009} that have either spectroscopic or photometric redshift information \citep{Civano2012}.  While we began such an effort with FMOS in the low-resolution (LR; $R=600$; $0.9<\lambda<1.8\mu$m) mode \citep{Matsuoka2013,Brightman2013}, we aim to improve the NIR coverage at higher resolution for the following purposes.  For instance, we can detect broad (FWHM $>$ 2000 km s$^{-1}$) H$\alpha$ and H$\beta$ for fainter AGNs, likely of lower black hole mass than present in the LR program due to the increase in sensitivity.  On the issue of measuring virial black hole masses, we can establish how resolution impacts our characterization of the velocity profile of the broad-line emitting gas for AGN that have been observed using both the low and high resolution gratings.  Closely related to this issue, higher resolution spectra permit us to determine how well we are able to disentangle the broad and narrow lines, especially with respect to the H$\alpha$ line when using spectra of lower resolution.  Of equal interest, we target narrow-line (FWHM $<$ 2000 km s$^{-1}$) AGNs, primarily in observations carried out in 2013 through early 2014, to provide a unique data set for studying the physics of the narrow-line region including photoionization conditions \citep{Kewley2013a,Kewley2013b} and extinction properties that can be directly related to the X-ray obscuring medium \citep{Maiolino2001}.  We reserve the study of the rest-frame optical emission lines of the X-ray AGN to subsequent works, particularly since the line profiles are more complex than the star-forming galaxies.

\subsection{Low-mass galaxies}
\label{lowmass}

When possible, we placed a limited number of fibers on low mass star-forming galaxies ($1.4<z_{\rm phot}< 1.7$) to fill all the free fibers primarily during the pilot program (S12A).  While these were selected from \citet{Ilbert2009} to have $10^9<$ M$_*<10^{10}$ M$_{\odot}$, the restriction on the expected H$\alpha$ flux to be above $4\times10^{-17}$ erg s$^{-1}$ cm$^{-2}$ induced an effective lower mass cut around $M_*\sim10^{9.4}$ M$_{\odot}$.  We discontinued the observations of the low-mass galaxies after S12A to achieve a higher completeness for the more massive ($M_{\rm stellar}\gtrsim10^{10}$ M$_{\odot}$) galaxies. 

\begin{deluxetable*}{llllllll}
\tabletypesize{\scriptsize}
\tablecaption{Summary of Subaru/FMOS high-resolution observations\label{observations} of COSMOS in S12 A-B}
\tablewidth{0pt}
\tablehead{
\colhead{Date}&\colhead{Program}&\colhead{Period}&\colhead{Pointing}&\colhead{Grating}&\colhead{Total exp.}&\colhead{Effective}\\
\colhead{UT}&\colhead{ID}&&&&time\tablenotemark{a} (hr)&\colhead{seeing\tablenotemark{b} ($\arcsec$)}}
\startdata
13/03/2012&UH-B3&S12A&HR4&H-long&5.0&1.1\\
14/03/2012&S12A-096&S12A&HR1&H-long&5.0&0.9\\
15/03/2012&S12A-096&S12A&HR2&H-long&4.5&0.9\\
16/03/2012&S12A-096&S12A&HR1&H-long&5.0&1.0\\
17/03/2012&S12A-096&S12A&HR3&H-long&4.0&1.2\\
19/03/2012&UH-B5&S12A&HR1&J-long&4.5&1.4\\
29/12/2012&UH-18A&S12B&HR2&J-long&3.5&1.4\\
19/01/2013&S12B-045I&S12B&HR3&H-long&3.0&1.4\\
20/01/2013&S12B-045I&S12B&HR4&H-long&3.5&0.8\\
21/01/2013&UH-18A&S12B&HR3&J-long&4.5&1.5\\
22/01/2013&UH-18A&S12B&HR4&J-long&3.5&1.3
\enddata
\tablenotetext{a}{The total exposure time given here is the amount of science-grade data.}
\tablenotetext{b}{Effective seeing reported here is the smoothing scale required to match the FMOS and ground-based photometry (see Section~\ref{fluxcal} for details)}
\end{deluxetable*}

\begin{deluxetable*}{lllllllllll}
\tabletypesize{\scriptsize}
\tablecaption{Target galaxy samples and statistics\label{target_table}}
\tablehead{\colhead{Sample}&\colhead{$z_{phot}$}&\colhead{All}&\colhead{$M_\ast>M_\mathrm{lim}$\tablenotemark{a}}&\colhead{$f^\mathrm{exp}_\mathrm{H\alpha}> f^\mathrm{lim}_\mathrm{H\alpha}$\tablenotemark{b}}&\colhead{Observed}&\colhead{\# w/ $z_\mathrm{spec}$}&\colhead{Flag=1}&\colhead{Fl=2}&\colhead{Fl=3}&\colhead{Fl=4}}
\startdata
`Very good' sBzK& 1.4--1.7 & 2894 & 2885 & 1440 & 524 & 282 & 26 & 51 & 40 & 165 \\
SED-selected\tablenotemark{c} &1.44--1.72& 8928 & 7201 & 4054 & 739 & 349 & 29 & 68 & 53 & 199 \\
Herschel-PACS&1.44--1.72& 96 &&& 74 & 47 & 7 & 7 & 4 & 29 \\
X-ray AGN&&&&&127&45&4&4&10&27\\
Low-mass galaxies &1.4--1.7&& 11807 & 2246 & 427 & 183 & 20 & 38 & 41 & 84 
\enddata
\tablenotetext{a}{The limiting masses are the following: sBzK-$\log~M_\ast / M_\odot> 10.0$, SED-selected - $\log~M_\ast / M_\odot> 9.8$, Low-mass galaxies - $9.0 < \log~M_\ast / M_\odot< 10.0$}
\tablenotetext{b}{The limiting expected H$\alpha$ flux is based on $f=0.44$ for the sBzK and low-mass galaxy samples while $f=0.66$ for the SED-selected sample.}
\tablenotetext{c}{These targets include 199 galaxies selected by stellar mass \citep{Ilbert2009} and observed during a single night of UH time (PI J. Zahid) using the $H$-long grating.}
\end{deluxetable*}

\section{Induced sample characteristics and biases}
\label{text:biases}

\begin{figure*}
\epsscale{1.0}
\plotone{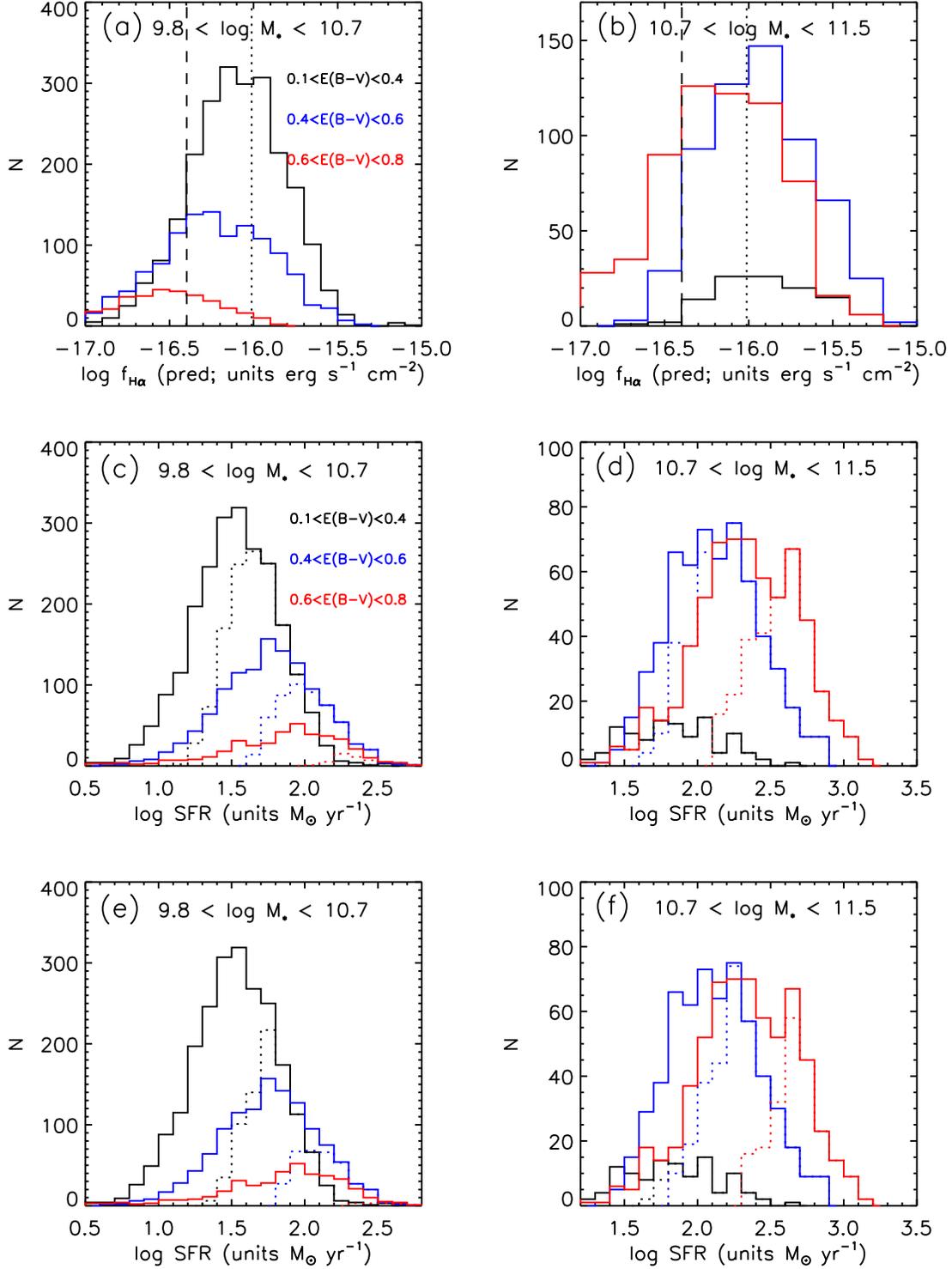}
\caption{Target selection and impact of a weight on the predicted emission-line flux based on the parent photometric galaxy sample between $1.46 < z < 1.67$: $(a)$ and $(b)$ Distribution of predicted H$\alpha$ emission-line flux (total; i.e., no consideration of the aperture size) split in bins of stellar mass and color excess as indicated.  The vertical dashed lines show our imposed limits on line flux (see text for details).  $(c-f)$ SFR distribution in bins of stellar mass and color excess.  The solid histograms are the full sample while the dashed histograms represent the sample with the imposition of a limit on H$\alpha$ emission-line flux (panels $c-d$: $f_{\rm H\alpha}>4\times10^{-17}$ erg s$^{-1}$ cm$^{-2}$; panels $e-f$: $f_{\rm H\alpha}>1\times10^{-16}$ erg s$^{-1}$ cm$^{-2}$) for each bin of color excess.}
\label{mass_select}
\end{figure*}

There are induced constraints on the characteristics of the observed sample due to the imposition of a limit on the H$\alpha$ flux expected to be detected with FMOS.  In particular, this will bias the sample by placing greater weight on galaxies having higher SFRs, and lower levels of extinction.  In Figure~\ref{mass_select}, we illustrate the magnitude of the effect by comparing the distributions of SFR (in bins of stellar mass and colour excess) for the parent and observed samples.  The top two panels ($a-b$) show the distribution of the expected total (i.e., galaxy-wide) H$\alpha$ flux split into two mass bins ($10<log~M_*<10.7$, $10.7<log~M_*<11.5$).  In each panel, the subsample is further divided in bins of color excess $E_{\rm star}(B-V)$, covering a range of 0.1 to 0.8.  In the low mass bin (panel $a$), the emission-line limit of $4\times10^{-17}$ erg s$^{-1}$ cm$^{-2}$ prevents most of the heavily dust-obscured galaxies ($0.6<E_{\rm star}(B-V)<0.8$) to be included as targets for FMOS observations and even more so when implementing a higher weighting for those with $f_{{\rm H}\alpha}>10^{-16}$ erg s$^{-1}$ cm$^{-2}$ (panels $e-f$).  While a similar effect is seen in the higher mass bin (panel $b$), the majority of the dust-obscured population ($0.4<E_{\rm star}(B-V)<0.8$) falls above the lower cut (dashed vertical line) thus present in the observed sample but to a lesser degree with the greater weighting at higher flux levels (dotted vertical line).  The inclusion of these galaxies is due to the fact that the SFRs are substantially higher, given the SFR - stellar mass relation, and that extinction increases with stellar mass.  The latter point is easily seen since the lower mass galaxy population (panel $a$) is clearly dominated by those with very little dust ($0.1<E_{\rm star}(B-V)<0.4$) while the trend reverses at higher masses (panel $b$).  Again, it is clear that the exclusion of dust-obscured galaxies is greater when placing more weight on targeting galaxies with a predicted emission-line flux greater than $1\times10^{-16}$ erg s$^{-1}$ cm$^{-2}$ that is implemented after the pilot observations to improve upon the redshift success rate especially for the more massive galaxies that can have a large fraction of their H$\alpha$ emission falling outside the FMOS fiber (see Section~\ref{fluxcal}).         
          
It is then imperative to establish the degree to which the emission-line flux limit shapes the distribution of SFR at a given stellar mass.  In Figure~\ref{mass_select}$c-f$, we show the SED-based SFR distributions in the equivalent mass bins as given above.  For this analysis, we illustrate the effect by simply plotting the distributions with (solid histogram) and without (dashed histogram) a cut on emission-line flux.  We show the results of this exercise for two different H$\alpha$ flux limits since we carried out the pilot survey with a flux limit of $4\times10^{-17}$ erg s$^{-1}$ cm$^{-2}$ (panels $c-d$) and then placed more weight on those with a limit of $1\times10^{-16}$ erg s$^{-1}$ cm$^{-2}$ for the remainder of the survey (panels $e-f$).  As in the top two panels, we compare the distributions for galaxies of varying extinction based on their color excess.  We see that the self-imposed flux limit has an impact on the sample selection, differently in each bin of mass and extinction.  All panels ($c-f$) clearly demonstrate that the emission-line limit has a stronger effect on those with higher levels of extinction.  The consequences are an increase in the mean SFRs and a narrower width of the SFR distribution within each given mass range.  In particular, the dispersion in SFR for the sample with $9.8< log~M_* < 10.7$ and a limit on the expected H$\alpha$ flux ($>4\times10^{-17}$ erg s$^{-1}$ cm$^{-2}$) is a factor of 1.4 smaller than the parent population.  A slightly smaller factor of 1.3 is measured for the higher mass bin.  This correction factor is likely applicable to a distribution of H$\alpha$-based SFRs since we find very good agreement between these SFR indicators \citep{Rodighiero2014} for MS galaxies.  Any assessment of the intrinsic characteristics of the SFR - stellar mass relation using our sample will need to consider the magnitude of these effects as done in \citet{Speagle2014}.  Although, it remains to be established whether these SFR distributions, shown in Figure~\ref{mass_select}, are accurate in terms of their width that can be assessed by using an alternative indicator of star-formation activity, namely the H$\alpha$ emission line that is one of the key science drivers of our program.  

\begin{figure}
\epsscale{1.2}
\plotone{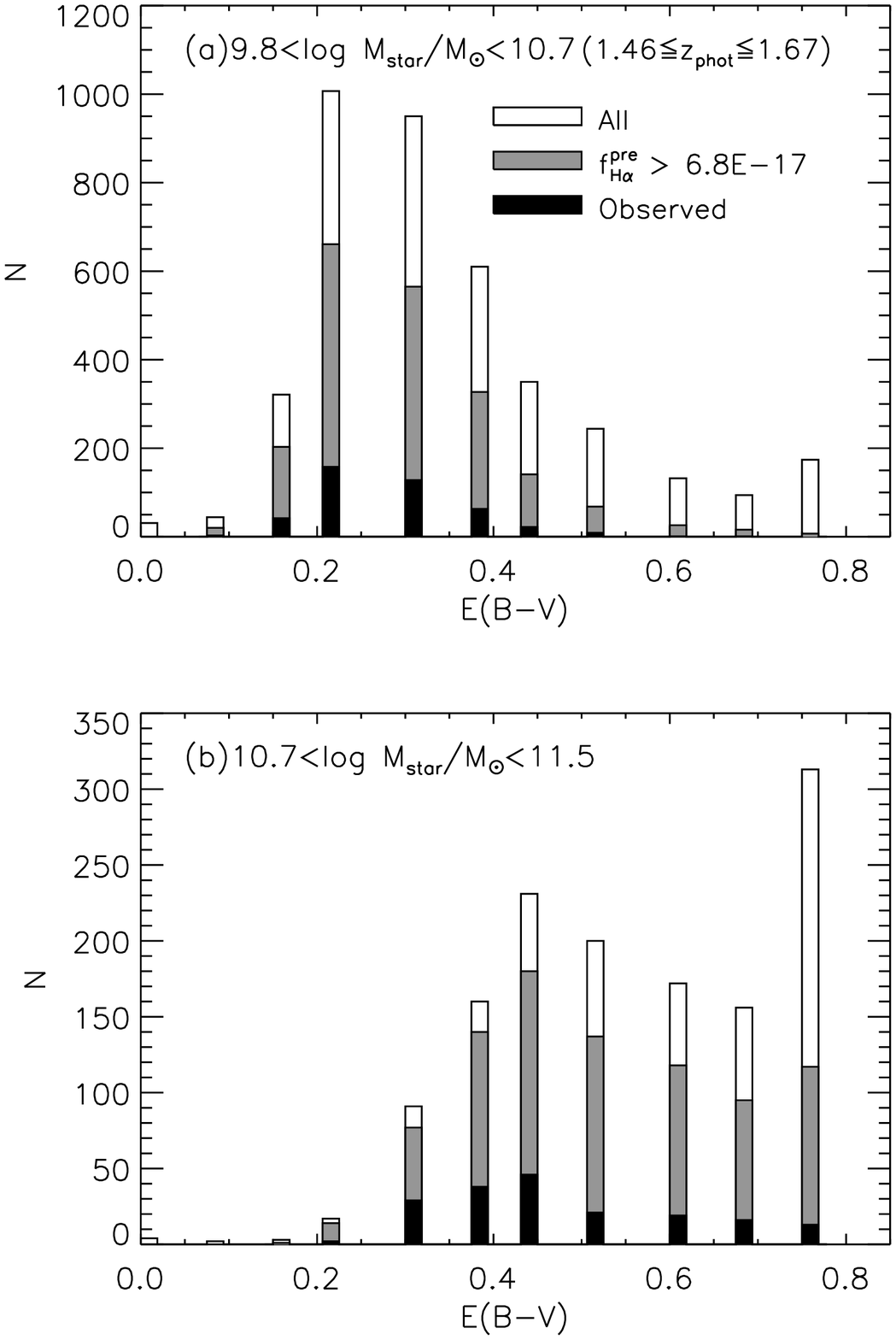}
\caption{Distribution of $E_{\rm star}(B-V)$ from SED-fitting split into two mass bins as indicated.  The open histogram is the full mass-selected star-forming population while the grey histogram highlights those with an predicted H$\alpha$ flux greater than $6.8\times10^{-17}$ erg s$^{-1}$ cm$^{-2}$.  Targets observed with FMOS are displayed by a filled black histogram.}
\label{cexcess}
\end{figure}

For further illustrative purposes, we plot in Figure~\ref{cexcess} the distribution of $E_{\rm star}(B-V)$ and isolate those expected to have an H$\alpha$ emission line above our cut and those actually observed through our FMOS program.  Both panels indicate that our sample is less representative of the intrinsic population with higher levels of extinction especially where $E(B-V)$ is above 0.5.  We note that our final sample with spectroscopic redshifts, provided by FMOS, will inevitably be biased by having detected H$\alpha$, hence it will be an H$\alpha$-selected sample; however, we provide limits on H$\alpha$ emission for an effective mass-selected subset having spectroscopic redshifts available by other means.  

\section{Observations and data reduction}

We report on observations carried out between March 2012 and January 2013 with Subaru/FMOS that represent the first installment of data to be acquired through our program.  The central square degree of COSMOS is effectively covered with four FMOS footprints, designated as HR1-HR4 (Figure~\ref{layout}), using both the high-resolution H-long and J-long gratings.  In Table~\ref{observations}, we provide a log of the observations with exposure times and notes on the effective seeing conditions (see section~\ref{fluxcal}).  

The assignment of fiber pairs to galaxies is done by using the software package $Spine-to-Object$\footnote{http://www.naoj.org/Observing/Instruments/FMOS/}.  For each fiber configuration, the CBS mode is implemented with sequential exposures taken by offsetting the telescope by 60$\arcsec$ while keeping the target within one of the two fibers; this reduces the number of science targets to $\sim200$ per FMOS configuration.  A priority is given to ensure that rarer objects of interest are assigned a fiber at a sufficient rate such as galaxies detected in either the far infrared or X-ray bands.  The star-forming galaxy population represents the majority of our targets (see Table~\ref{target_table}).  Specifically, we place the $Herschel$/PACS targets as the top priority followed by the X-ray detected AGN ($Chandra$ and XMM-$Newton$) and then the larger star-forming population.   At least, three fibers per spectrograph are placed on bright stars for flux calibration (see section~\ref{fluxcal}).  Flat-field images are taken at the beginning and end of each night using a screen within the dome and exposure times are 20 seconds in duration.  Arc lamp images are acquired for wavelength calibration having exposure times of 30 and 60 seconds taken at two position angles separated by 180 degrees.  

Our aim is to maximize the on-source exposure time that can be achieved for each fiber configuration while observing in a single night and below an airmass of 2.  At optimal times of the year when COSMOS has a high visibility, we effectively reach an exposure time of $\sim$5 hours per observation.  This integration time is sufficient to detect an emission line with an integrated flux $f=4\times10^{-17}$ erg s$^{-1}$ cm$^{-2}$ at a signal-to-noise (S/N hereafter) level of 3 (see section~\ref{sn_texp}).  An increase of the total integration time beyond five hours per night is constrained by the considerable amount of time needed to setup a field ($\sim1$ hour) and re-align fibers every 30 minutes.    

We begin our spectroscopic observations using the H-long grating that covers a wavelength range of 1.6-1.8 $\mu$m.  Seven FMOS configurations have led to the acquisition of 1153 galaxies with H-long spectra.  Using the Fibre-pac reduction software (see below) at the summit, we are able to identify galaxies with positive emission-line detections (i.e., H$\alpha$ and [NII]) to generate an input catalog for subsequent FMOS observations using the J-long grating covering a wavelength range of 1.1-1.35 $\mu$m.  This lower wavelength range is ideally suited to supplement the detection of H$\alpha$ with H$\beta$ and [OIII]$\lambda$5008.  Of the 1153 galaxies, we have acquired spectra using the J-long grating for 313 of them from four FMOS configurations that also cover the full one square degree.  Targets are prioritized for J-long followup dependent on the reliability of a spectroscopic redshift based on the centroid of the H$\alpha$ line (see below for details on redshift measurements).  An additional 410 galaxies have J-long but no H-long coverage as a result of assigning many remaining fibers primarily to lower redshift ($0.7 \lesssim z \lesssim1.1$) galaxies and AGNs (Kartaltepe et al. in preparation).        

\subsection{Data reduction}

All raw data is processed to obtain science-quality spectra both 1D and 2D using the publicly available pipeline FIBRE-pac \citep[FMOS Image-Based REduction package][]{Iwamuro2012}.  Here, we provide an outline of the data reduction procedure.

First, we process the calibration data taken before and after the science exposures.  Dome-flat images are used to determine two-dimensional transformations to correct for spatial ($y$-axis) and spectral ($x-axis$) distortions.  After correcting the Th-Ar lamp images for such distortions, we fit the Th-Ar lines and determine the relation between detector pixel and wavelength.

For science exposures, the thermal background is removed through the $A-B$ image subtraction, where A and B are the two positions of the field center (separated by 60$\arcsec$) in CBS mode.  For this procedure, two B-frames, taken just before and after the considered A-frame, are subtracted with a optimized weight to minimize residual sky errors.  Corrections are applied to account for the detector cross talk and the bias difference between quadrants.  Pixel-to-pixel variations across both detectors (IRS1 and IRS2) are minimized using the flat field image of a uniform light source taken at the beginning of each night.  After rejecting bad pixels, the science frame is corrected for distortions, as mentioned above, by using transformation coefficients, resulting from the analysis of the dome-flat images.  The residual OH airglow emission lines are then subtracted.  All science frames are combined into a single averaged image with an additional subtraction of the residual background and correction for bad pixels.  Since one target is observed in each of the two fibers in CBS mode, the corresponding spectra of position A and B are combined.  In the combine process,  the relative throughput between fibers is measured using the dome-flat image taken with the fibers in the same position as arranged for the science exposure.

For the final steps, we correct for the broad atmospheric absorption features by using a flux calibration star, whose spectral type is known or can be assessed through broad-band photometry.  A wavelength scale is applied based on the pixel--wavelength relation obtained from the Th-Ar lamp image.  The typical uncertainty is less than 1 pixel, corresponding to $1.2~\mathrm{\AA}$ in high-resolution mode.

\subsection{Flux calibration and aperture correction}
\label{fluxcal}

An initial conversion of detector counts to physical flux units (Jy) is applied based on calibration data taken during engineering observations and under an assumption of ideal conditions.  This step is built into the FIBRE-pac data reduction software.  A second-order correction is obtained by using bright stars taken simultaneously with the science targets in the COSMOS field.  These bright stars are selected based on their color using SDSS DR9 and 2MASS photometry as described in \citet{Covey2007}.  We further restrict their brightness in the infrared bands ($J>15$, $H<18$).  There is a preference for stars with existing optical spectroscopy from SDSS that provides an accurate spectral classification.  For those lacking such classification, a determination of their spectral type is based on correlations between optical and NIR colors \citep{Covey2007}.  When using spectral types determined in this manner, we confirm that it does not critically impact the flux calibration due to the fact that the FMOS spectral window in high-resolution mode is relatively narrow so any inaccuracy with respect to spectral shape results in minimal errors in the actual fluxes.  Three to four flux standards are chosen per spectrograph to ensure that a suitable single star can be used for this correction.  The uncertainty of the relative flux calibration is $\sim 10\%$.      

At this stage, the absolute flux calibration is systematically off due to factors that include fiber positioning errors ($\sigma \sim 0''\!.2$; \citealt{Kimura2010}), variable seeing conditions, and atmospheric throughput that can cause significant light loss.  With a typical size ($\sim 0''\!.5$) of our science targets and seeing conditions of $\sim1\arcsec$, there is a considerable amount of flux falling outside the FMOS fiber of $1''\!.2$--diameter.  

We implement a scheme to achieve an improvement in the absolute flux calibration for each individual galaxy by dealing with these effects collectively since it is difficult to account for each of them separately.  We estimate the fraction of the light sampled by the fiber by using {\it Hubble Space Telescope}/Advanced Camera for Surveys (ACS) $I_\mathrm{F814W}$--band images (Koekemoer et al. 2007) of individual targets of a specific observation.  This requires an assumption that the rest-frame UV and line emission (e.g., H$\alpha$) have a similar spatial extent.  Justification of this assumption comes from the tight correlation between the half-light radii in H$\alpha$  and the ACS $I$-band for the SINS/zC-SINF galaxies (Mancini et al. 2011; C. Mancini and N. M. F{\" o}rster-Schreiber, private communication).  For each FMOS observation, we smooth the HST/ACS $I$-band (F814W) image of each galaxy by convolving with a point spread function of an effective seeing as determined below.  We then perform aperture photometry using {\it SExtractor} (Bertin \& Arnouts 1996) to derive the ratio of the observed flux (i.e., rest-frame UV emission within the FMOS aperture) to the total object flux to derive an aperture correction factor for each galaxy.  This factor is then applied to each emission line flux measurement to derive a value for the total flux. 

\begin{figure*}
\hspace{1.2cm}
\includegraphics[width=15cm,angle=0]{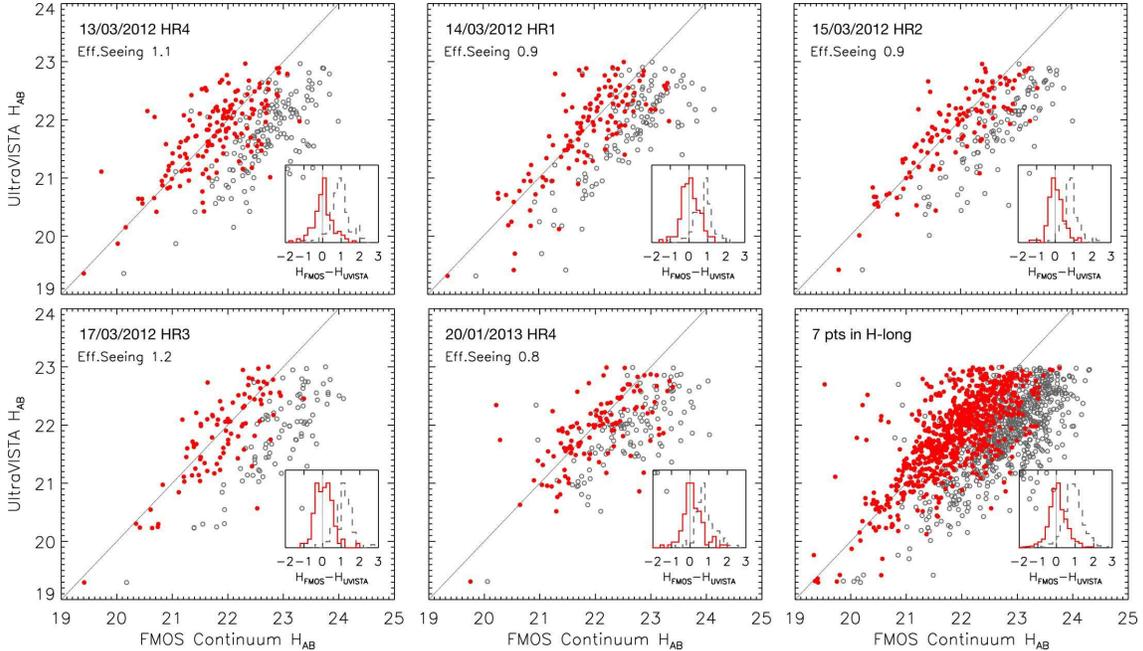}
\caption{Comparisons of continuum flux observed by FMOS with ground-based photometry (UltraVISTA) for five different H-long observations with varying effective seeing sizes ($0''\!.8$-$1''\!.2$).  The color of the symbols indicate whether an aperture correction has been applied (red: with correction; grey: no correction).  The diagonal lines are linear fits to the corrected photometry.  In each panel, a histogram of the difference is shown.  The bottom right panel shows the combined sample of seven observations.}
\label{aper_corr}
\end{figure*}

Prior to obtaining aperture corrections based on the HST/ACS imaging, we need to determine an effective seeing for each FMOS observation.  We elect to use the brighter science targets that have the continuum detected in the FMOS spectral window for this purpose.  We compare the ground-based magnitudes from UltraVISTA in the respective band (either $J$ or $H$) to that determined from the FMOS spectra.  For all observations, only galaxies for which the continuum is detected with FMOS at a S/N greater than 5 and $H_{UVISTA}<23$ are used for this purpose.  In Figure~\ref{aper_corr}, we plot the distribution of FMOS and ground-based UltraVISTA photometry for five different observations.  It is apparent that there is considerable flux loss for all objects in each observation.  It is clear that without any aperture correction the FMOS magnitudes are offset towards fainter magnitudes.  We then proceed to derive a correction to achieve unity between the FMOS and UltraVISTA photometry as expressed as an effective seeing for each observation separately.  This is in essence an average of the seeing conditions over the duration of a full exposure.  We report these values in Table~\ref{observations}.  The effective seeing of each observation is then used to smooth the HST/ACS images of each galaxy observed within the corresponding FMOS configuration.  

\begin{figure}
\epsscale{1.2}
\plotone{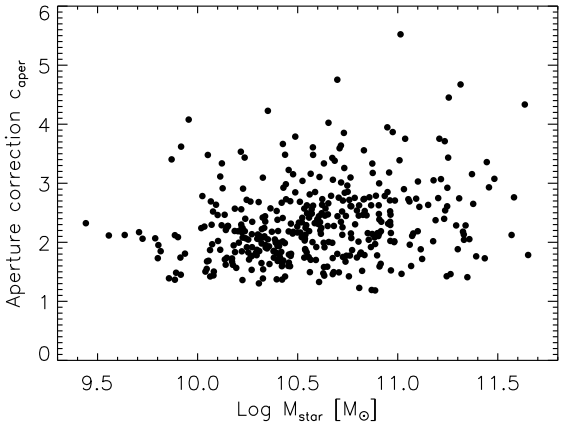}
\caption{Aperture correction factor as a function of stellar mass.  Galaxies are shown for those observed by FMOS, yield positive H$\alpha$ detections, and have aperture correction factors not suffering from blending issues in $I_\mathrm{F814W}$--band imaging.}
\label{apcorrect}
\end{figure}

Aperture correction factors are then measured using these smoothed HST/ACS images as described above.  The aperture correction factors are distributed from $\sim1.2$ to $\sim5$ (Figure~\ref{apcorrect}), with a typical value of $\sim2.2$.  It is important to recognize that there is inherent dispersion in these corrections as shown for the galaxies with detected continuum.  In the bottom right panel of Figure~\ref{aper_corr}, the level of dispersion is $\sigma=$0.42 (based on a Gaussian fit to the distribution of magnitude offsets) that corresponds to an uncertainty on the flux for an individual galaxy to be within a factor of 1.5.  Such dispersion in the flux calibration should be considered when addressing issues such as the width of the star-forming main sequence that is likely to be intrinsically narrower than the observed width.  We apply the same procedure to the $J$-band observations as well.

\begin{turnpage}

\begin{figure*}
\epsscale{0.8}
\includegraphics[width=22cm,angle=0]{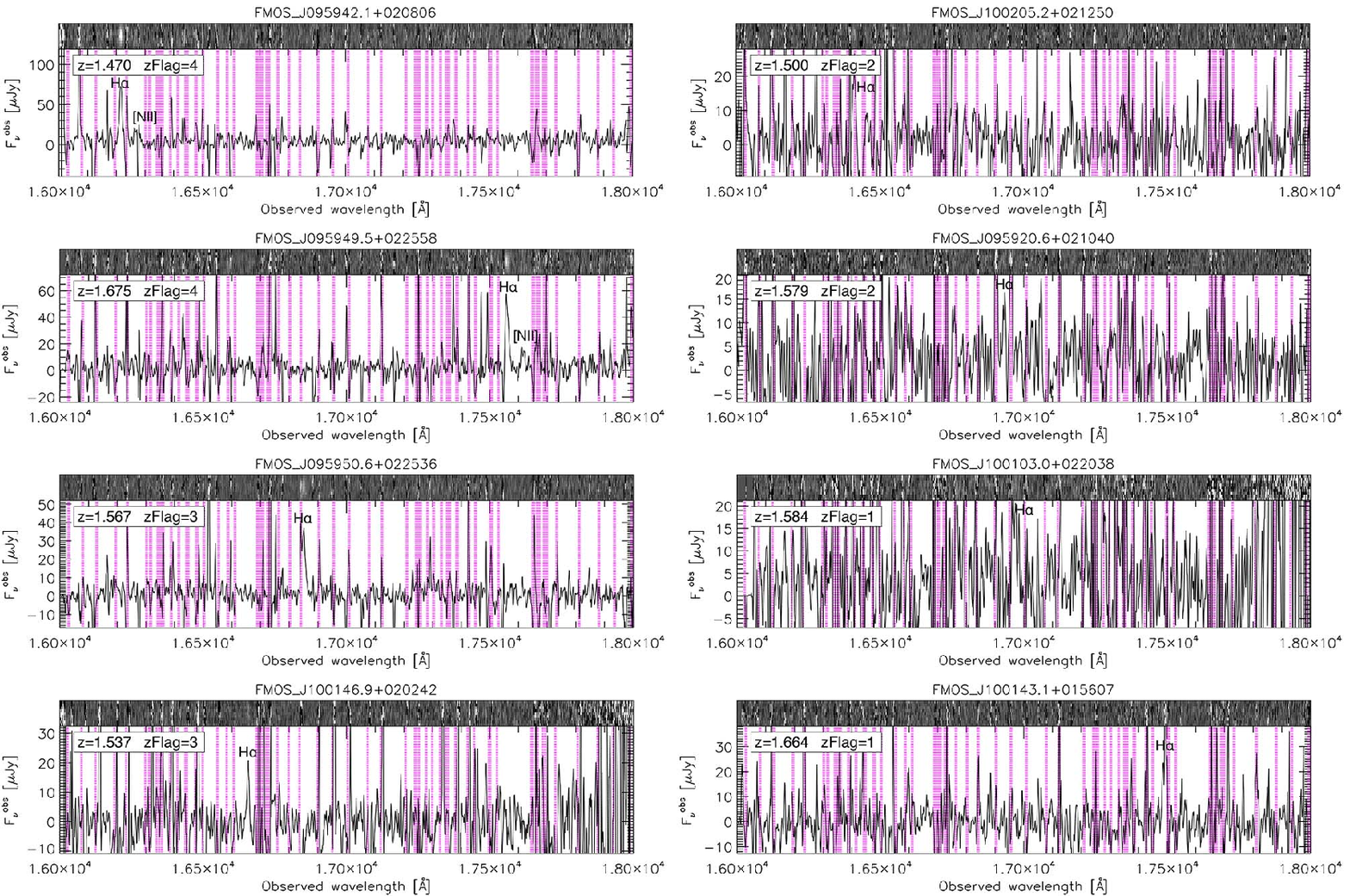}
\caption{Examples of FMOS H-long spectra (1D and 2D).  The H$\alpha$ emission line used to measure the spectroscopic redshift and quality flag is indicated.  The vertical bands mask the spectral regions impacted by OH emission.}
\label{spectra_examples}
\end{figure*}
\end{turnpage}

Final flux calibrated spectra both 1D and 2D are displayed in Figure~\ref{spectra_examples}.  We overlay regions impacted by OH lines and indicate emission lines with positive detections. 

\begin{figure*}
\epsscale{1.1}
\plotone{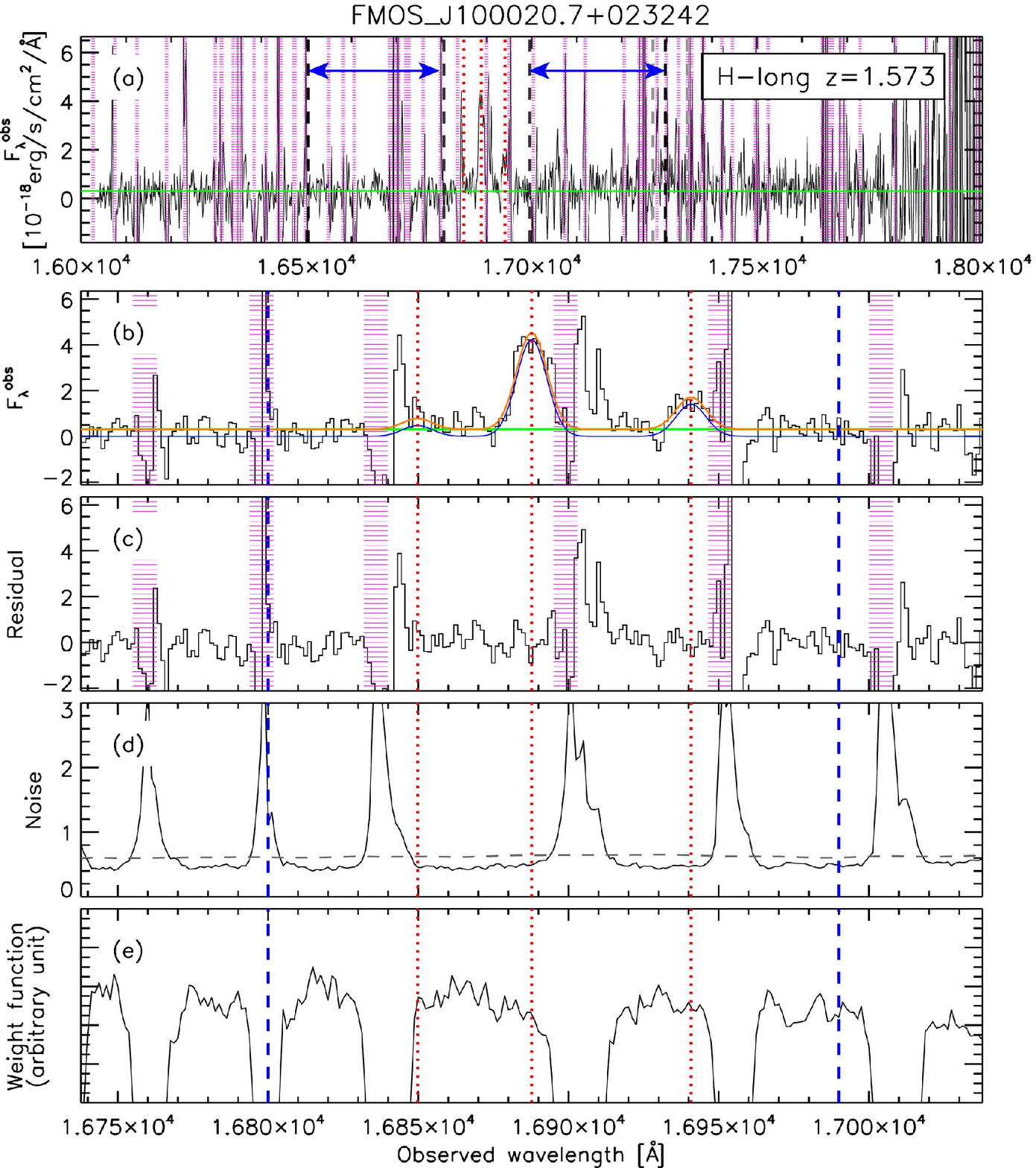}
\caption{Example spectrum of a galaxy at $z=1.573$ with a significant detection of H$\alpha$ and [NII]$\lambda6585$ as indicated by the vertical dotted lines (red).  Individual panels are as follows:  (a) Entire observed spectrum with regions impacted by OH lines shown in magenta.  Two blue arrows indicate the regions used to determine the continuum level by fitting a linear function shown in green.  (b) Observed spectrum over a narrower spectral range and centered on H$\alpha$.  The overall fit to the spectrum is indicated in yellow with the components marked in green (continuum) and blue  (emission lines).  (c) Residual after subtracting the best-fit function to the emission lines and continuum. (d) RMS noise level and threshold (horizontal dashed line) for avoiding the wavelength range impacted by OH lines. (e) Weight function applied to each pixel for fitting both the continuum and emission lines.}
\label{fit_example1}
\end{figure*}

\section{Emission-line fitting}
\label{sec:efit}

We measure the position and strength of emission lines (i.e., H$\alpha$, [NII]$\lambda6585$, [OIII]$\lambda$5008, H$\beta$) present in our near-infrared spectra to derive spectroscopic redshifts, star formation rates, levels of dust extinction, and gas-phase metallicities.  We specifically measure the centroid, amplitude and full-width at half maximum (FHWM) of each emission line thus the total flux using a line fitting routine.  For the majority of individual spectra, we do not detect the stellar continuum;  although, we do fit the continuum for the limited number of bright galaxies with a linear function and remove this component prior to fitting the emission lines.

\begin{figure}
\epsscale{1.5}
\plotone{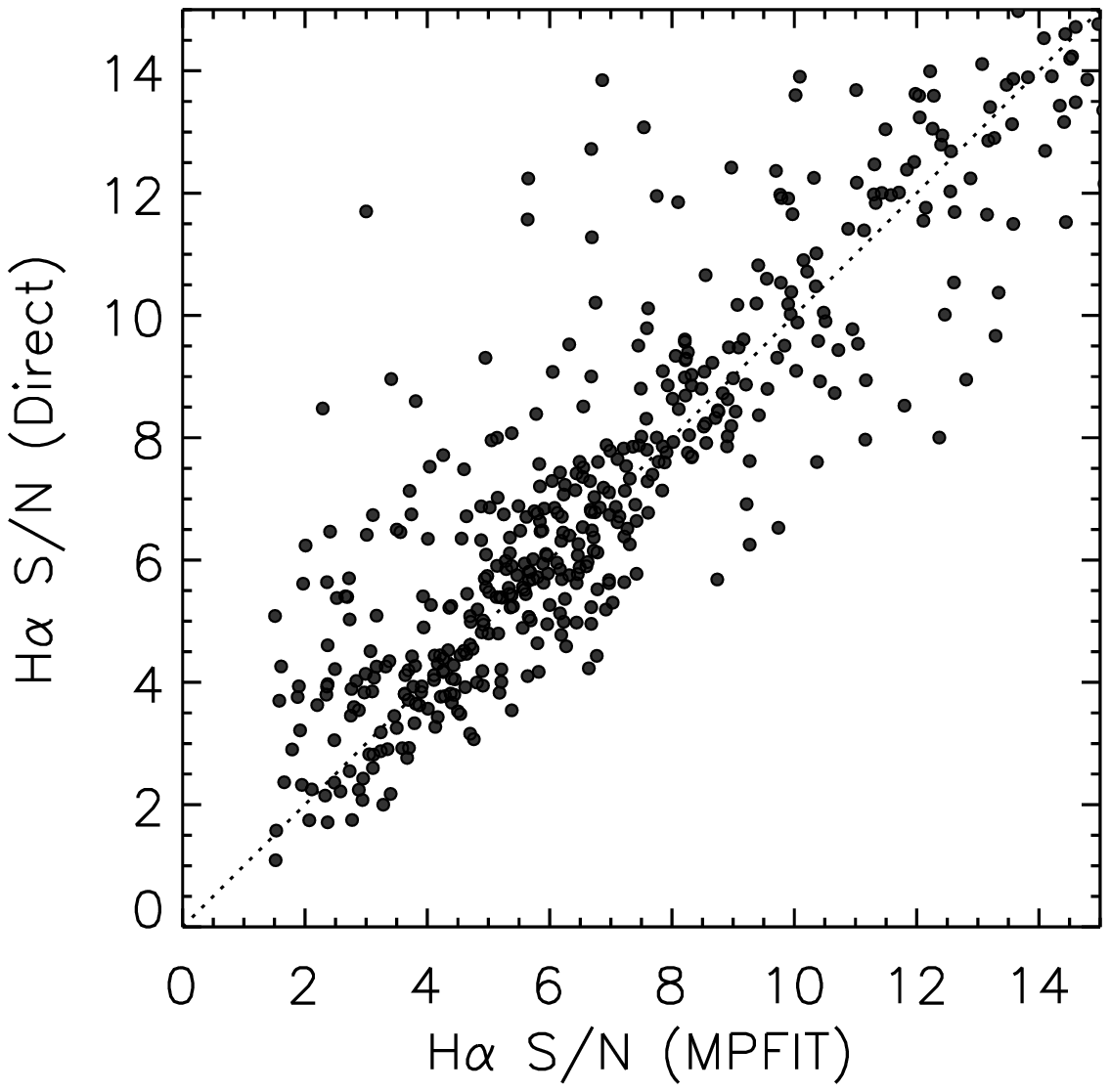}
\caption{Signal-to-noise estimate of H$\alpha$ using our direct method compared to that based on the errors returned by `MPFITFUN' for an assumed Gaussian line flux profile.}
\label{error_comp}
\end{figure}

Our line-fitting procedure utilizes `MPFITFUN', a Levenberg-Marquardt least squares minimization algorithm.  Each emission line (H$\alpha$, [NII]$\lambda$6550, 6585) is fit as a single\footnote{For our purpose, a single component fit to the H$\alpha$ line is sufficient.  Although, there remains the possibility that, in some cases, an additional component may be required possibly indicative of galactic outflows or underlying AGN.  At this time, we reserve such investigations to subsequent studies.} Gaussian function to characterize the line shape less impacted by pixel-to-pixel noise variations.  The relative amplitude of the [NII] lines is fixed to the laboratory value of 2.96.  The width of the [NII] lines are fixed to that of H$\alpha$.  The width of each component is constrained to have $\sigma > \sigma_\mathrm{min}\sim45$ km s$^{-1}$ that corresponds to the spectral resolution of FMOS in HR mode at $\lambda\sim1.6\mu$ (R=2600).  The value of $\sigma_\mathrm{min}$ is determined for each observation based on the width of the Th-Ar emission lines detected in the arc lamp calibration images taken just before the science exposures.  The routine returns best-fit parameters, errors, and a measure of the overall quality of the fit.  We visually inspect a subset of the line fits to gauge the overall accuracy of our fitting routine.  

As an example of our fitting procedure, we show the FMOS spectrum of a galaxy at $z=1.573$ (Figure~\ref{fit_example1}) with the best-fit Gausssian functions to the H$\alpha$ and [NII] emission lines overlaid.  Various panels illustrate the data quality, fitting results, and inherent issues such as residuals from regions blocked by the OH suppression mask.  Based on the initial spectroscopic redshift (see Section~\ref{sec:zspec}), we fit a section of the spectrum centered on the emission lines and spanning a range that enables an accurate determination of the continuum level (panel $b$).  A weighting for each pixel ($1/N^2$; panel $e$) is determined from the noise spectrum ($N$; panel $d$) that essentially indicates the regions free of OH emission.  Our fitting routine takes into consideration this weighting scheme.  Any pixel that is affected by an OH line, where the noise is generally large, is masked by setting the weight to zero.  

Throughout, we use the error on the fit parameters (i.e., amplitude and width), returned by 'MPFITFUN' as a covariance matrix, to estimate the error on the total line flux.  The errors are scaled to match the residual noise of the object spectra since we find that the noise spectra, output by FIBRE-pac, are typically underestimated by $\sim20\%$.  We confirm that the errors returned by `MPFITFUN' do actually represent the data quality of the line detections.  For this check, we directly measure the total flux within $\pm2.5\sigma$ of the line center, as determined by the fitting routine, and the sum of the rms dispersion ($\Sigma \sigma_{rms}$) from the regions in the object spectra local to the H$\alpha$ emission line.  We mask out regions impacted by OH emission.  Based on a comparison of both sets of error estimates (Figure~\ref{error_comp}), there is very good agreement between the two with some scatter as expected since, in many cases, the OH emission reduces the number of pixels over which the line can be detected.

\begin{figure}
\epsscale{2.3}
\plottwo{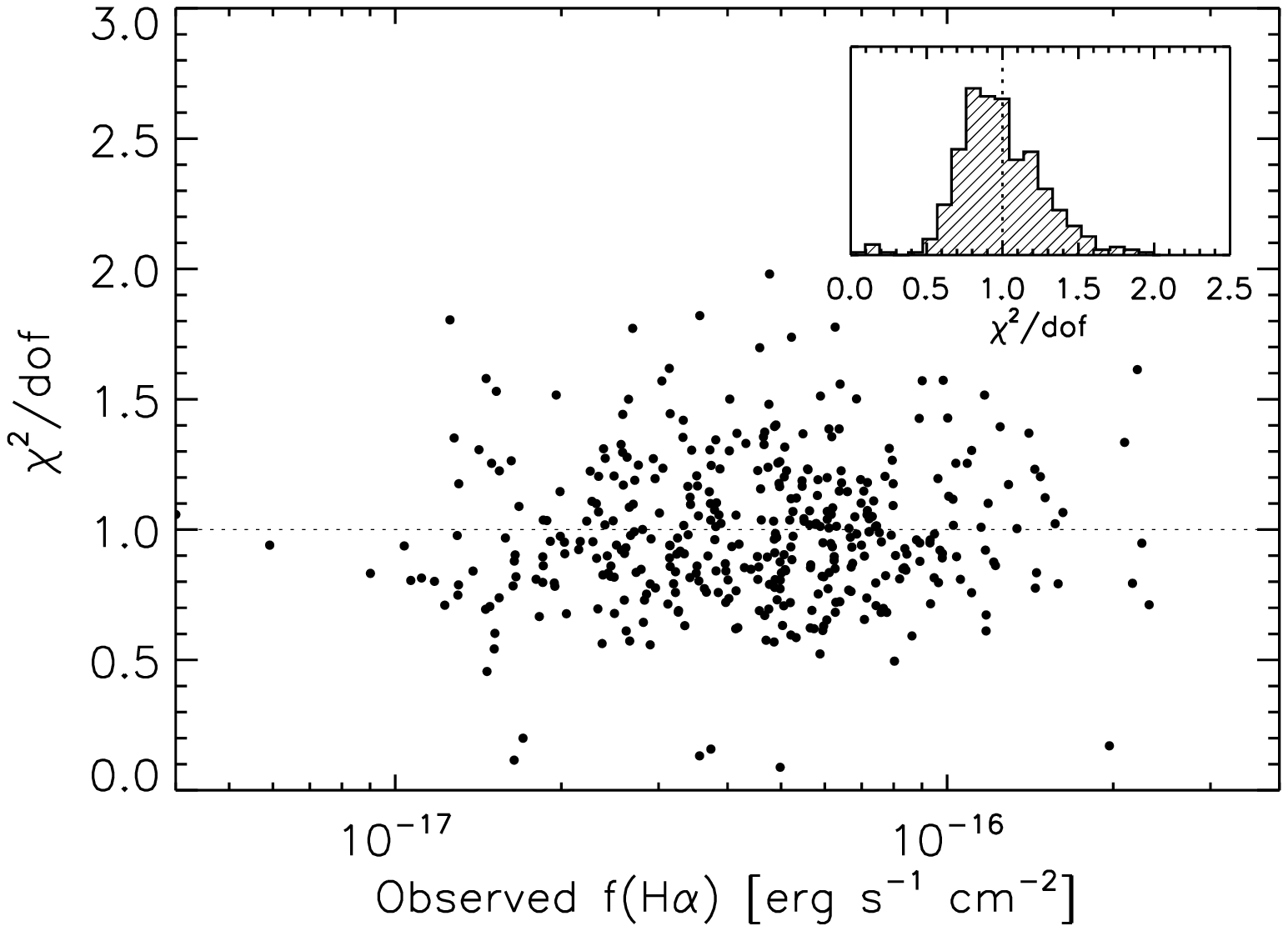}{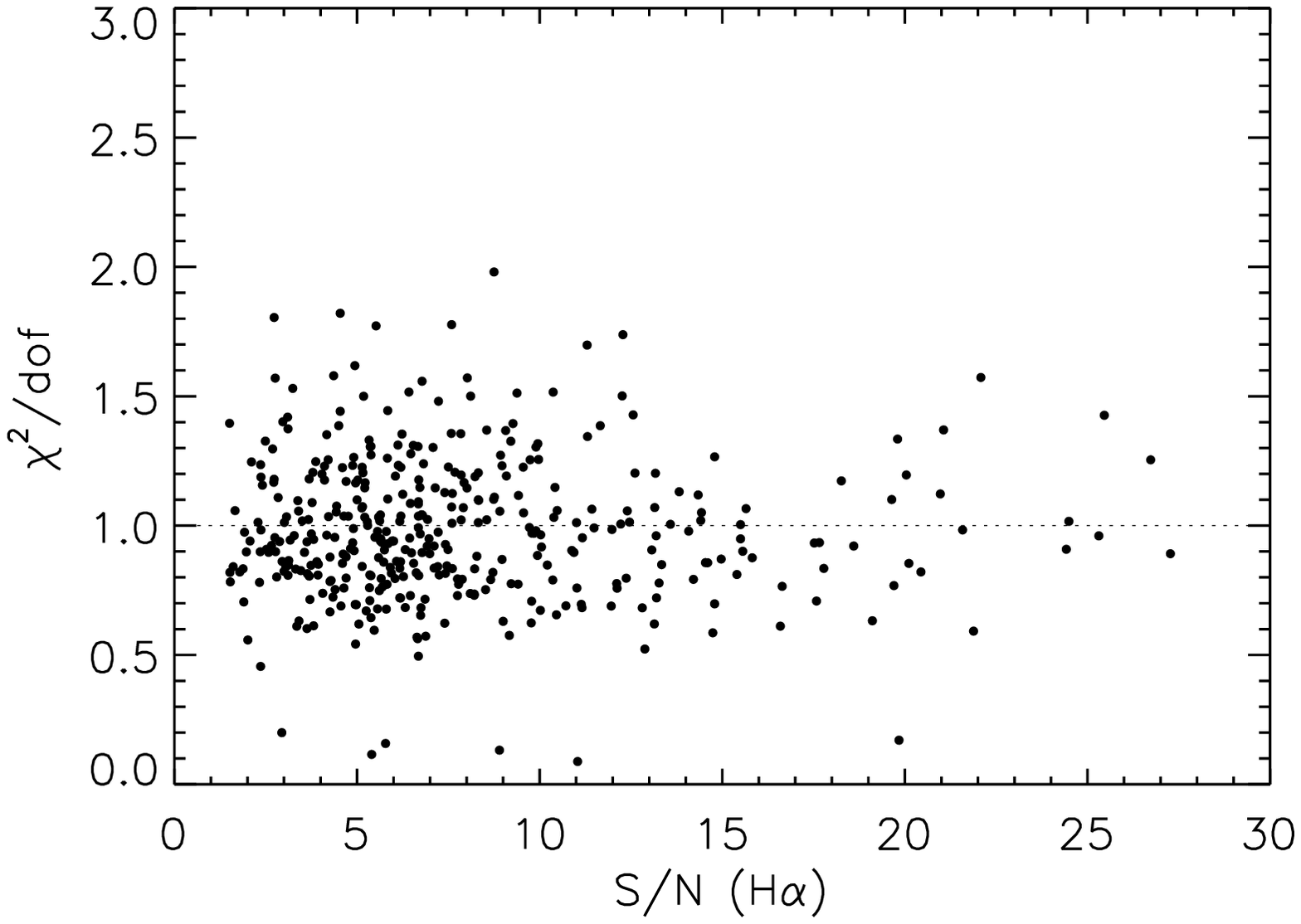}
\caption{Assessing the accuracy of our line-fitting routine.  The reduced chi-square statistic of each fitted H$\alpha$ line as a function of observed H$\alpha$ flux (top panel) and S/N (bottom panel).}
\label{fig:rchisq}
\end{figure}

\begin{figure*}
\epsscale{0.9}
\plotone{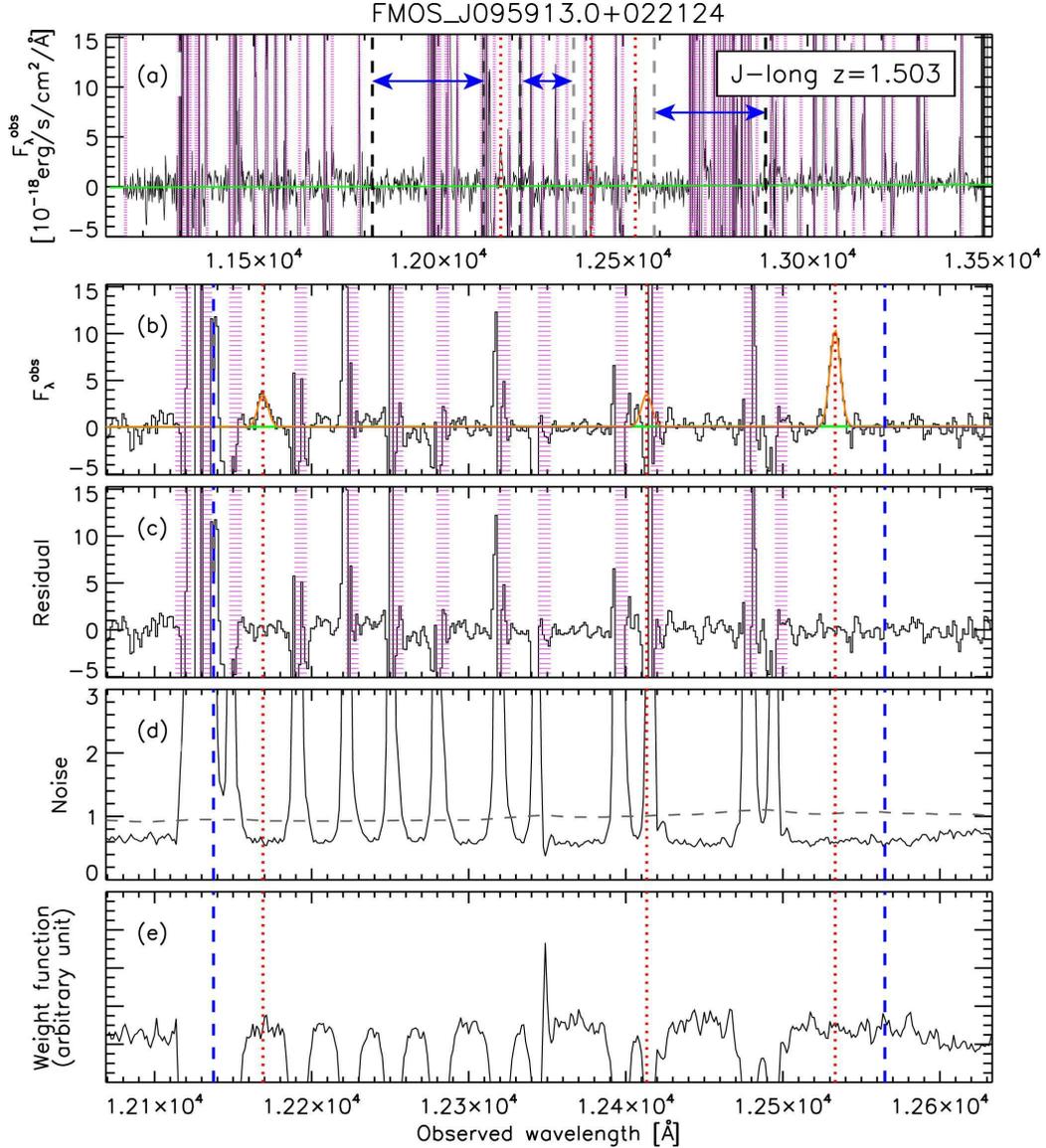}
\caption{Fitting the H$\beta$ and [OIII]$\lambda$5008 emission lines.  Panels are equivalent to those shown in Figure~\ref{fit_example1}.}
\label{fit_example2}
\end{figure*}

The goodness of our line fits is assessed using the reduced chi-square statistic.  In Figure~\ref{fig:rchisq}, it is evident that the distribution of the reduced chi-square statistic is close to being centered at 1.  There appears to be no significant deviations of reduced chi-square statistic with either line strength or S/N.

We carry out the equivalent fitting procedure for the J-long spectra (Figure~\ref{fit_example2}) to measure the strength of both H$\beta$ and [OIII]$\lambda5008$.  We remark that is uncommon to have detections of all four lines.  Of the 408 galaxies with H$\alpha$ detections at $S/N>1.5$, we significantly detect [NII]$\lambda6585$ in 62\% of the cases, H$\beta$ in 21\% and [OIII]$\lambda5008$ in 30\%.  As summarized in Table~\ref{lines_table}, the detection rate of having additional lines only slightly increases with a higher requirement on the S/N of H$\alpha$.  

\subsection{Line significance as a function of exposure time}
\label{sn_texp}

\begin{figure}
\includegraphics[width=9cm]{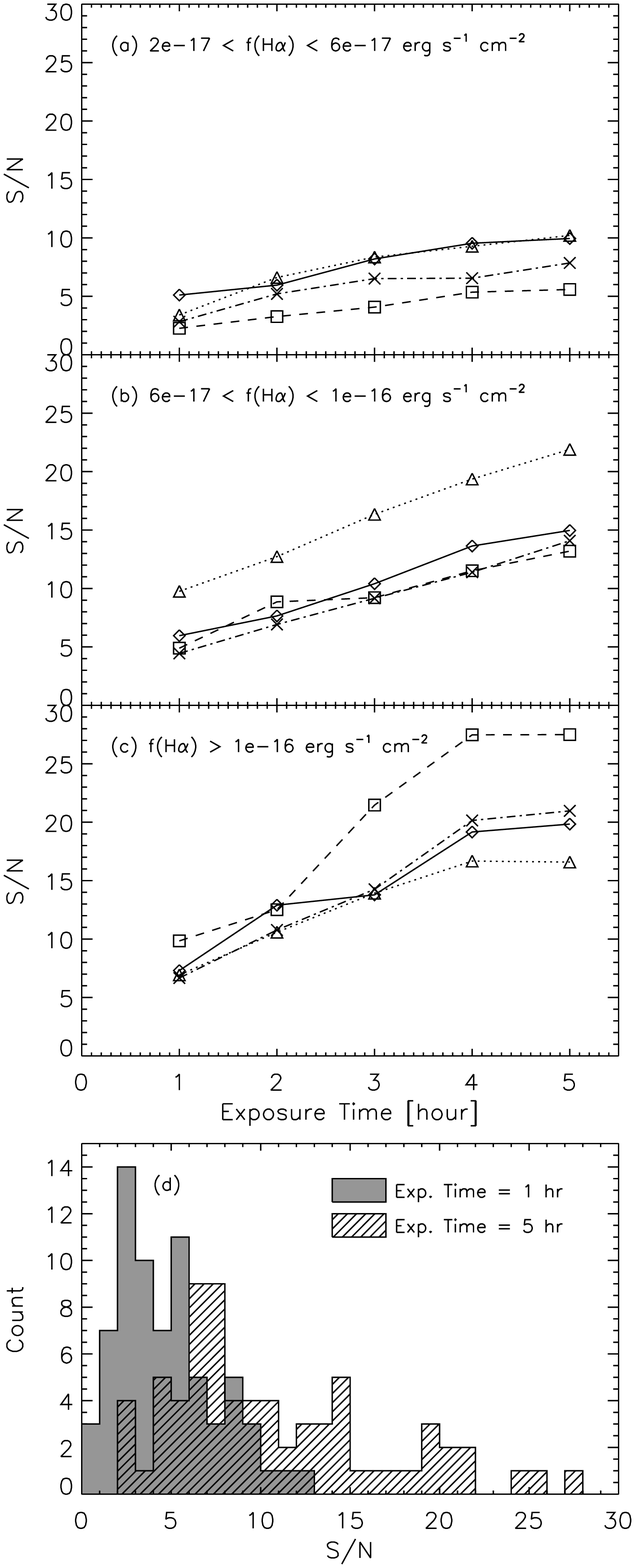}
\caption{Signal-to-noise ratio of H$\alpha$ as a function of exposure time in bins of observed emission-line flux (panels a-c) for a single FMOS observation with a total exposure time of five hours.  Symbol types discriminate between four different sources in each panel.  In panel $d$, a histogram shows the number distribution of line detections of a given S/N for both a 1 hour and 5 hour integration time.}
\label{lines_sn}
\end{figure}

We empirically determine whether our FMOS observations are capable of reaching the required depth with respect to detecting emission lines down to fluxes of $4\times10^{-17}$ erg s$^{-1}$ cm$^{-2}$.  We measure the improvement in S/N gained with each hour of integration by analyzing the data in five segments of increasing exposure time up to the full five-hour exposure (Figure~\ref{lines_sn}).  In panels a-c, the S/N is shown to increase as a function exposure time as expected at all flux levels.  This is demonstrated using four objects with H$\alpha$ detections in each bin of line flux.  Even at the lowest fluxes (panel a), the S/N is greater than 5 after the full 5-hr exposure.  At the brightest fluxes ($f_{H\alpha}\gtrsim1\times10^{-16}$ erg s$^{-1}$ cm$^{-2}$), the S/N reaches above 15.  In the lower panel (d), the number distribution of objects is shown for H$\alpha$ detections of a given S/N for both 1-hour and 5-hour exposures.  An integration time of $>$2 hours is needed to guarantee that the majority of the emission-lines are detected with significance (i.e., S/N $>$ 3) at our faintest flux levels (panel a) while a full 5 hours improves the significance (S/N $>$ 5; panel d).

\begin{deluxetable}{lllll}
\tabletypesize{\small}
\tablecaption{Detection rates of key emission lines\tablenotemark{a} \label{lines_table}}
\tablehead{\colhead{S/N}&\colhead{H$\alpha$}&\colhead{+[NII]}&\colhead{+H$\beta$}&\colhead{+[OIII]5007}\\
&&\colhead{(\%)}&\colhead{(\%)}&\colhead{(\%)}}
\startdata
$\ge 1.5$ & 408 & 251 (62)& 86 (21)& 122 (30)\\
$\ge 3$ &  370 & 239 (65)& 82 (22)&112 (30)\\
$\ge 5$ & 289 & 199 (69)& 69 (24)& 88 (30)
\enddata
\tablenotetext{a}{The first entry is the number of galaxies with an H$\alpha$ line detection above the given threshold.  The last three columns are the subset of the first column with S/N $>$ 1.5 of the line of interest.}
\end{deluxetable}

\subsection{Upper limits}

We estimate upper limits on non-detections of all emission lines of interest for which we have a highly significant spectroscopic redshift determination (see following section).  Upper limits are calculated at twice the noise level (N) of the observed spectrum as follows.

\begin{equation}
F < 1.25\times 2 \times \frac{N^\mathrm{all}_\mathrm{pix}}{N^\mathrm{good}_\mathrm{pix}} \int_{\pm 2.5\sigma} N (\lambda) d\lambda \end{equation}

\noindent where $N^\mathrm{all}_\mathrm{pix}, N^\mathrm{good}_\mathrm{pix}$ are the numbers of all or good (i.e., not impacted by OH mask or residual emission) pixels falling within $\pm 2.5\sigma$ around the expected position of the line.  The line width, $\sigma$, is determined from the detected emission line(s).  The factor of 1.25 is included since the noise spectra are typically underestimated by $\sim20\%$ as previously mentioned.  We report an upper limit for each emission line only if ${N^\mathrm{good}_\mathrm{pix} (\pm 2.5\sigma)} \ge 7$ pixels and $N^\mathrm{good}_\mathrm{pix} / N^\mathrm{all}_\mathrm{pix} >50\%$.

\section{Spectroscopic redshift determination}
\label{sec:zspec}

We measure spectroscopic redshifts from the presence of the H$\alpha$ and [NII] emission lines through a two-step process.  First, we inspect all NIR spectra using the graphical interface SpecPro\footnote{http://specpro.caltech.edu/} \citep{Masters2014}.  By viewing both the one- and two- dimensional spectra, a spectroscopic redshift is assigned manually based on the presence of the aforementioned emission lines.  This tool has been updated to work with FMOS spectra including the overlay of OH emission, essential to determine the likelihood of  line identification.  During this process, the full broad-band photometry is displayed with a best-fit spectral template.  In many cases, the clear presence of a discontinuity between two photometric bands (i.e., 4000 ${\rm \AA}$ break) redshifted to a value in close agreement with that determined by the presence of an emission line lends assurance to the realization of a successful H$\alpha$ detection.  As described above, a line-fitting routine returns the centroid of key emission lines thus providing a measure of the final spectroscopic redshift.

\begin{figure}
\epsscale{1.2}
\plotone{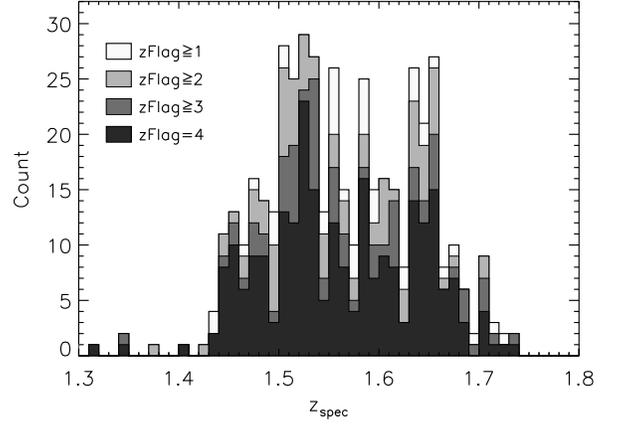}
\caption{Spectroscopic redshift distribution in various histograms colored by their final quantitive assessment of their reliability (flags 1-4; see Section~\ref{zflags}).}
\label{zdistribution}
\end{figure}

In total, we measure spectroscopic redshifts for 460 out of 1153 star-forming galaxies, targeted from either our sBzK, SED, $Herschel$-PACS, or low-mass samples (Table~\ref{target_table}).  The redshift distribution is shown in Figure~\ref{zdistribution} split by a quality flag indicative of the S/N of line detection as described below (Section~\ref{zflags}).  Various factors influence the overall distribution such as errors in photometric redshift estimates, OH emission and presence of large-scale structure within the central area of COSMOS.

\subsection{Quantitative assessment of the quality of the spectroscopic redshifts}
\label{zflags}

We assign a quality flag on each redshift measurement based on the S/N estimates of the emission-line detections.  The flags are determined using the following criteria:
\\

\begin{itemize}

\item Flag 0: No emission line detected.

\item Flag 1 Presence of a single emission line with S/N between 1.5 and 3.

\item Flag 2 One emission line having  S/N greater than 3 and less than 5.

\item Flag 3 One emission line having  S/N greater than 5.

\item Flag 4: One emission line having S/N greater than 5 (usually H$\alpha$) and a second line that both confirms the redshift and has S/N greater than 1.5.

\end{itemize}

\noindent The increase in the quality flag from a '3' to a '4' is usually the result of having a detection of either [NII]$\lambda6585$, [OIII]$\lambda5008$ or H$\beta$ (in addition to the H$\alpha$ detection).  Followup observations using the J-long grating are especially important for identifying this highly secure sample.

In total, our spectroscopic redshift sample contains 460 star-forming galaxies of which 418 (329) are highly reliable (flag $\ge$ 2 (3)).  We provide a breakdown on the distribution of quality flags for the limited subsamples  (i.e., sBzK, $Herschel$/PACS, AGN and low-mass galaxies; Table~\ref{target_table}).  As previously mentioned, the detection rates of other lines are given in Table~\ref{lines_table}.  

\begin{figure}
\epsscale{2.4}
\plottwo{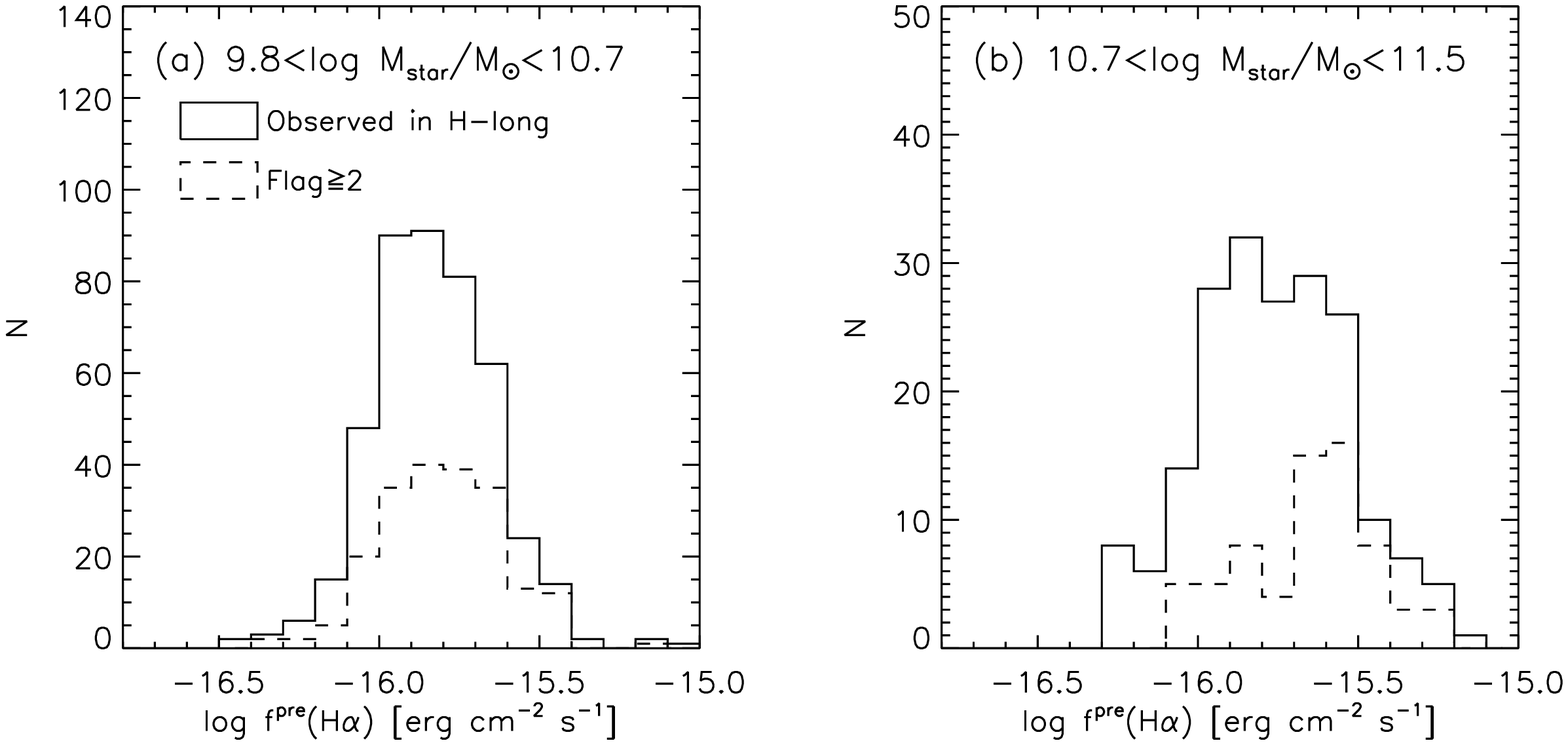}{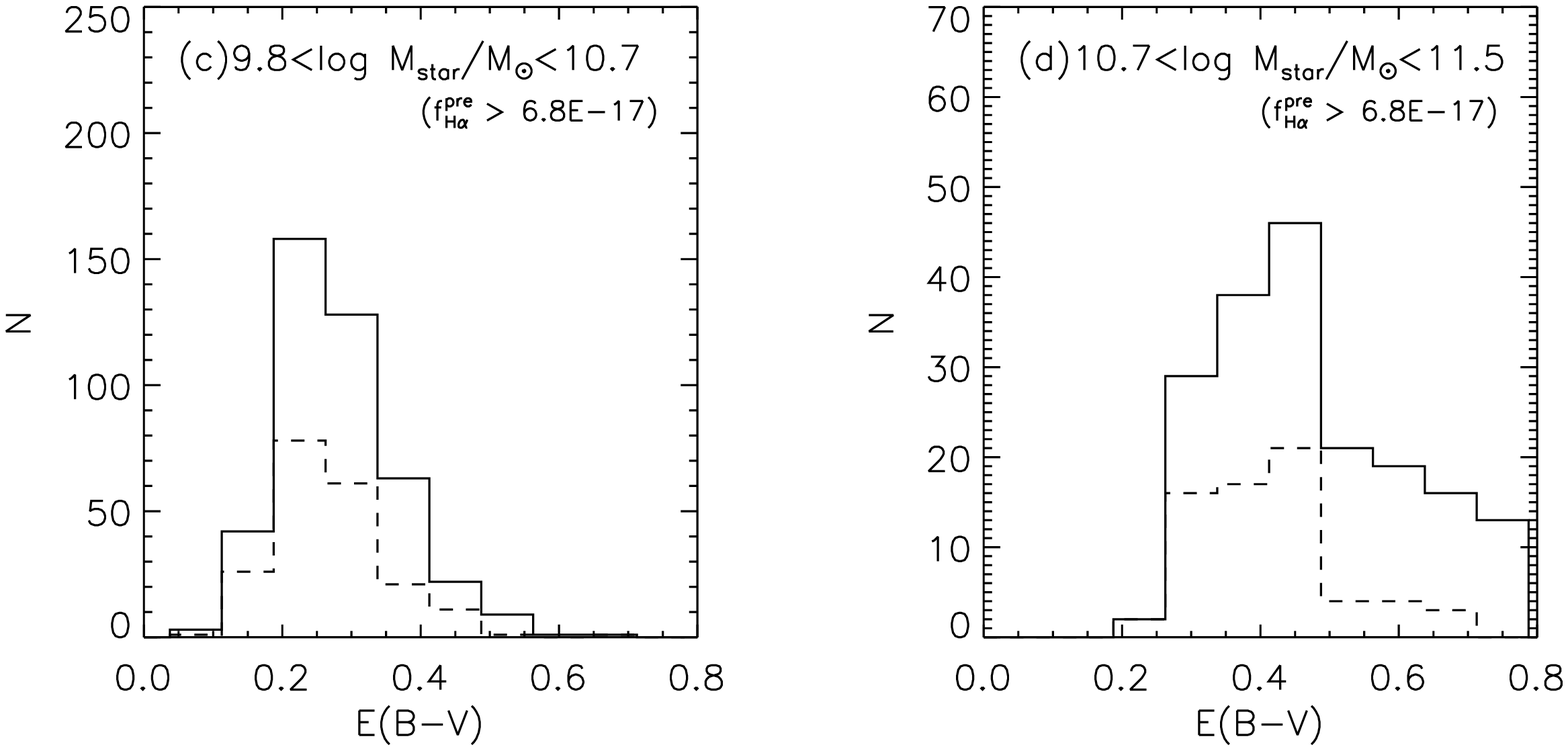}
\caption{Distribution of spectroscopic redshift successes (quality flag $\ge$ 2; dashed histogram) as a function of predicted H$\alpha$ flux (total; i.e., no consideration of the aperture size; panels a and b) and color excess ($E_{\rm star}(B-V)$; panels c and d).  The solid black histogram represents the distribution of the full galaxy sample that has been observed with FMOS.}
\label{zsuccess1}
\end{figure}

\subsection{Redshift success rate}
\label{sec:success}
For the primary star-forming galaxy sample with $9.8<log~M_{\rm stellar}<11.5$, we have a 44\% success rate of acquiring a spectroscopic redshift.  Our level of success with respect to predicting the redshift and line strength of individual objects is likely to be higher given the fact that $\sim25\%$ of the spectral window is blocked by the OH suppression system.  The remaining failures are likely due to less-than-ideal weather conditions, errors on the photometric redshift estimates, inaccurate estimates of the level of dust extinction, and surface brightness variations among the galaxy population.  Even so, our results are a significant improvement over those using FMOS in low-resolution mode ($\sim20\%$).  In Table~\ref{success_rate_table}, we provide estimates of our success rate for assigning redshifts above a given quality flag.      

\begin{deluxetable}{lllllll}
\tabletypesize{\scriptsize}
\tablecaption{Redshift success rates\tablenotemark{a}\label{success_rate_table}}
\tablehead{\colhead{Log}&\colhead{Log}&\colhead{N}&&\colhead{Flag:}&\colhead{N(\%)}\\
\colhead{mass}&\colhead{$f_\mathrm{H\alpha}$\tablenotemark{b}}&&\colhead{$\ge$ 1}&\colhead{$\ge$ 2}&\colhead{$\ge$ 3}&\colhead{$\ge$ 4}\\
&&&\colhead{N (\%)}&\colhead{N (\%)}&\colhead{N (\%)}&\colhead{N (\%)}
}
\startdata
9.8-10.7   & -16.17 & 427 & 222 (52) & 200 (47) & 158 (37) & 122 (29)\\
9.8-10.7   &  -15.7 & 104 & 65 (63) & 60 (58) & 50 (48) & 45 (43)\\
10.7-11.5 &-16.17 & 184 & 71(39) & 67 (36) & 50 (27) & 44 (24)\\
10.7-11.5 &  -15.7 &  78 & 46 (59) & 45 (58) & 38 (49) & 34 (44)
\enddata
\tablenotetext{a}{SED-selected sample with $1.46\le z_\mathrm{phot} \le 1.67$}
\tablenotetext{b}{The predicted H$\alpha$ flux limit in $\mathrm{erg~s^{-1}~cm^{-2}}$.}
\end{deluxetable}

In Figure~\ref{zsuccess1}, we display the distribution of targets (using our SED-selected sample) with successful spectroscopic redshift measurements as a function of predicted H$\alpha$ flux and color excess $E_{\rm star}(B-V)$ for two ranges of stellar mass.  In comparison with the full observed FMOS sample (solid histograms), we have a higher level of success detecting H$\alpha$ at lower masses (panel $a$: 47\%, panel $b$: 36\%).  For both mass bins, our success rate is significantly higher at $f_{H\alpha}^{\rm pred}\gtrsim2\times10^{-16}$ erg s$^{-1}$ cm$^{-2}$, although, it is lower for the higher mass bin (panel $b$) above this flux limit.  The drop in the success rate at higher masses is most likely due to their larger size that results in more light falling outside the FMOS aperture (see Figure~\ref{apcorrect}) and high levels of extinction $E_{\rm star}(B-V)>0.5$ causing an especially noticeable drop (Fig.~\ref{zsuccess1}$d$).

\begin{figure}
\epsscale{1.0}
\plotone{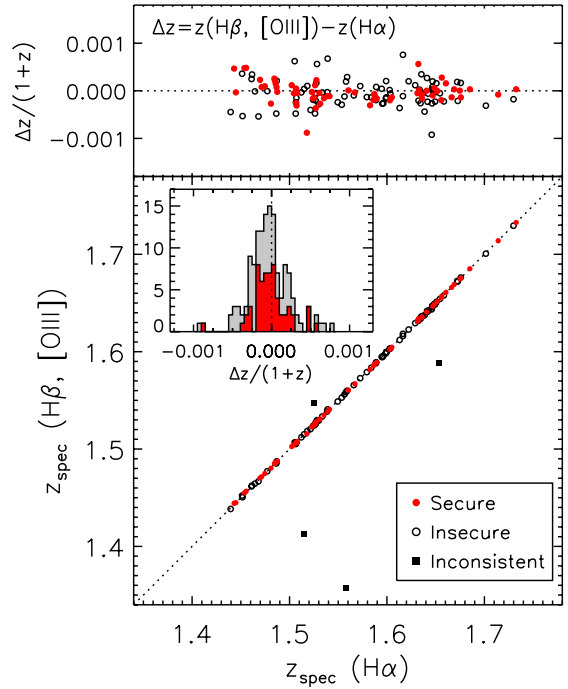}
\caption{FMOS $J$-long confirmation of H$\alpha$-based spectroscopic redshifts.  Symbols indicate whether at least one line is detected with a $S/N\ge3$ (secure) or not (insecure).  Filled squares indicate the insecure redshifts that are in disagreement.}
\label{zcomp2}
\end{figure}

\subsection{Redshift confirmation using FMOS/J-long observations}

The effort to supplement the H-long observations with followup J-long coverage gives us an opportunity to check the reliability of our emission-line identification and the accuracy of our redshift measurements.  For this exercise, we first search for emission lines (H$\beta$ and [OIII]$\lambda$5008) at the expected positions based on the redshift derived from H$\alpha$.  We then identify additional lines if present.  As a result of our procedure, the measurements are not completely independent.  In Figure~\ref{zcomp2}, we plot the spectroscopic redshift measured from H$\alpha$ compared to that for either H$\beta$ or [OIII]$\lambda$5008.  Symbols are color-coded depending on whether we detect at least one or more emission lines having $S/N\ge3$ (secure) or not (insecure).  Overall, we find confirmation of the redshifts for many of our galaxies, even those with H$\alpha$ detected at low S/N.  We find perfect agreement for all the secure cases and the few outliers are based on low $S/N$ detections.  Furthermore, we measure an almost non-existent systematic offset (mean $\Delta z/(1+z)=-2.5e-5$ for the secure sample) that corresponds to 0.43 ${\rm \AA}$ at 1.6 $\mu$m which is less than the pixel resolution ($\sim1 {\rm \AA}$).  This indicates that the wavelength calibration of the H-long and J-long spectra are in excellent agreement.  The typical error on our redshift measurements is $\sigma (\Delta z / (1+z))=1.8e-4$ based on the standard deviation of the ensemble for the secure cases.  The redshift accuracy is about a factor of two worse for the low-quality detections  ($\sigma (\Delta z / (1+z))=2.2e-4$).  From this comparison, it is apparent that many of our low quality flags, initially based on S/N of a single emission line, can be raised based on confirmation with the detections of other lines.

\begin{figure}
\epsscale{2.3}
\plottwo{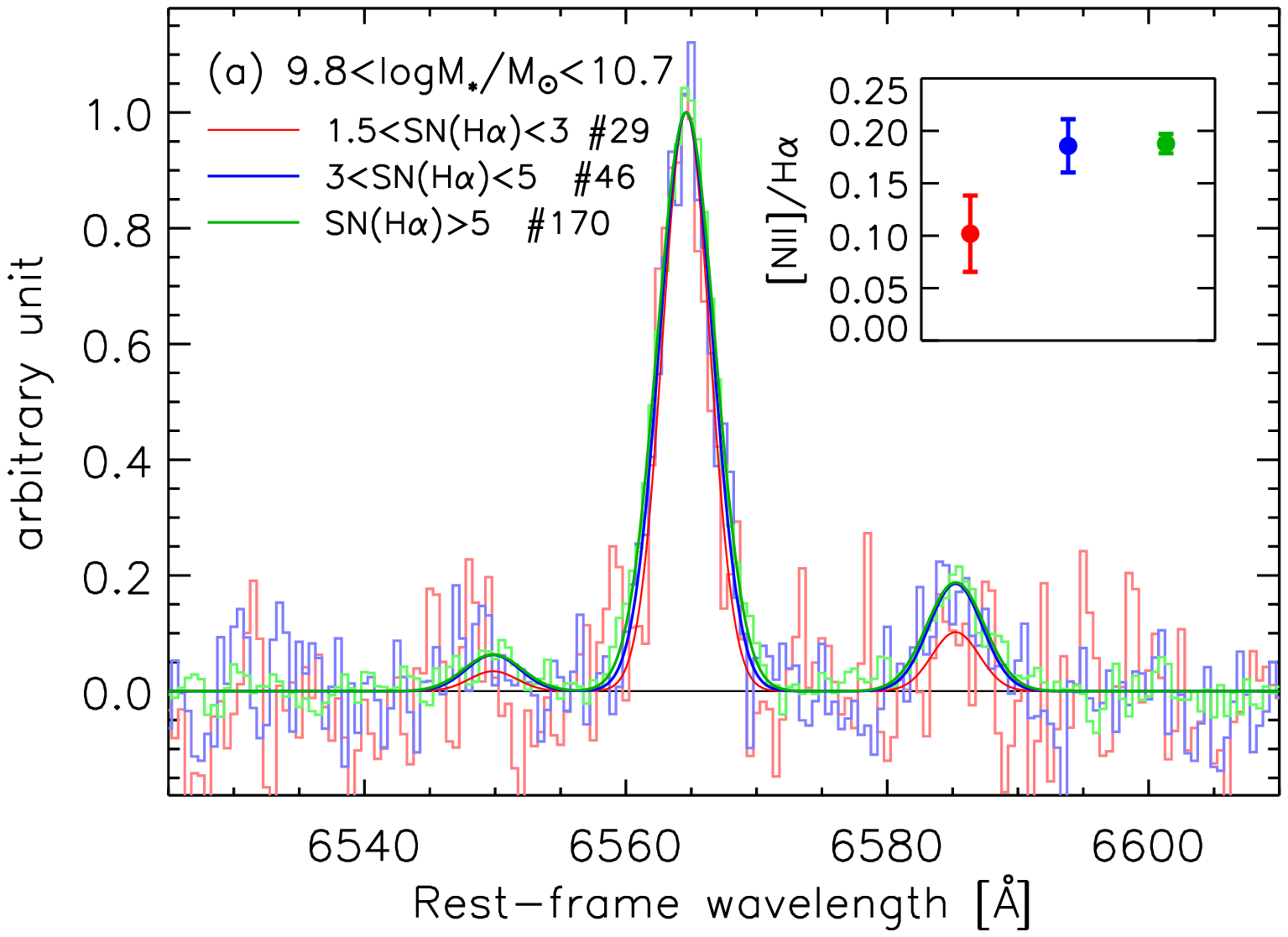}{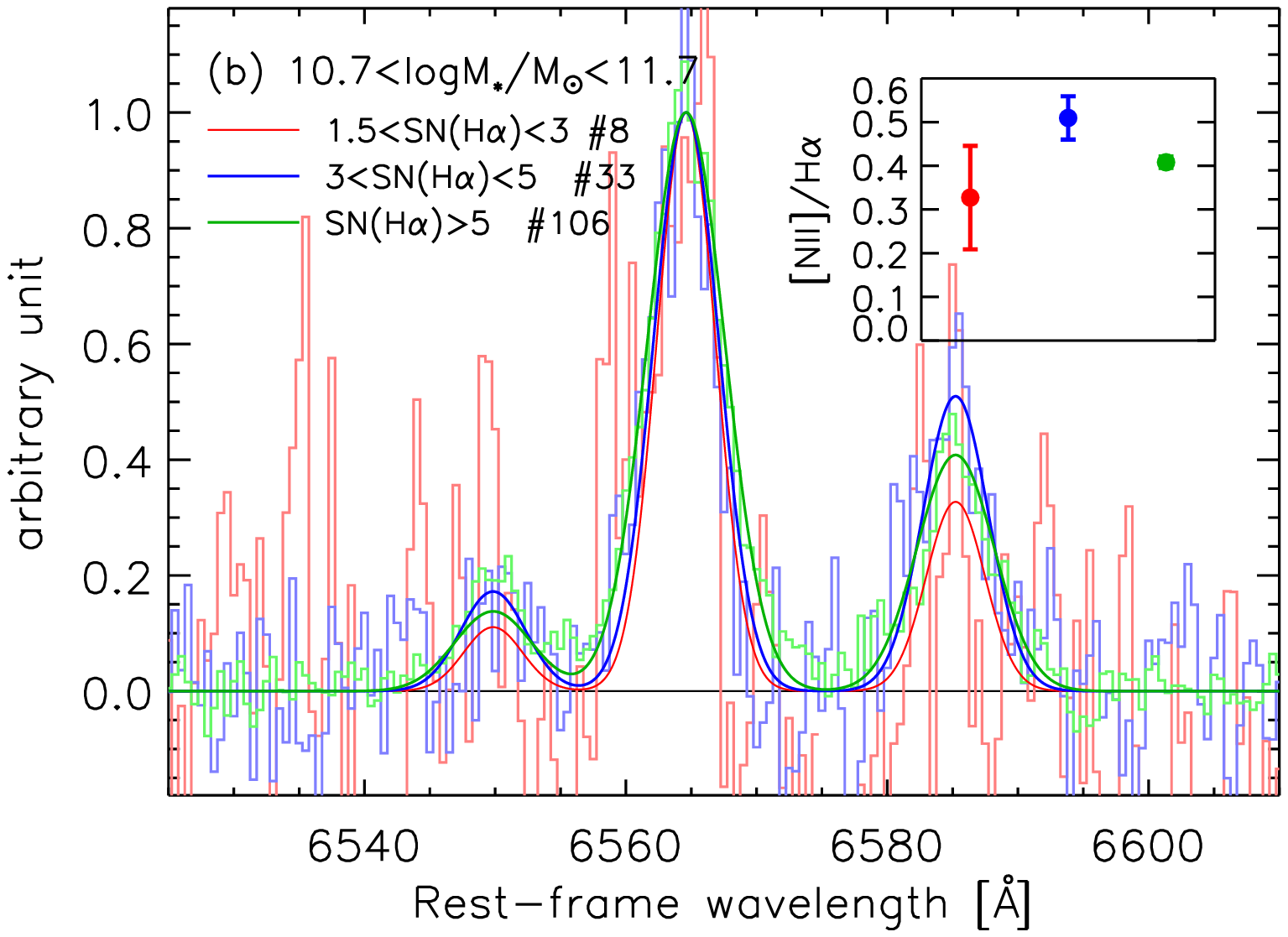}
\caption{An evaluation of the reliability of line identification using the average (stacked) H$\alpha$+[NII] emission line profiles in bins of S/N of the H$\alpha$ line.  The two panels represent different ranges in stellar mass.  Best-fit gaussian profiles are shown and used to match the amplitude of the H$\alpha$ emission line.  The inset figure is a measure of the flux ratio of [NII] to H$\alpha$ based on the line fit.  The sample size for each category is provided.}
\label{stacked_spec}
\end{figure}

\subsection{Stacked FMOS H-long spectra}
\label{sec:stack}
We further assess the reliability of our spectroscopic redshifts by generating average (stacked) H-long spectra.  The idea is to check whether the stacked spectra of those with H$\alpha$ detections of low S/N show a clear emission line profile of [NII]$\lambda6585$.  If there are a significant number of false H$\alpha$ detection, there should be a noticeable degradation in the strength of the accompanying [NII] line.  

In Figure~\ref{stacked_spec}, we compare average spectra split into three bins of S/N and two bins of stellar mass.  All binned spectra have the amplitude of the H$\alpha$ line set to 1.  To aid in our assessment, we plot the ratio [NII]/H$\alpha$ (see insert) for each average spectrum.  It is reassuring that there is a clear detection of [NII] for each bin of stellar mass and S/N thus indicating a high level of accuracy for our redshifts, especially those with H$\alpha$ detections at $S/N> 3$.  There is not much significant difference between the lines ratios for the two bins with S/N $>$ 3.  We do find that for the lowest mass bin (Fig.~\ref{stacked_spec}$a$) there is a significant drop in the strength of the [NII] line by about 50\% for the low S/N bin (1.5 $<$ S/N $<$ 3).  The drop in the higher mass bin is less severe ($\sim20\%$).  From this, we can conclude that the lowest S/N class likely includes false identifications.  If significant [OIII]$\lambda5008$ emission lurks within our stacked H$\alpha$ profile, we should find evidence for the [OIII]$\lambda4960$ line.  Upon inspection of our stacked spectra, there is no signs of such emission thus indicating that a large fraction of misidentifications is unlikely.  We also note that there is no difference in the mean E$_{star}$(B-V) between these bins thus lending support that the change seen here in line ratio [NII]/H$\alpha$ is not impacted by variations in dust that can manifest itself as a metallicity dependence.  We conclude by stressing that the redshifts having a quality flag of 1 in our catalog likely have some misidentifications and should be treated with care.

\subsection{Comparison with zCOSMOS deep}

We compare the FMOS spectroscopic redshifts to those from the zCOSMOS Deep survey \citep{Lilly2007}.  The zCOSMOS survey targets sBzK galaxies over our redshift range of interest and higher.  There are 383 galaxies in our full sample (SED, sBzK, $Herschel$-PACS, and low-mass galaxies) that have been observed by both programs.  Of the 383 matches, 160 have a spectroscopic redshift measured by both programs.  For this comparison, we select only those galaxies having a FMOS measurement based on an H$\alpha $ detection and a secure redshift available through zCOSMOS (zc\_flag $\ge$ 3).  This results in a sample of 35 galaxies to assess the reliability of our FMOS spectroscopic redshifts.  In Figure 20, we show the comparison for FMOS galaxies split by the FMOS quality flag.  Overall, there is very good agreement for the majority (32 out of 35, 91\%).  In particular, our most secure FMOS sample (Flag = 4) shows complete agreement.  If the zCOSMOS redshifts are correct, it appears that we have misidentified [O{\sc iii}]$\lambda5008$ as H$\alpha$ for two cases, only 6\% (2 out of 35) of the sample with flag equal to one or higher.

\begin{figure}
\epsscale{1.2}
\plotone{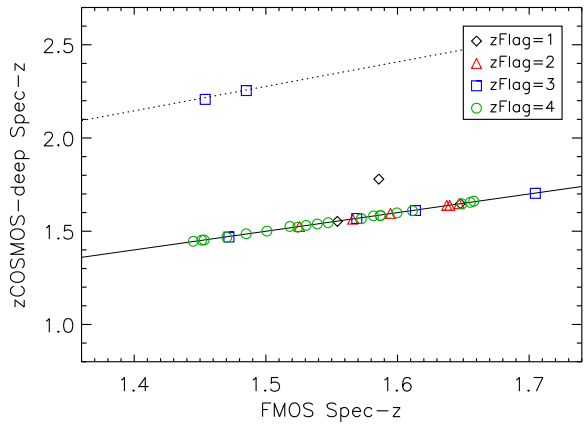}
\caption{Comparison of spectroscopic redshifts between FMOS-COSMOS and zCOSMOS Deep surveys.  FMOS redshifts are assessed separately by their quality flag as indicated.  Only zCOSMOS deep redshifts having a secure (flag $\ge$ 3) redshift are used for this comparison.  The solid line represents a unity relation while the dashed line identifies where measurements will lie if [OIII]$\lambda$5008 was mistaken for H$\alpha$.}
\label{zcompare1}
\end{figure}

\begin{figure}
\epsscale{0.55}
\plotone{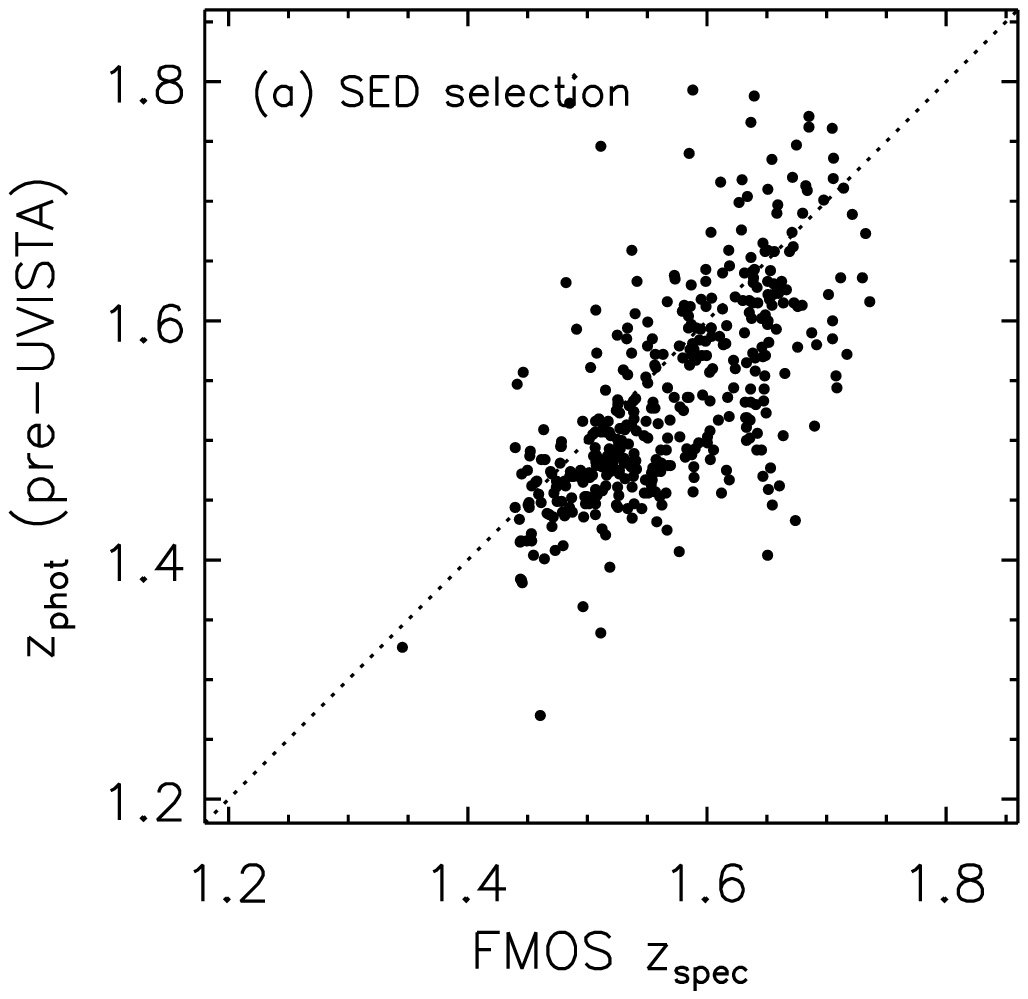}
\plotone{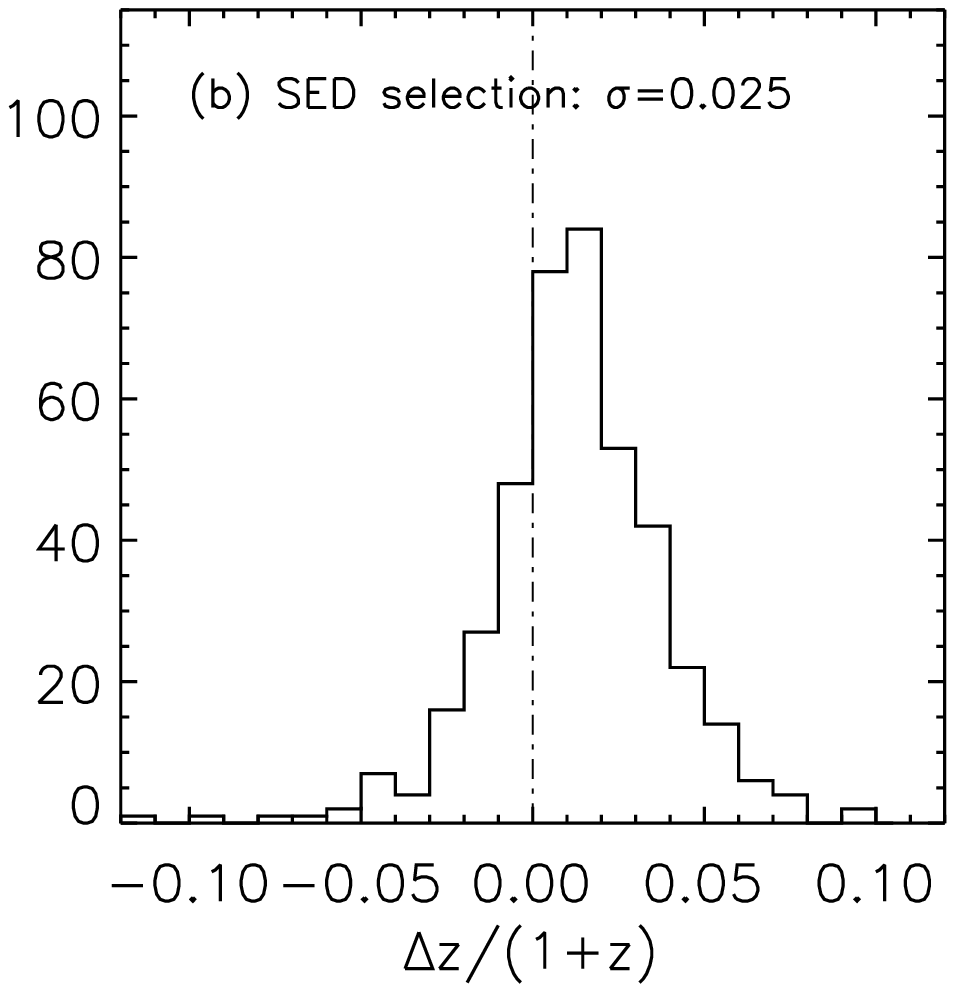}
\plotone{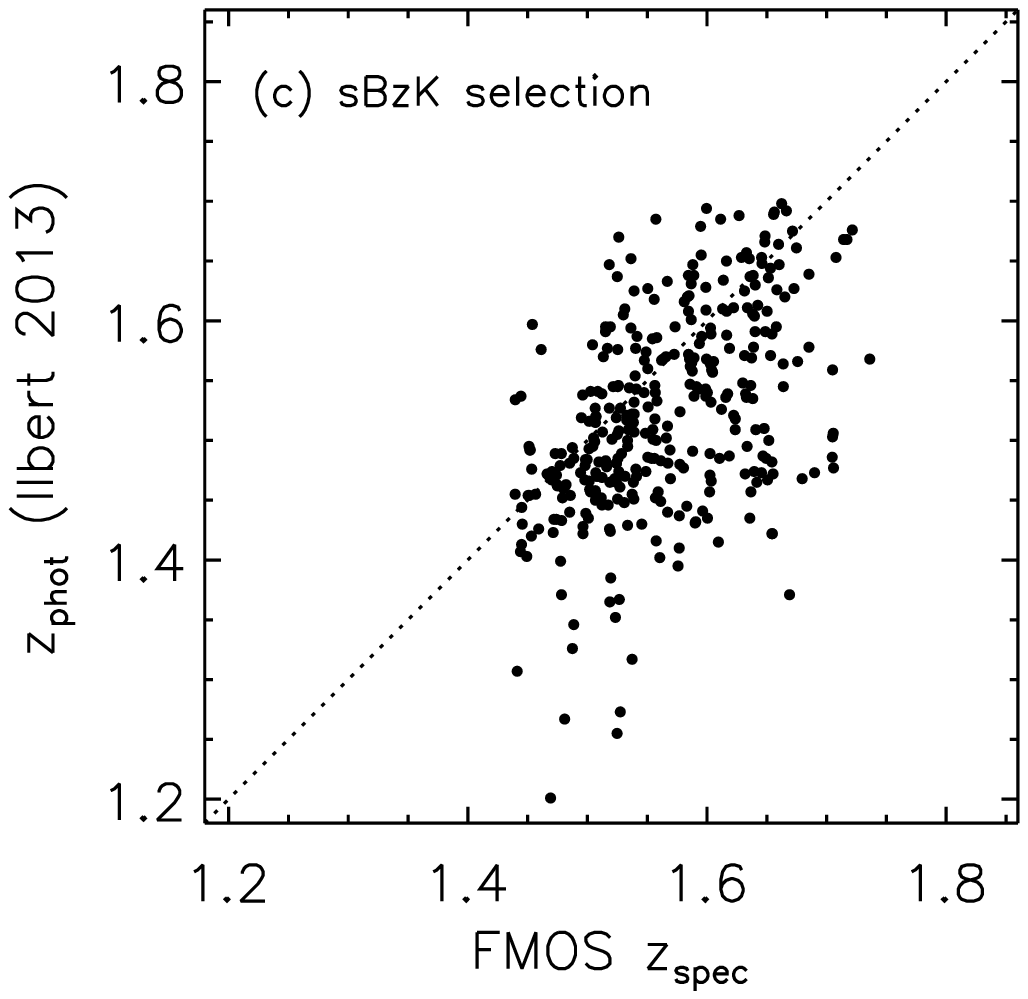}
\plotone{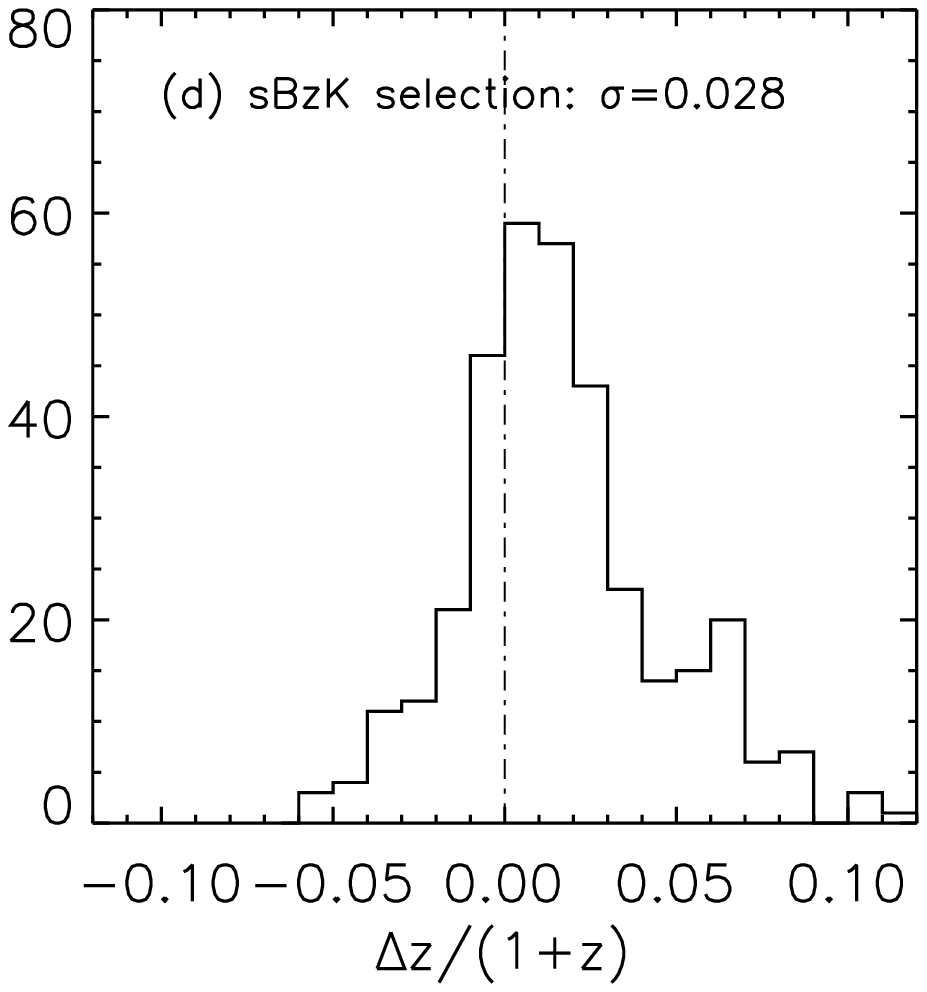}
\caption{Comparison of spectroscopic and photometric redshifts.  The top two panels are based on the SED-selected sample with the latest photometric redshifts \citep{Ilbert2013}.  The bottom two panels make the same comparison using the sBzK sample and photometric redshifts derived prior to the availability of the UltraVISTA photometry.  For both histograms, the quantity on the abscissa is $(z_{spec}-z_{phot})/(1+z_{spec})$ with the standard deviation, including the clipping of 5$\sigma$ outliers, of the distribution given in the respective panel for each distribution.}
\label{zcomp}
\end{figure}

While we are especially interested in the fraction of FMOS galaxies with `insecure' (flag=1) redshifts that are likely to be correct, the number of such targets having secure zCOSMOS redshifts is small.  Even so, out of the three such sources, two of them have FMOS redshifts in agreement with zCOSMOS.  While also considering the prior stacking analysis (Section~\ref{sec:stack}), we conclude that a fair fraction of the redshifts having a low quality flag are likely correct.

It is important to highlight that the FMOS and zCOSMOS spectroscopic programs are complementary.  To illustrate this, we have 130 (97) FMOS redshifts and flag $\ge$ 2 (3) that lack a secure zCOSMOS redshift (zc\_flag $\ge$ 3).  On the other hand, the zCOSMOS deep survey has succeeded in measuring a redshift for 26 (79) galaxies with zc\_flag $\ge$ 3 (2) for which we failed with FMOS (flag $<$ 2).  In total, the FMOS program has a success rate of 44$\%$ (see Section 9.2) as compared to 36$\%$ (106/291) for the zCOSMOS deep program over the redshift interval $1.4\lesssim z \lesssim 1.7$.  These success rates clearly exemplify the challenges with working at a difficult redshift regime either with UV spectra that lack strong features or the NIR where the sky emission is challenging to overcome.

\subsection{Accuracy of the photometric redshifts}

We test the accuracy of the photometric redshifts using our FMOS spectroscopic catalog.  In Figure~\ref{zcomp}, we present such a comparison using the two photometric redshift catalogs \citep{Ilbert2009,Ilbert2013} generated before and after the availability of the UltraVISTA NIR photometry.  In the top panels, we present the latest photometric redshifts from the recent catalog v1.2 of \citet{Ilbert2013}.  While the input target catalog for FMOS followup has been fixed with photometric redshifts from the v1.0 catalog, there is little difference between the two (v1.0 vs. v1.2) that are relevant for our purposes.  We find that there is very good agreement between the two with $\sigma(\Delta z/1+z) = 0.025$, marginally reduced from the value of 0.028 found with the previous photo-z catalog.  This is similar to the accuracy reported in \citet{Ilbert2013} using a smaller subset of the FMOS sample.  We still see a small shift of the distribution towards higher spectroscopic redshifts by $\Delta z/(1+z)\sim 0.01$.  It should be noted that such level of accuracy is expected since the photometric redshifts have been trained using the initial spectroscopic redshift catalog from the FMOS observations and the identification of an H$\alpha$ emission line is aided by knowledge of the photometric redshifts in some cases.  Furthermore, an improvement in the accuracy of the photometric redshifts between the 2009 and 2013 catalogs is evident based on the distribution of offsets being less asymmetric when comparing panels $b$ and $d$.

\section{High-level data products and catalog}
\label{sec:catalog}
 
We provide to the community a set of measurements that includes a spectroscopic redshift for each galaxy and emission-line properties (Table~\ref{tab:catalog}).  As detailed above, the spectroscopic redshift is primarily determined from centroid of the H$\alpha$ emission line.  In a fair number of cases, a confirmation is available through our own followup spectroscopy at lower NIR wavelengths (J-long) or optical spectra from the zCOSMOS deep program.  Each redshift is accompanied by a quality flag indicative of a level of assurance based on the FMOS observations.  Here, we provide a description of each item listed in the catalog.

\begin{itemize}

\item Column 1 (ID): Identifier

\item Columns 2 - 3: Right Ascension and Declination (J2000).

\item Column 4: Spectroscopic redshift .

\item Column 5: Quality flag on the spectroscopic redshift measurement based on the S/N of the H$\alpha$ detection and corroborative information as described in Section~\ref{sec:zspec}.  

\item Columns 6 - 7: H$\alpha$: observed flux (no aperture correction) and S/N; units of erg s$^{-1}$ cm$^{-2}$.

\item Columns 8 - 9: H$\alpha$: FWHM and error (1$\sigma$); units of km s$^{-1}$.

\item Columns 10 - 11 [NII]$\lambda6585$; observed flux (no aperture correction) and S/N; units of erg s$^{-1}$ cm$^{-2}$.

\item Column 12: Correction factor to compensate for the effect of the aperture size and should be multiplied to the flux measurements of H$\alpha$ and [NII] reported above.

\item Columns 13 - 14 H$\beta$; observed flux (no aperture correction) and S/N; units of erg s$^{-1}$ cm$^{-2}$.

\item Columns 15 - 16 [OIII]$\lambda5008$; observed flux (no aperture correction) and S/N; units of erg s$^{-1}$ cm$^{-2}$.

\item Column 17: Correction factor to compensate for the effect of the aperture size and should be multiplied to the flux measurements of H$\beta$ and [OIII] reported above.

\item Column 18: zCOSMOS spectroscopic redshift (if available)

\item Column 19: zCOSMOS quality flag on spectroscopic redshift measurement.

\end{itemize}   

In addition to the catalog, we make available the 1D and 2D spectra in fits format\footnote{All high-level data products can be found at http://member.ipmu.jp/fmos-cosmos/.}.  Individual spectra (object + error) in fits format are available for every observed galaxy irrespective of whether a redshift measurement was attained.  A larger measurement set is available at the public web address but omitted here to provide the quantities of most interest.

\section{Emission-line characteristics of the FMOS-COSMOS sample}

\begin{figure}
\epsscale{2.2}
\plottwo{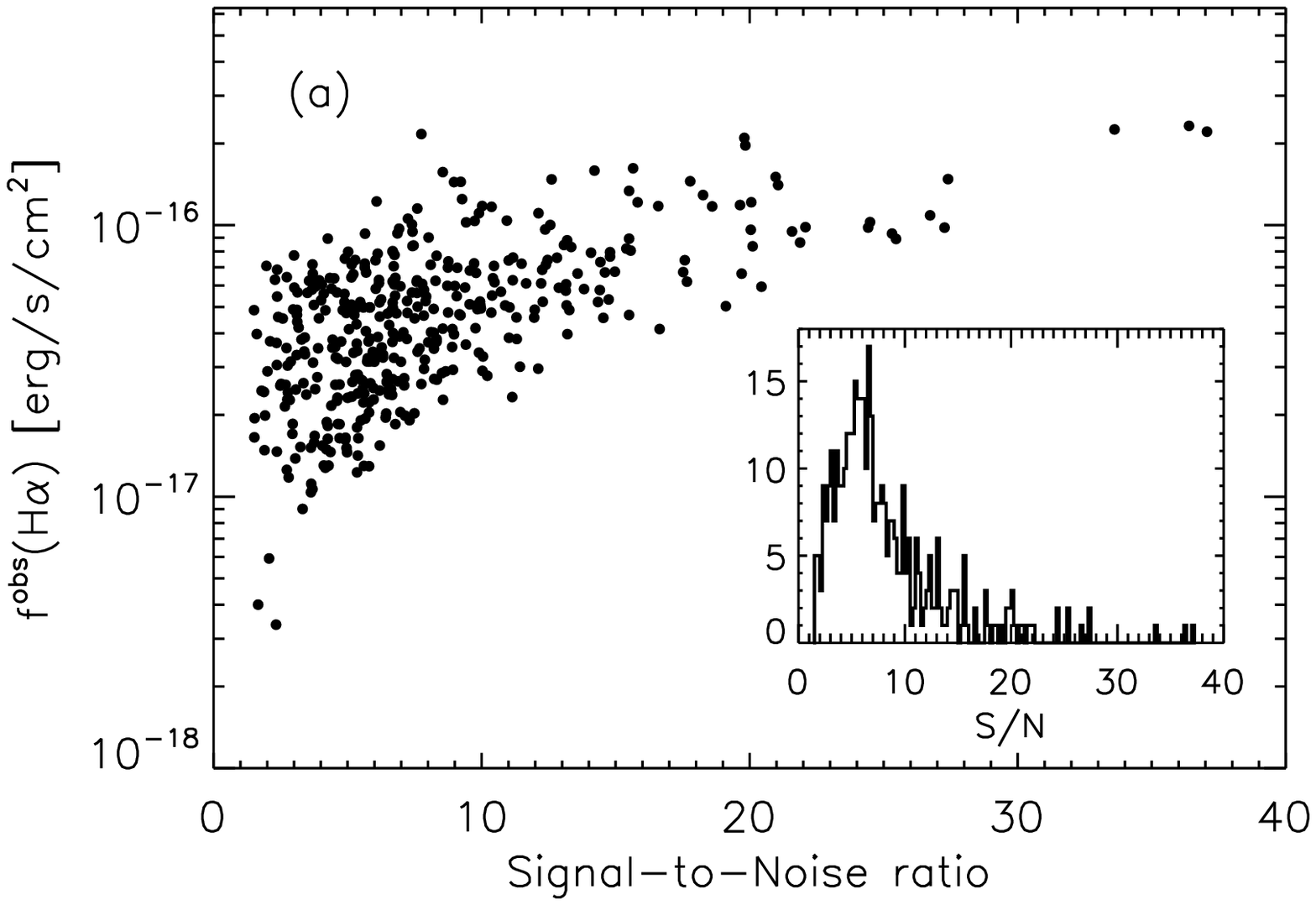}{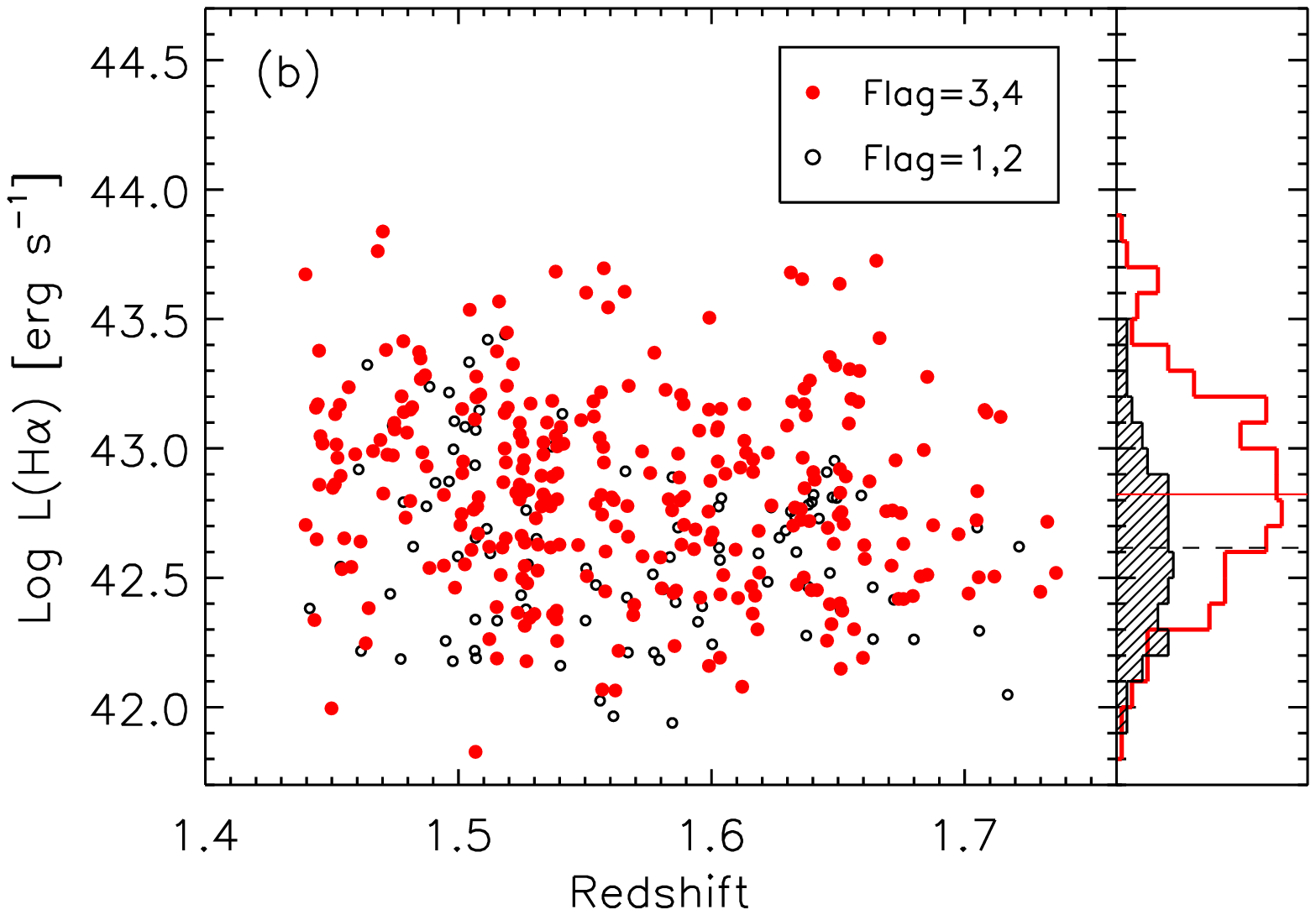}
\caption{H$\alpha$ line strengths: $(a)$ observed flux (not corrected for the aperture size) as a function of S/N for individual galaxies.  $(b)$ logarithm of the luminosity corrected for extinction as a function of spectroscopic redshift.  In the lower panel, points are colored based on their S/N.  The median values are indicated on the right by the horizontal lines that match by color the distributions shown.}
\label{sec:result1}
\end{figure}

\begin{figure}
\epsscale{1.1}
\plotone{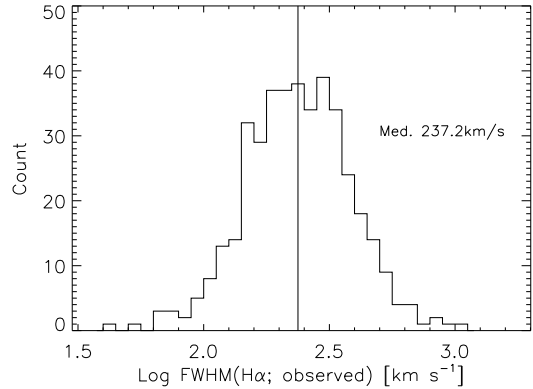}
\caption{Distribution of the observed log FWHM of the H$\alpha$ line (units of km s$^{-1}$).}
\label{ha_fwhm}
\end{figure}

\subsection{H$\alpha$}
\label{sec:results}

We present the emission-line properties of our sample of star-forming galaxies by starting with H$\alpha$.  In Figure~\ref{sec:result1}$a$, we plot the observed line flux and corresponding S/N estimate for each galaxy in our sample.  The emission-line fluxes are those as detected by FMOS which means that there has been no additional correction for the aperture size.  Here our intention is to demonstrate the performance of the instrument, limits of our survey and quality of the individual detections.  Noise estimates are determined from the error on the fit parameters output by the fitting routine (see Section~\ref{sec:efit}).  

Emission lines are detected down to flux levels of $\sim10^{-17}$ erg s$^{-1}$ cm$^{-2}$ with $S/N\sim3$ at this faint limit.  For targets expected to yield significant detections with $f_{H\alpha}\gtrsim4\times10^{-17}$ erg s$^{-1}$ cm$^{-2}$, it is evident that the majority have $S/N > 3$ thus meeting the expectations of our program.  In Figure~\ref{sec:result1}$b$, the H$\alpha$ luminosities of our sample span a range of $42\lesssim log~L_{\rm H\alpha} \lesssim 43.5$; these estimates include aperture and extinction corrections as discussed in \citet{Kashino2013}.  Not surprising, it is apparent from the inset histogram that the secure sample (flag = 3 or 4) is slightly more luminous than those of having lower quality redshifts (flag = 1 or 2).  We remind the reader that many of the sources with low S/N detections are likely to be correct (see Section~\ref{sec:zspec}).  In addition, we report on the FWHM distribution of H$\alpha$ (Figure~\ref{ha_fwhm}) that has a median value of 237.2 km s$^{-1}$. 

\begin{figure}
\epsscale{1.1}
\plotone{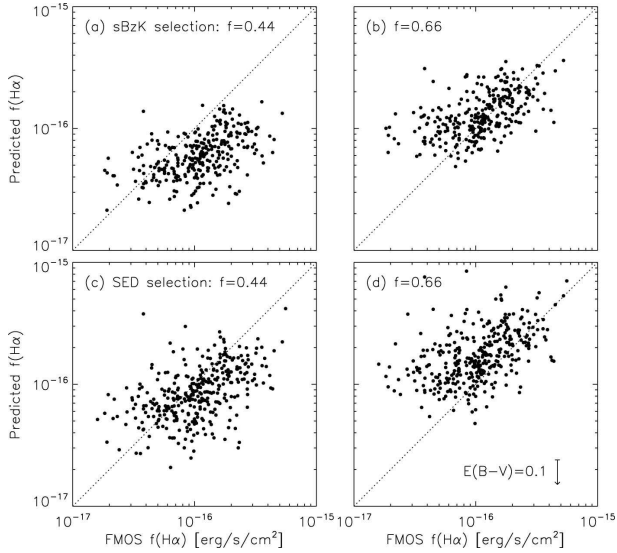}
\caption{Predicted versus observed H$\alpha$ emission-line flux ($f_{H\alpha}$) with both given as galaxy-wide quantities (i.e., including an aperture correction applied to the observed value).  The sample is split into whether the target is included in the sBzK- ($a-b$) or SED-selected ($c-d$) target catalogs.  Predicted values are based on the original $f$-factor of 0.44 ($left~panels$) and our new value $f=0.66$ ($right~panels$).}
\label{sec:result2}
\end{figure}

Our program relies on the ability to accurately predict not only the redshift but also the H$\alpha$ emission line flux to ensure that we effectively target galaxies that will result in a positive line detection given the sensitivity limits of FMOS.  Our aim is to maximize the number of spectroscopic redshifts to improve our characterization of the large-scale distribution of galaxies.  In Figure~\ref{sec:result2}, we indicate how well we can predict the line flux from the UV and broad-band photometry that includes a correction for extinction for galaxies for both the sBzK- and SED-selected parent catalogs.  The two columns represent a different factor implemented to convert extinction of stellar light to that of the nebular regions (panel $a, c$: $f=0.44$; panel $b, d$:$ f=0.66$).  The higher $f$-factor was determined by minimizing the offset between the predicted and observed fluxes using an early subset of the sBzK-selected galaxies shown in panels $a$ and $b$.  This is a crude approximation of the more proper exercise of directly comparing the extinction to the nebular regions (based on the Balmer decrement) to the color excess of the stellar light as carried out in \citet{Kashino2013} that resulted in a comparable value of 0.69.  As discussed in \citet{Kashino2013}, this value is dissimilar to the canonical factor of 0.44 (typically used for local starburst galaxies) that resulted in predictions of emission line flux for the preliminary FMOS sample that were substantially lower than their observed strengths (Fig.~\ref{sec:result2}$a$).  To remedy this discrepancy, we argue that less extinction to the nebular regions, given their stellar extinction, is evident thus a change in the $f$-factor appears to be effective.  After the pilot survey, target selection for all observations incorporated this change in the $f$-factor.  We remark that the similarity between the stellar and nebular extinction is applicable for the sample observed with FMOS that resulted in positive line detections.  Whether this holds for the fainter and/or more obscured star-forming population remains to be determined.

Furthermore, we confirm that a similar ratio of nebular-to-stellar extinction is applicable for the SED-selected sample (Figure~\ref{sec:result2}~panels~$c-d$).  Based on $f=0.44$, the predicted estimates fall short as compared to the observed values (panel $c$).  With the higher $f$-factor, the predicted flux estimates appear to overshoot the one-to-one relation (panel $d$).  We attribute this discrepancy to the fact that the broad-band SED fitting results in a slight underestimate of the color excess (Figure~\ref{sfr_ext_comparison}).  We indicate in Figure~\ref{sec:result2}$d$ the magnitude (0.1) of the offset in $E_{\rm neb}(B-V)$.

\begin{figure}
\epsscale{2.3}
\plottwo{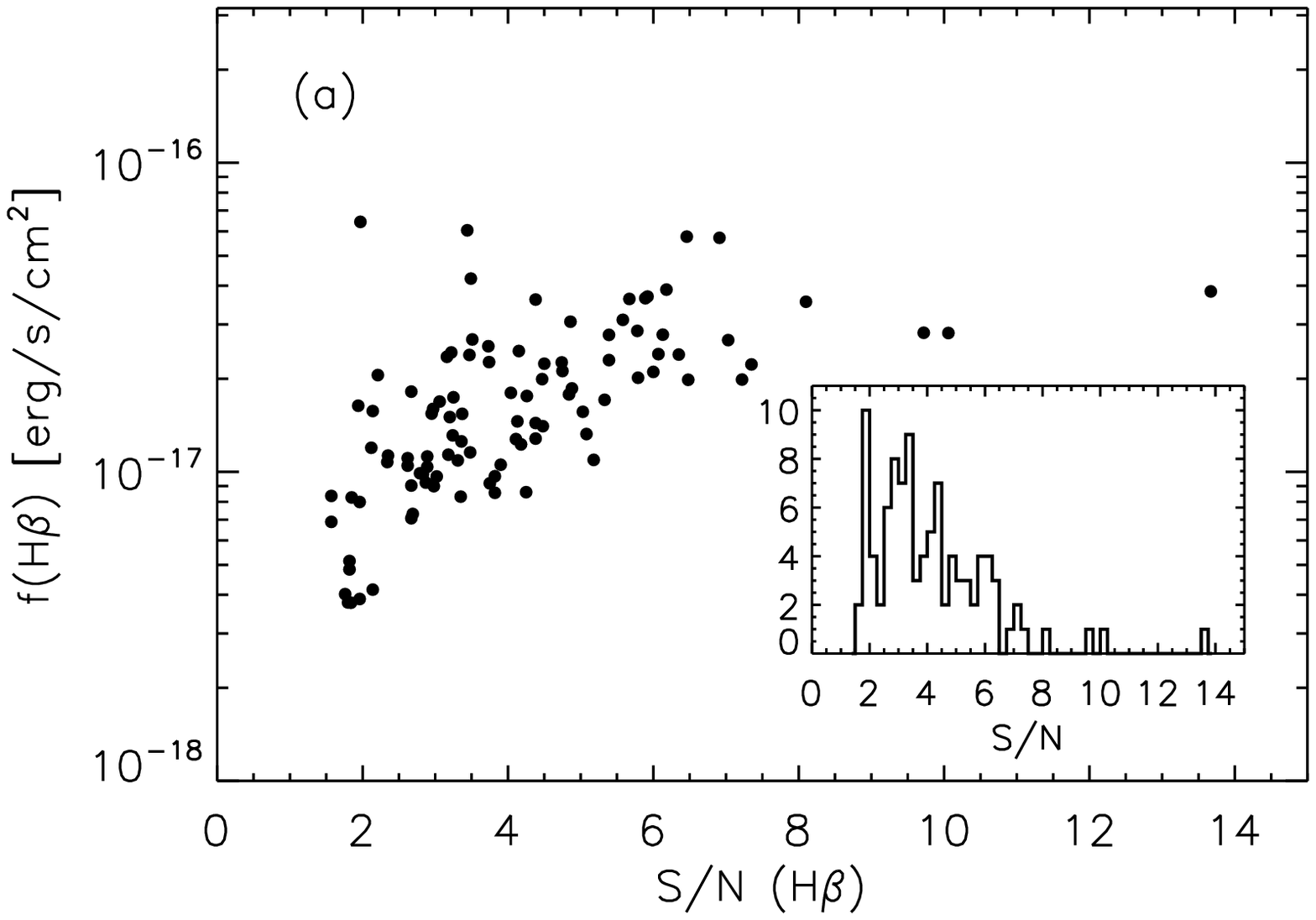}{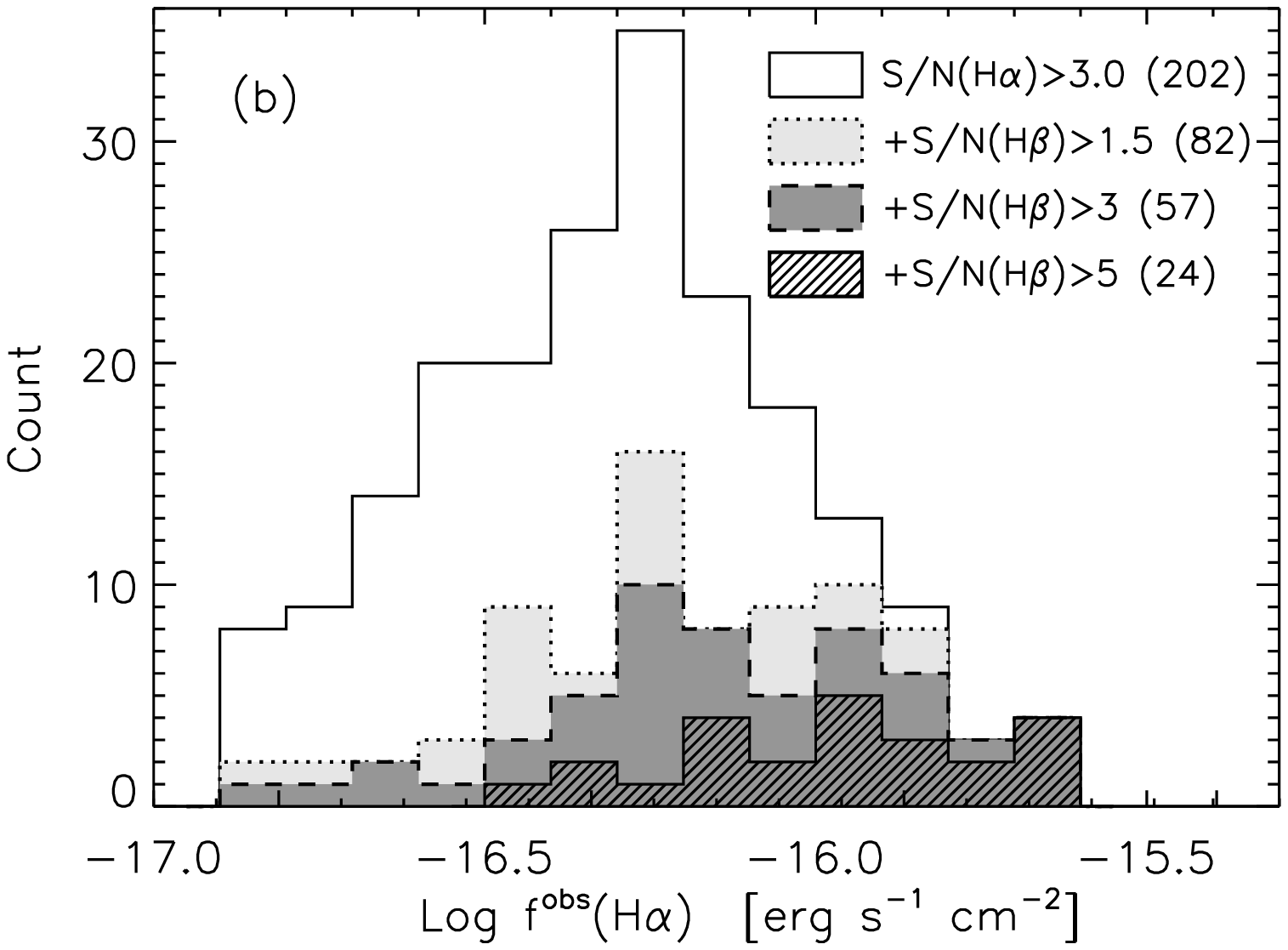}
\caption{H$\beta$ line strengths: $(a)$ Logarithm of the observed flux (not corrected for the aperture size) as a function of S/N of the line detection.  $(b)$ Number distribution of galaxies with H$\alpha$ detections with $S/N>3$ and J-long coverage.  Additional histograms indicate those that resulted in individual H$\beta$ detections above a S/N as indicated.} 
\label{hbeta}
\end{figure}

\subsection{Individual H$\beta$, [OIII]$\lambda$5008 and [NII]$\lambda$6585 detections}
\label{sec:otherlines}
With an extension of our NIR spectral coverage to the $J$-band, we detect key diagnostic emission lines including H$\beta$ for a number of individual cases with rates given in Table~\ref{lines_table}.  As previously described, these galaxies are targeted due to their existing H$\alpha$ detections, from previous FMOS observations, that yields a spectroscopic redshift thus ensuring H$\beta$ to fall within the FMOS J-long window.

In Figure~\ref{hbeta}$a$, we plot the flux (not corrected for the aperture size) of the H$\beta$ line for individual detections that constitute 21\% of our H$\alpha$-detected sample.  This success rate is clearly indicative of the challenges with detecting H$\beta$ at high redshift especially for the population with faint lines.  Such difficulty can be seen by comparing the flux levels and S/N estimates of those of H$\beta$ (Fig.~\ref{hbeta}$a$) with H$\alpha$ (Fig.~\ref{sec:result1}$a$).  The H$\beta$ fluxes are over a factor of four fainter than the H$\alpha$ fluxes.  Even so, it is evident that we are able to detect H$\beta$ with $S/N>2$ down to a limit $\sim10^{-17}$ erg s$^{-1}$ cm$^{-2}$.  This depth is consistent with expectation given the long integration times ($\sim3-4$ hours) and throughput estimates of FMOS.  While some detections reach fainter flux levels, these are mostly at $S/N \sim 2$.     

We further illustrate the incompleteness with respect to detecting H$\beta$ in Figure~\ref{hbeta}$b$.  Here, we plot the number distribution of galaxies observed with the J-long grating because H$\alpha$ was previously detected.  We include a subset of our sample with a $S/N>3$ (202 galaxies) for the detection of H$\alpha$ (solid histogram).  The distribution is shown as a function of the logarithm of H$\alpha$ flux (not corrected for the aperture size).  From this sample, we then identify those with positive H$\beta$ detections having a S/N greater than 1.5, 3, and 5 each with a different shading.  Above $S/N$ of 1.5 (3), we have detections of H$\beta$ for 21 (22)\% of the FMOS sample.  Our low success rate for detecting H$\beta$ is also due to the fact that we did not neglect to target galaxies for which H$\beta$ may be significantly ($\gtrsim$50\%) blocked by the OH suppression mask.  In such cases, there would be the possibility to detect [OIII]$\lambda$5008.  Given the low level of individual detections, we resort to using stacked spectra for much of our scientific analysis and are pursuing deeper J-band spectroscopic followup to increase the sample with significant individual detections.   

\begin{figure}
\epsscale{1.4}
\plotone{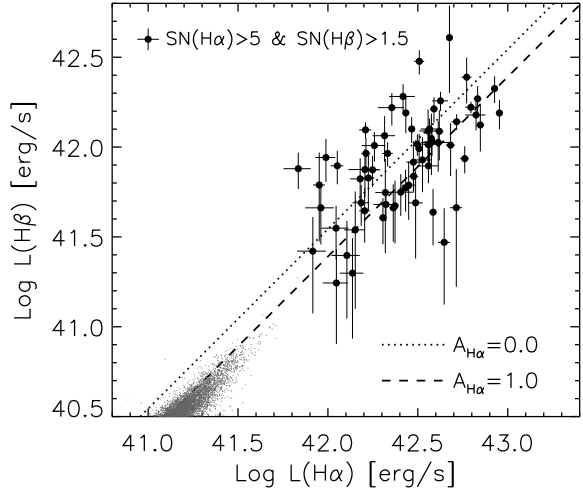}
\caption{Emission-line luminosities (H$\alpha$ versus H$\beta$) of our secure sample (H$\alpha$ S/N $>$ 5; H$\beta$ S/N $>$ 1.5; OH impact $<$ 50\%).  For reference, we plot values from SDSS as small grey dots.  We indicate the expected values with ($A_{H\alpha}=1$) and without extinction.}
\label{lum_ha_hb}
\end{figure}

In Figure~\ref{lum_ha_hb}, the luminosities of H$\alpha$ and H$\beta$ (both corrected for the aperture size) are compared against each other for galaxies with H$\alpha$ (S/N $>$ 5) and H$\beta$ (S/N $>$ 1.5).  While many sources are spread about a relation, expected for an $A_{H \alpha}\sim1$, there are galaxies with a ratio of H$\alpha$/H$\beta$ lower than the theoretical value of 2.86 (assuming case B recombination with no extinction).  This is likely an indication that there is some uncertainty in the H$\alpha$, or more likely, the H$\beta$ line fluxes for individual detections that may be further affected by being observed separately possibly under varying sky conditions that result in having to apply different aperture corrections with their own level of uncertainty.

\begin{figure}
\epsscale{2.3}
\plottwo{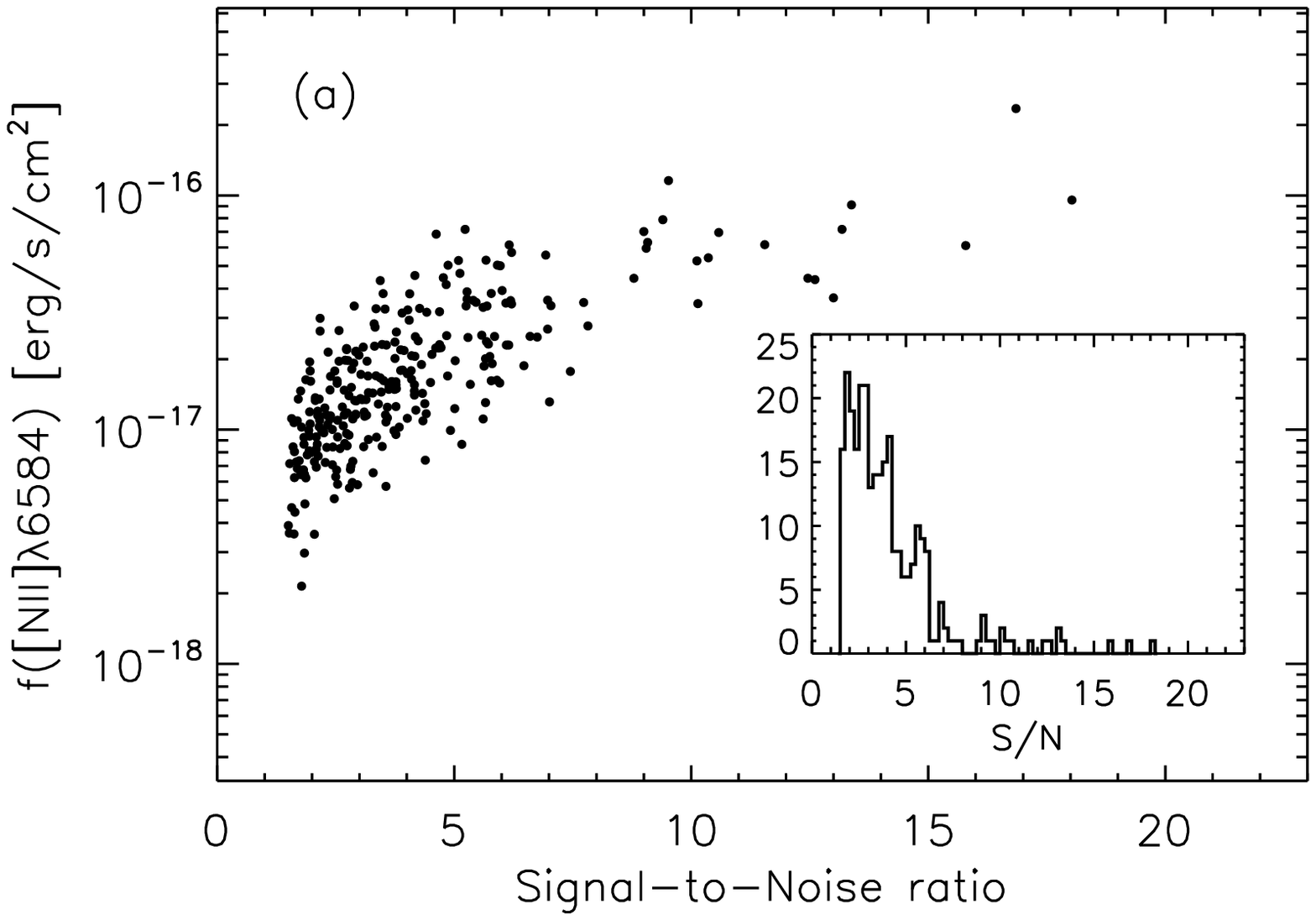}{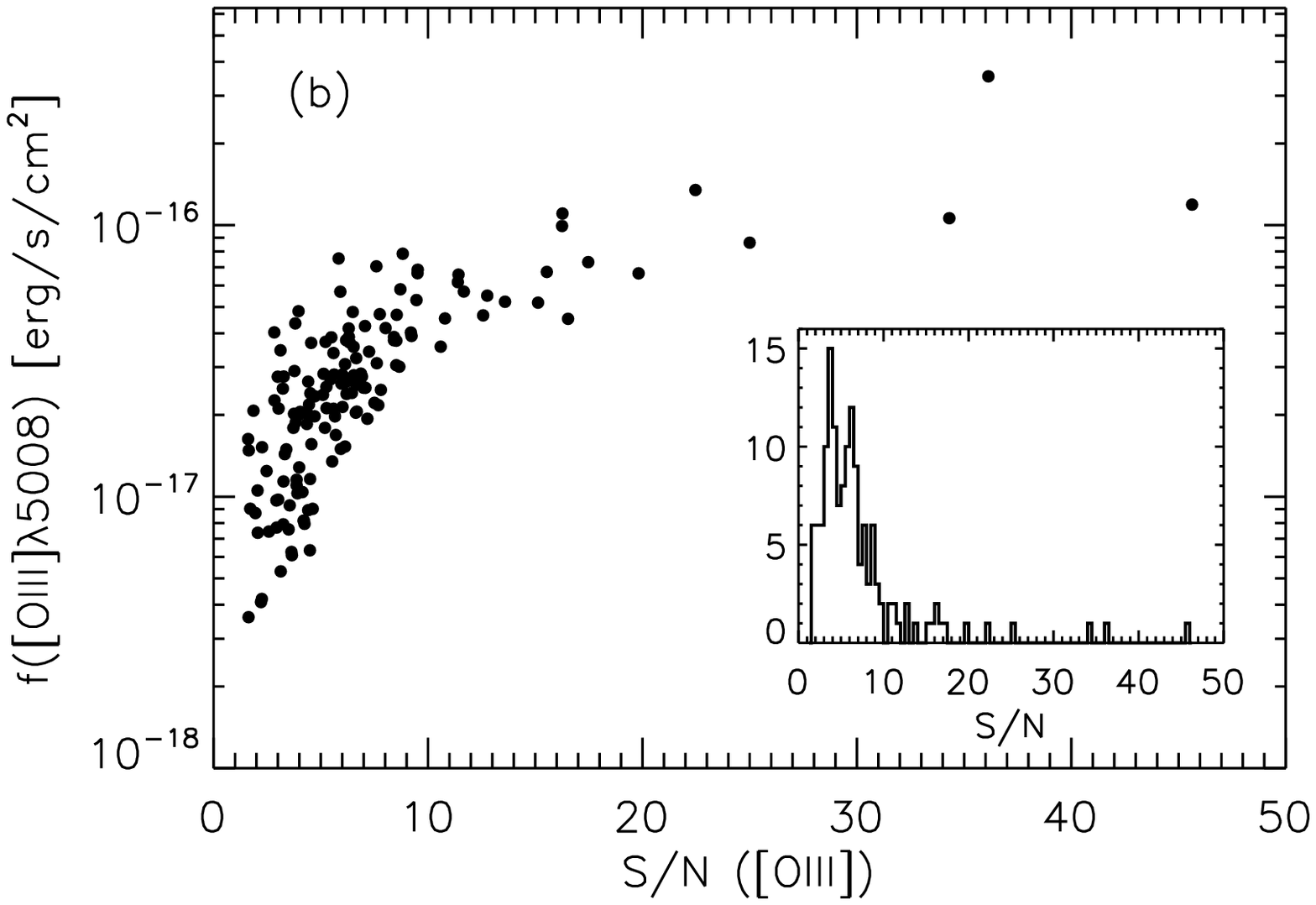}
\caption{Emission-line strength (not corrected for the aperture size) versus S/N of each detections for  [NII]$\lambda$6585 ($a$) and [OIII]$\lambda$5008 ($b$).}
\label{n2_o3}
\end{figure}

Other lines of interest, namely [NII]$\lambda$6585 and [OIII]$\lambda$5008, are shown in Figure~\ref{n2_o3}.  While [NII] is present in many spectra with secure H$\alpha$ detections (62\%; Table~\ref{lines_table}), the significance of individual detections is fairly low with most having $2< S/N < 5$ as a result of their having faint line fluxes characteristic of galaxies with low metallicity \citep{Zahid2014b}.  As expected, the strength of the [OIII] lines are lower than those of H$\alpha$ but do include a handful of bright cases ($f_{[OIII]}> 1\times10^{-16}$ erg s$^{-1}$ cm$^{-2}$) at high S/N.  Both of these lines permit a study of the ionization conditions in the ISM using line ratio diagnostic ([NII]/H$\alpha$, [OIII]/H$\beta$).  While such studies will be fully explored in subsequent works, we have reported higher ionization levels of the ISM for galaxies in our sample \citep{Zahid2014b,Kartaltepe2015} as compared to those at low redshift from SDSS.

\section{Summary}

The FMOS multi-fiber NIR spectrograph mounted on the Subaru Telescope is a unique instrument at this time.  The ability to carry out spectroscopic surveys of distant galaxies ($z>1$) by detecting key rest-frame optical diagnostic emission lines over a wide sky area opens a new window on our understanding of galaxy evolution.  In particular, preliminary results have already shed new light on stellar mass growth, chemical enrichment of the ISM and role of AGNs.  

Here, we present the details of a Subaru Intensive Program with FMOS to carry out a large galaxy survey in the central area of the COSMOS field aimed at acquiring a sample of over one thousand galaxies with spectroscopic redshifts over a previously challenging redshift regime.  We specifically focus on the redshift window $1.4 \lesssim z \lesssim 1.7$ for which we can detect the H$\alpha$ emission line using the H-long grating that covers a wavelength range $1.6<\lambda<1.8 \mu$m.  Followup observations using the J-long grating ($1.11<\lambda<1.35 \mu m$) allow us to complete the spectral coverage in the NIR to detect all four key diagnostic lines (H$\alpha$, H$\beta$, [NII]$\lambda$6585 and [OIII]$\lambda$5008) required to assess the dust content, metallicity and ionization state of our high redshift galaxies.    

We discuss in full detail the selection function for the star-forming galaxy sample that spans the full main sequence at stellar masses above $10^{9.8}$ M$_{\odot}$.  The selection is based on the expectation to detect the H$\alpha$ emission line.  This puts constraints on the extinction properties of the sample by excluding the most heavily dust-obscured galaxies.  To alleviate this bias, we specifically target galaxies detected by $Herschel$ that lie on and off the star-forming main sequence.  In addition, we observe $Chandra$ X-ray selected AGNs (both type 1 and 2) to study black hole accretion at a cosmic epoch of maximal growth.  Both $Chandra$ and $Herschel$ sources will be presented in subsequent studies.  This paper is mainly focused on presenting a detailed analysis of the sample, information on the observing program, data analysis, redshift measurements and characteristics of the observed and final sample with spectroscopic redshifts.  Finally, we provide data products (i.e., individual spectra in fits format both 1D and 2D) and a catalog of the first installment of our sample that includes 1153 entries and 460 spectroscopic redshifts with a quality flag and emission line properties such as line flux and width (see http://member.ipmu.jp/fmos-cosmos/).

Our broad aim for the program is to build a large enough spectroscopic sample in the COSMOS field to provide a map of large-scale structure at $z\sim1.6$ and elucidate whether the environment plays a role in shaping the properties of galaxies as seen at low redshift.  Furthermore, the rich multi-wavelength resources (i.e., HST, $Spitzer$, VLA, $Chandra$, etc) of the COSMOS field provide a wealth of information to supply the means to answer many fundamental questions as to how galaxies and supermassive black holes grow, possibly in concert, with cosmic time.  With FMOS, we have started to open this spectral window not yet fully explored and already being complemented with results coming from other multi-aperture NIR instruments such as Keck/MOSFIRE, and the VLT/KMOS.  The wide-field capability of FMOS will not be matched for a few years until the operation of a new generation of multi-fiber spectrographs such as MOONS, or Prime-Focus-Spectrograph \citep[PFS;][]{Takada2014}.

\acknowledgments

JDS is supported by JSPS KAKENHI Grant Number 26400221 and the World Premier International Research Center Initiative (WPI), MEXT, Japan.  DK receives support through the Grant-in-Aid for JSPS Fellows (No. 26-3216).  AR acknowledges support by a grant PRIN-INAF 2012.



Facilities: \facility{Subaru(FMOS)}.

\nocite{*}
\bibliography{myrefs}



\clearpage

\begin{turnpage}
\begin{table}[htdp]
\caption{Observed FMOS-COSMOS galaxies and measurements\label{tab:catalog}}
\begin{center}
\tiny
\begin{tabular}{lllllllllllllllllll}
\hline \hline
ID & RA & DEC & $z_\mathrm{spec}$ & zFLAG & $f_\mathrm{H\alpha}$\tablenotemark{a} & H$\alpha$ & FWHM &Error & $f_\mathrm{[NII]}$\tablenotemark{a} & [NII] & Aper. & $f_\mathrm{H\beta}$\tablenotemark{a} & H$\beta$ & $f_\mathrm{[OIII]}$\tablenotemark{a} & [OIII] & Aper. & zCOS & zCOS \\
& (J2000) & (J2000) &&&  & S/N & (km s$^{-1}$) &FWHM & & S/N & correct\tablenotemark{b}& & S/N & & S/N & correct\tablenotemark{b} & $z_\mathrm{spec} $ & flag \\
\hline
FMOS\_J095943.6+014211 & 149.931790 & 1.70292 & 1.671 & 4 &   3.148 &  5.73 &  325.84 &   56.48 &   1.038 &  2.66 & 2.15 &         &       &  -3.742 &       & 3.06 &       &     \\
FMOS\_J095934.7+014251 & 149.894500 & 1.71419 &       & 0 &         &       &         &         &         &       &      &         &       &         &       &      &       &     \\
FMOS\_J095936.5+014251 & 149.902000 & 1.71425 &       & 0 &         &       &         &         &         &       &      &         &       &         &       &      &       &     \\
FMOS\_J095925.2+014254 & 149.854830 & 1.71506 &       & 0 &         &       &         &         &         &       &      &         &       &         &       &      &       &     \\
FMOS\_J095933.0+014258 & 149.887290 & 1.71603 &       & 0 &         &       &         &         &         &       &      &         &       &         &       &      &       &     \\
FMOS\_J095956.4+014311 & 149.985130 & 1.71967 &       & 0 &         &       &         &         &         &       &      &         &       &         &       &      &       &     \\
FMOS\_J095914.4+014328 & 149.810120 & 1.72444 & 1.504 & 1 &   5.459 &  2.37 &  423.56 &  167.14 &         &       & 1.73 &  -7.891 &       &  -8.459 & & 1.96 &       &    0 \\
FMOS\_J095937.9+014360 & 149.908000 & 1.73325 &       & 0 &         &       &         &         &         &       &      &         &       &         &       &      & 1.612 & 2.5\\
FMOS\_J095910.7+014405 & 149.794750 & 1.73475 & 1.712 & 3 &   5.648 &  7.40 &  181.78 &   30.75 & -13.442 &       & 1.42 &         &       &         &      &      &       & \\    
FMOS\_J095950.1+014409 & 149.958830 & 1.73583 & 1.646 & 4 &   2.042 &  5.80 &  136.35 &   26.61 &   1.515 &  2.83 & 2.08 &   1.438 &  4.38 & -1.871 &       & 3.28 &       &\\     
FMOS\_J100123.3+014409 & 150.346960 & 1.73586 &       & 0 &         &       &         &         &         &       &      &         &       &         &       &      &       &     \\
FMOS\_J095923.9+014412 & 149.849750 & 1.73653 &       & 0 &         &       &         &         &         &       &      &         &       &         &       &      &       &     \\
FMOS\_J100003.3+014419 & 150.013830 & 1.73847 & 1.651 & 4 &   3.707 &  8.01 &  155.93 &   22.02 &   0.947 &  2.78 & 1.81 &         &       &         &    &      &       &\\     
FMOS\_J100114.8+014427 & 150.311790 & 1.74083 &       & 0 &         &       &         &         &         &       &      &         &       &         &       &      &       &     \\
FMOS\_J100009.2+014428 & 150.038460 & 1.74097 &       & 0 &         &       &         &         &         &       &      &         &       &         &       &      &       &    0\\
FMOS\_J100108.0+014428 & 150.283500 & 1.74111 &       & 0 &         &       &         &         &         &       &      &         &       &         &       &      &       &     \\
FMOS\_J095940.2+014450 & 149.917670 & 1.74714 & 1.494 & 4 &   4.658 &  9.21 &  150.04 &   18.79 &   0.840 &  2.31 & 1.97 &         &       &         &    &      &       &\\     
FMOS\_J100120.7+014451 & 150.336330 & 1.74756 &       & 0 &         &       &         &         &         &       &      &         &       &         &       &      &       &     \\
FMOS\_J095953.3+014458 & 149.972130 & 1.74931 &       & 0 &         &       &         &         &         &       &      &         &       &         &       &      & 1.434 & 1.5\\
FMOS\_J100005.2+014501 & 150.021830 & 1.75014 & 1.674 & 3 &   1.642 &  5.40 &   99.08 &   20.71 &  -2.061 &       & 2.19 &         &       &         &       &      &       & \\    
\hline
\tablenotetext{a}{units of $10^{-17}$ erg s$^{-1}$ cm$^{-2}$}
\tablenotetext{b}{Aperture corrections, based on HST I-band (F814W) images, are reported for galaxies that are not significantly blended.}
\end{tabular}
\end{center}
\label{default}
\end{table}

\end{turnpage}
	\clearpage
	\global\pdfpageattr\expandafter{\the\pdfpageattr/Rotate 90}




\end{document}